\documentclass[onecollarge]{svjour2}                    
\smartqed  
\usepackage{graphicx}
\usepackage{color}
\usepackage{amsmath}
\usepackage{amsfonts}
\usepackage{amssymb}
\usepackage{epsfig}
\usepackage{bm}

\newcommand{\be}{\begin{equation}}
\newcommand{\ee}{\end{equation}}
\newcommand{\bea}{\begin{eqnarray}}
\newcommand{\eea}{\end{eqnarray}}

\newcommand{\la}{\langle}
\newcommand{\ra}{\rangle}

\newcommand{\beq}{\begin{equation}}
\newcommand{\eeq}{\end{equation}}

\newcommand{\vrr}{{\bf{r}}}
\newcommand{\vnn}{{\hat{\bf{n}}}}
\newcommand{\vxi}{{\bm{\xi}}}
\newcommand{\vnabla}{{\bm{\nabla}}}
\newcommand{\vj}{{\bf{j}}}
\newcommand{\vQ}{{\bf{Q}}}
\newcommand{\vE}{{\bf{E}}}

\newcommand{\vJJ}{\mathbf{J}}
\newcommand{\cvJ}{\bm{{\cal J}}}
\newcommand{\cvA}{\bm{{\cal A}}}

\newcommand{\vecep}{\bm{\epsilon}}
\newcommand{\ecep}{\bm{\epsilon}}
\newcommand{\vpi}{\bm{\pi}}

\newcommand{\vvep}{\bm{\varepsilon}}
\newcommand{\vla}{\bm{\lambda}}
\newcommand{\vnu}{\bm{\nu}}
\newcommand{\vqq}{\mathbf{q}}

\newcommand{\pp}{{\text{P}}}

\def\(({\left(}
\def\)){\right)}
\def\[[{\left[}
\def\]]{\right]}

\def\mint{\displaystyle\int}

\def\mprod{\displaystyle\prod}

\journalname{jsp}
\begin{document}

\title{Thermodynamics of currents in nonequilibrium diffusive systems: theory and simulation\thanks{This work has been supported by Spanish MICINN project FIS2009-08451, University of Granada, and Junta de Andaluc\'{\i}a projects P07-FQM02725 and P09-FQM4682}}
\subtitle{Dedicated to Herbert Spohn on the occasion of his $65^{\textrm{th}}$ birthday}

\author{Pablo I. Hurtado        \and
 Carlos P. Espigares\and
Jes\'us J. del Pozo  \and
Pedro L. Garrido }


\institute{              Pablo I. Hurtado
              \email{phurtado@onsager.ugr.es}\and
              Carlos P. Espigares \email{cpespigares@onsager.ugr.es}\and
              Jes\'us J. del Pozo \email{jpozo@onsager.ugr.es} \and
              Pedro L. Garrido \email{garrido@onsager.ugr.es}\at
              Institute Carlos I for Theoretical and Computational Physics, \\
Universidad de Granada, Granada 18071, Spain
}

\date{Received: date / Accepted: date}

\maketitle

\begin{abstract}
Understanding the physics of nonequilibrium systems remains as one of the major challenges of modern theoretical physics. We believe nowadays that this problem can be cracked in part by investigating the macroscopic fluctuations of the currents characterizing nonequilibrium behavior, their statistics, associated structures and microscopic origin.
This fundamental line of research has been severely hampered by the overwhelming complexity of this problem. However, during the last years two new powerful and general methods have appeared to investigate fluctuating behavior that are changing radically our understanding of nonequilibrium physics: a powerful macroscopic fluctuation theory (MFT) and a set of advanced computational techniques to measure rare events.
In this work we study the statistics of current fluctuations in nonequilibrium diffusive systems, using macroscopic fluctuation theory as theoretical framework, and advanced Monte Carlo simulations of several stochastic lattice gases as a laboratory to test the emerging picture.
Our quest will bring us from (1) the confirmation of an additivity conjecture in one and two dimensions, which considerably simplifies the MFT complex 
variational problem to compute the thermodynamics of currents, to (2) the discovery of novel isometric fluctuation relations, which opens an unexplored route 
toward a deeper understanding of nonequilibrium physics by bringing symmetry principles to the realm of fluctuations, and to (3) the observation of coherent 
structures in fluctuations, which appear via dynamic phase transitions involving a spontaneous symmetry breaking event at the fluctuating level. 
The clear-cut observation, measurement and characterization of these unexpected phenomena, well described by MFT,
strongly support this theoretical scheme as the natural theory to understand the thermodynamics of currents in nonequilibrium diffusive media, 
opening new avenues of research in nonequilibrium physics.
\end{abstract}

\section{Introduction}
\label{intro}
Nonequilibrium phenomena characterize the physics of many natural systems. Despite their ubiquity and importance, no general bottom-up approach exists yet capable of predicting nonequilibrium macroscopic behavior in terms of microscopic physics, in a way similar to equilibrium statistical physics. This is due to the difficulty in combining statistics and dynamics, which always plays a key role out of equilibrium \cite{Bertini}. This lack of a general framework is a major drawback in our ability to manipulate, control and engineer many natural and artificial systems which typically function under nonequilibrium conditions. Consequently, this challenging problem has been subject to a intense study during the last 30 years. However, it has not been until recently that nonequilibrium physics has undergone a true revolution. At the core of this revolution is the realization of the essential role played by macroscopic fluctuations, with their statistics and associated structures, to understand nonequilibrium behavior \cite{Bertini,Derrida,Seifert}. The language of this revolution is the theory of large deviations, with large-deviation functions (LDFs) measuring the probability of fluctuations and optimal paths sustaining these rare events as central objects in the theory. In fact, LDFs play in nonequilibrium systems a role akin to the equilibrium free energy \cite{Derrida,Bertini}. In this way, the long-sought general theory of nonequilibrium phenomena is currently envisaged as a theory of macroscopic fluctuations, and the calculation, measurement and understanding of LDFs  and their associated optimal paths has become a fundamental issue in theoretical physics. This paradigm has led to a number of groundbreaking results valid arbitrarily far from equilibrium, many of them in the form of fluctuation theorems \cite{Bertini,Derrida,Seifert}.

\begin{figure}
\centerline{
\includegraphics[width=13cm]{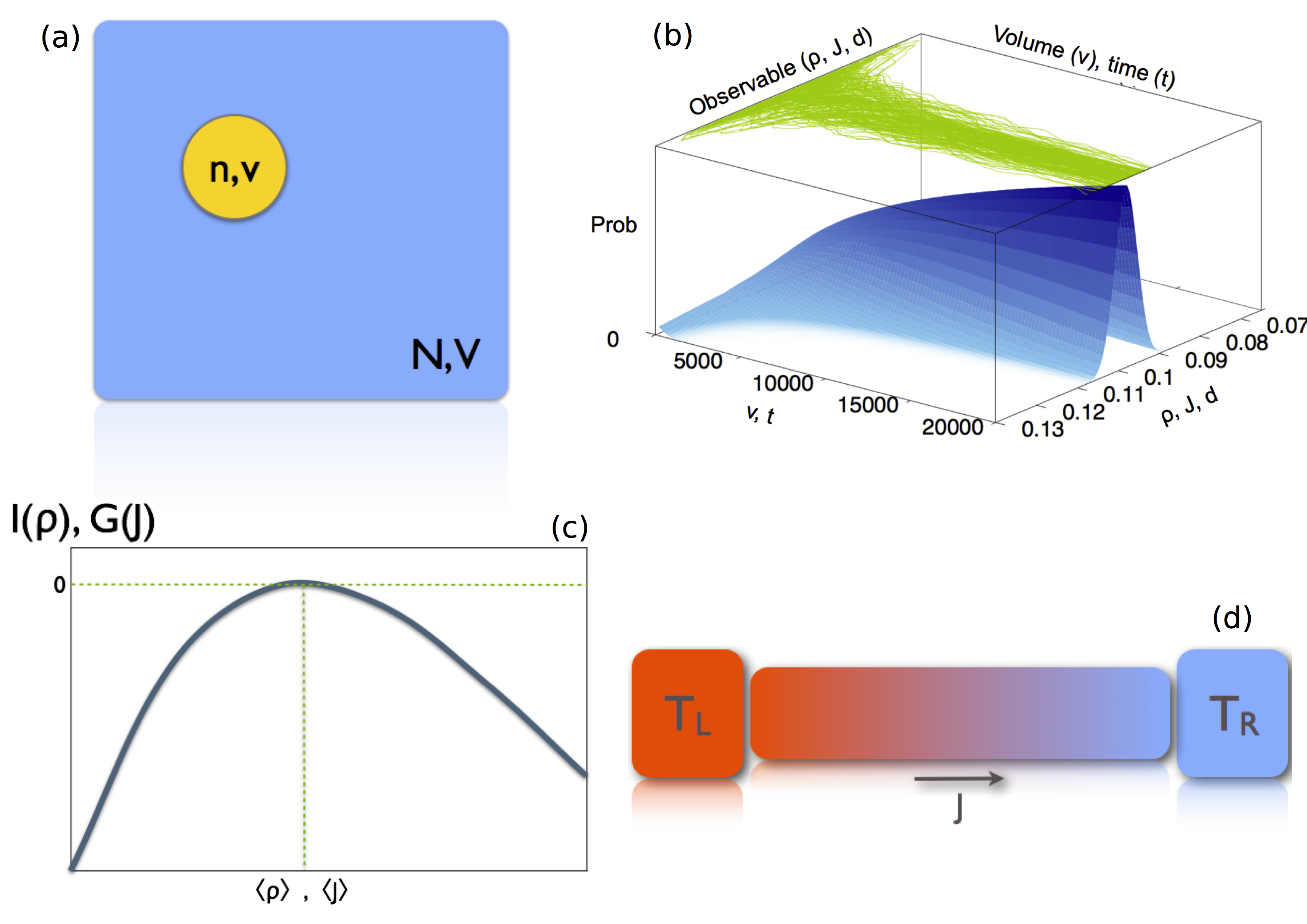}
}
\caption{(Color online) (a) Density fluctuations in a subsystem with volume $v$ for an equilibrium system. (b) Convergence of the time-averaged current to its ensemble value $\la q\ra$ for many different realizations (top line cloud), and sketch of the probability concentration as time increases, associated with the large deviation principle, Eq. (\ref{ch1:pj}). (c) Typical shape of a large deviation function. (d) Current fluctuations out of equilibrium as a result of an external temperature gradient.}
\label{LDs}
\end{figure}

To better grasp the physics behind a large deviation function, consider the example of density fluctuations in an equilibrium system \cite{Derrida}. In particular, let us consider an isolated box of volume $V$ with $N$ particles, as in Fig. \ref{LDs}.a. The probability of finding $n$ particles in a subvolume $v$ of our system scales asymptotically as $\pp_v(n)\sim\exp[+v{\cal I}(n/v)]$. This scaling defines a \emph{large-deviation principle} \cite{Ellis,Touchette}, and the function ${\cal I}(\rho)\leq 0$ is the density large deviation function \cite{Derrida,Bertini}.
The previous scaling means that the probability of observing a density fluctuation $\rho=n/v$ appreciably different from the average density $\la \rho\ra=N/V$ decays exponentially fast with the volume $v$ of the subregion. In this way, the LDF ${\cal I}(\rho)$ measures the \emph{rate}\footnote{For this reason, LDFs are known as rate functions in the mathematical literature \cite{Ellis,Touchette}.} at which the probability measure $\pp_v(\rho)$ concentrates around the average $\la \rho\ra$ as $v$ grows, see Fig. \ref{LDs}.b. In general, LDFs are negative in all their support except for the average value of the observable of interest, where the LDF is zero, see Fig. \ref{LDs}.c. Moreover, the typical LDF is quadratic around the average (where it is maximum), a reflection of the central limit theorem for small fluctuations.

For equilibrium systems, it is easy to show \cite{Derrida,Ellis} that the density LDF ${\cal I}(\rho)$ described above is univocally related with the free energy. Furthermore, the LDF of the density \emph{profile} in equilibrium systems can be simply related with the free-energy functional, a central object in the theory \cite{Ellis,Touchette,Derrida}.

It is therefore natural to use LDFs in systems far from equilibrium to define the nonequilibrium analog of the free-energy functional.
Key for this emerging paradigm is the identification of the relevant macroscopic observables characterizing nonequilibrium behavior. The system of interest often conserves locally some magnitude (a density of particles, energy, momentum, charge, etc.), and the essential nonequilibrium observable is hence the current or flux the system sustains when subject to, e.g., boundary-induced gradients or external fields, see Fig. \ref{LDs}.d. In this way, the understanding of current statistics in terms of microscopic dynamics has become one of the main objectives of nonequilibrium statistical physics, triggering an enormous research effort which has led to some remarkable results \cite{Bertini,Derrida,BD,PabloPRL,GC,ECM,LS,K,PabloIFR}.

Computing LDFs from scratch, starting from microscopic dynamics, is a daunting task which has been successfully accomplished only for a handful of oversimplified, low-dimensional stochastic lattice gases \cite{Derrida,Bertini}. This overwhelming complexity has severely hampered progress along this fundamental research line. However, during the last few years, two new powerful and general methods have appeared to investigate fluctuating behavior
that are changing radically our understanding of nonequilibrium physics.
On one hand, advanced computational methods have been recently developed to directly measure in simulations LDFs and the associated optimal paths for complex many-particle systems \cite{sim,sim2,sim3}.
On the other hand, a powerful and general macroscopic fluctuation theory (MFT) has been developed during the last ten years to understand dynamic fluctuations and the associated LDFs in driven diffusive systems arbitrarily far from equilibrium \cite{Bertini}, starting from their macroscopic evolution equation.
The application of these new tools to oversimplified models has just started to provide deep and striking results evidencing the existence of a rich and fundamental structure in the fluctuating behavior of nonequilibrium systems, crucial to crack this long-unsolved problem.

With this paper we want to describe our recent work in this direction. In order to do so, we first provide a brief description of macroscopic fluctuation theory in \S\ref{sMFT}, together with its application to understand the thermodynamics of currents out of equilibrium. Section \ref{smodels} introduces some paradigmatic stochastic lattice gases to model transport out of equilibrium. We will use these models along the paper as laboratories to test and validate the hypotheses underlying the formulation of MFT, as well as its detailed predictions. The analysis of these models will also provide deep insights into the fluctuating behavior of nonequilibrium diffusive systems. In \S\ref{ssim} we describe in detail the novel Monte Carlo techniques which allow us to measure the statistics of rare event in many-body systems. These methods are based on a modification of the underlying stochastic dynamics so that the rare events responsible of a given large deviation are no longer rare. Once the main tools for our work have been introduced, we set out to describe some recent results. In \S\ref{saddit} we introduce the additivity conjecture, an hypothesis on the time-independence of the optimal path responsible for a given current fluctuation which greatly simplifies the complex, spatiotemporal variational problem posed by MFT. In this section we also provide strong numerical evidence supporting the validity of this additivity conjecture in one- and two-dimensional diffusive systems for a wide interval of current fluctuations. Moreover, we show that the optimal path solution of the MFT problem is in fact a well-defined physical observable, and can be interpreted as the path that the system follows in phase space in order to facilitate a particular current fluctuation. In \S\ref{sIFR} we show that, by demanding invariance of these optimal paths under symmetry transformations, new and general fluctuation relations valid arbitrarily far from equilibrium are unveiled. In particular, we derive an isometric fluctuation relation which links in a strikingly simple manner the probabilities of any pair of isometric current fluctuations, confirming its validity in extensive simulations. We further show that the new symmetry implies remarkable hierarchies of equations for the current cumulants and the nonlinear response coefficients, going far beyond Onsager's reciprocity relations and Green-Kubo formulae.
The additivity conjecture assumes that optimal paths are time-independent for a broad range of current fluctuations. In \S\ref{sSSB} we show however that this additivity scenario eventually breaks down in isolated periodic diffusive systems for large fluctuations via a dynamic phase transition at the fluctuating level involving a symmetry-breaking event. Moreover, we report compelling evidences of this phenomenon in two different one-dimensional stochastic lattice gases. Finally, \S\ref{sDO} contains a discussion of the results presented here and some outlook regarding the work that remains to be done in a near future. We leave for the appendices some technical details that we prefer to  omit from the main text for the sake of clarity.

\section{Macroscopic fluctuation theory and thermodynamics of currents}
\label{sMFT}

In a series of recent works \cite{Bertini}, Bertini, De Sole, Gabrielli, Jona-Lasinio, and Landim have introduced a macroscopic fluctuation theory (MFT) which describes dynamic fluctuations in driven diffusive systems and the associated LDFs starting from a macroscopic rescaled description of the system of interest (typically a hydrodynamic-like equation), where the only inputs are the system transport coefficients. This is a very general approach which leads however to a hard variational problem whose solution remains challenging in most cases. Therefore a main research path has been to explore different solution schemes and simplifying hypotheses. This is the case for instance of the recently-introduced additivity conjecture, which can be justified under certain conditions and used within MFT
to obtain explicit predictions, opening the door to a systematic way of computing LDFs in nonequilibrium systems. As usual in physics, it is as important to formulate a sound hypothesis as to know its range of validity. The additivity conjecture may be eventually violated for large fluctuations, but quite remarkably this additivity breakdown, which is well characterized within MFT, proceeds via a dynamic phase transition at the fluctuating level involving a symmetry breaking. More on this below.

We now proceed to describe MFT in detail. Our starting point is a continuity equation that describes the mesoscopic evolution of a broad class of systems characterized by a locally conserved magnitude (e.g. energy, particles, momentum, charge, etc.) \cite{Bertini,Derrida,PabloAIP}
\beq
\partial_t \rho(\vrr,t) = -\vnabla \cdot  \left( \vQ_{\vE}[\rho(\vrr,t)] + \vxi(\vrr,t) \right) \, .
\label{ch1:langevin}
\eeq
This equation is obtained after an appropriate scaling limit in which the microscopic time and space coordinates, ${\tilde t}$ and ${\tilde \vrr}$, respectively, are rescaled diffusively: $t={\tilde t}/N^2$, $\vrr={\tilde \vrr}/N$, where $N$ is the linear size of the system \cite{Spohn}. The macroscopic coordinates are then $(\vrr,t)\in \Lambda \times [0,\tau]$, where $\Lambda\in [0,1]^d$ is the spatial domain and $d$ the dimensionality of the system. In Eq. (\ref{ch1:langevin}), $\rho(\vrr,t)$ represents the density field, and $\vj(\vrr,t) \equiv \vQ_{\vE}[\rho(\vrr,t)] + \vxi(\vrr,t)$ is the fluctuating current field, with a local average given by $\vQ_{\vE}[\rho]$ which includes in general the effect of a conservative external field $\vE$,
\beq
\vQ_{\vE}[\rho]=\vQ[\rho] + \sigma[\rho] \vE \, .
\label{ch1:curr}
\eeq
The field $\vxi(\vrr,t)$ is a Gaussian white noise characterized by a variance (or \emph{mobility}) $\sigma[\rho(\vrr,t)]$, i.e.,
\be
\la \vxi(\vrr,t)\ra=0 \qquad ; \qquad \la \xi_i(\vrr,t)\xi_j(\vrr',t') \ra=N^{-d}\sigma[\rho]\delta_{ij}\delta(\vrr-\vrr')\delta(t-t') \, ,
\label{ch1:noise}
\ee
being $i,j\in [0,d]$ the components of the spatial coordinates
and $d$ the spatial dimension. This (conserved) noise term accounts for microscopic random fluctuations at the macroscopic level. This noise source represents the many fast microscopic degrees of freedom which are averaged out in the coarse-graining procedure resulting in Eq. (\ref{ch1:langevin}), and whose net effect on the macroscopic evolution amounts to a Gaussian random perturbation according to the central limit 
theorem. Since $\vxi(\vrr,t)$ scales as $N^{-d/2}$, in the limit $N\rightarrow \infty$ we recover the deterministic hydrodynamic equation, but as we want to study the fluctuating behavior, we consider large (but finite) system sizes, i.e., we are interested in the weak noise limit.

Examples of systems described by Eq. (\ref{ch1:langevin}) range from diffusive systems  \cite{Bertini,Derrida,BD,PabloPRL,PabloPRE,BD2,PabloSSB}, where $\vQ[\rho(\vrr,t)]$ is given by Fourier's (or equivalently Fick's) law,
\beq
\vQ[\rho(\vrr,t)]=-D[\rho] \vnabla \rho(\vrr,t),
\label{ch1:fourier}
\eeq
with $D[\rho]$ the diffusivity, to most interacting-particle fluids \cite{Spohn,Newman}, characterized by a Ginzburg-Landau-type theory for the locally-conserved particle density.
To completely define the problem, the above evolution equations (\ref{ch1:langevin})-(\ref{ch1:curr}) must be supplemented with appropriate boundary 
conditions, which can be for instance periodic, when $\Lambda$ is the $d$-dimensional torus, or non-homogeneous with
\beq
\varphi (\rho(\vrr,t))=\varphi_0 (\vrr),~~~~\vrr \in \partial\Lambda
\label{ch1:dirichcond}
\eeq
in the case of boundary-driven systems in which the driving is due to an external gradient. Here $\partial \Lambda$ is the boundary of $\Lambda$ and $\varphi_0$ is the chemical potential of the boundary reservoirs.
The few transport coefficients that enter into the hydrodynamic Eq. (\ref{ch1:langevin}) can be readily measured in experiments or simulations, which thus offers a close description of the macroscopic fluctuating behavior of the system of interest.
For diffusive systems governed by Fourier's law (\ref{ch1:fourier}), the diffusion coefficient $D[\rho]$ and the mobility $\sigma[\rho]$ satisfy a local Einstein relation
\beq
D[\rho]=\frac{\sigma[\rho]}{\kappa[\rho]}
\label{ch1:einsrel}
\eeq
where $\kappa[\rho]$ is the compressibility, $\kappa[\rho]^{-1}=f_0''[\rho]$, $f_0[\rho]$ being the equilibrium free energy of the system. The above 
equations describe an equilibrium model when either (a) $\Lambda$ is the torus 
and there is no external field, or (b) in the case of boundary-driven diffusive systems (i.e. $\vQ[\rho]=-D[\rho]\vnabla \rho$) in which the 
external field in the bulk matches the driving from 
the boundary \cite{Bertini}. We are also in equilibrium when the chemical potentials of the boundaries are the same. In any other case the resulting stationary state 
sustains a non-vanishing current and the system is out of equilibrium.

The probability of observing a history $\{\rho(\vrr,t),\vj(\vrr,t)\}_0^{\tau}$ of duration $\tau$ for the density and current fields, which can be different from the average hydrodynamic trajectory,
can be written as a path integral over all possible noise realizations, $\{\vxi(\vrr,t)\}_0^{\tau}$, weighted by its Gaussian measure and restricted 
to those realizations compatible with Eq. (\ref{ch1:langevin}) (and the associated boundary conditions) at every point of space and time\footnote{Note that the 
path integral formalism here described is based on a discretized Langevin equation of Ito-type \cite{Zinn}}
\beq
\text{P}\left(\{\rho,\vj\}_0^{\tau} \right)=\mint {\cal D} \vxi \, \exp\left [-N^{d}\mint_{0}^{\tau}dt \mint_{\Lambda} d\vrr\dfrac{\vxi^2}{2 \sigma[\rho]} \right ]
\, \mprod_{t} \mprod_{\vrr}\delta \Big[\vxi-\left(\vj-\vQ_{\vE}[\rho] \right) \Big] ,
\label{ch1:probtraj1}
\eeq
with $\rho(\vrr,t)$ and $\vj(\vrr,t)$ coupled via the continuity equation,
\beq
\partial_t \rho + \vnabla\cdot \vj = 0.
\label{ch1:cont2}
\eeq
Notice that this coupling does not determine univocally the relation between $\rho$ and $\vj$. For instance, 
the fields $\tilde \rho(\vrr,t)=\rho (\vrr,t)+\chi (\vrr)$ and $\tilde \vj(\vrr,t)=\vj(\vrr,t) + \mathbf{g}(\vrr,t)$, with $\chi(\vrr)$ arbitrary 
and $\mathbf{g}(\vrr,t)$ divergence-free, satisfy the same continuity equation. In other words, this means that from a density field we can 
determine the current field up to a divergence-free vector field. 
This non-uniqueness in the macroscopic description is the price we pay for the information 
lost when coarse-graining the deterministic microscopic degrees of freedom.
Eq. (\ref{ch1:probtraj1}) naturally leads to \cite{PabloAIP}
\beq
\text{P}\left(\{\rho,\vj\}_0^{\tau} \right)\sim \exp \left( +N^d I_{\tau}[\rho,\vj]\right),
\label{ch1:LDpple}
\eeq
which has the form of a large deviation principle. The rate functional $I_{\tau}[\rho,\vj]$ is given by
\beq
I_{\tau}[\rho,\vj]=-\mint_{0}^{\tau}dt \mint_{\Lambda} d\vrr\dfrac{(\vj (\vrr,t)-\vQ_{\vE}[\rho])^2}{2 \sigma[\rho]} \, .
\label{ch1:functional}
\eeq
This functional plays a pivotal role in MFT and its extensions, as it contains all the information needed to compute LDFs of any relevant macroscopic 
observable via standard contraction principles  in large deviation theory \cite{Ellis,Touchette}.
For instance, it has been used by Bertini and collaborators to study fluctuations of the density field out of 
equilibrium. Using this approach, a Hamilton-Jacobi equation for the nonequilibrium density LDF has been derived \cite{Bertini} 
showing that this LDF is usually non-local out of equilibrium, a reflection of the long-range correlations typical of nonequilibrium situations. 
Moreover, MFT shows that the optimal path leading to a macroscopic density fluctuation is the time-reversal of the relaxation path from 
this fluctuation according to some \emph{adjoint} hydrodynamic laws (not necessarily equal to the original) \cite{Bertini}. This general 
result, valid arbitrarily far from equilibrium, reduces to the well-known Onsager-Machlup theory when small deviations from equilibrium are considered.

\subsection{Thermodynamics of currents}
\label{sTC}

We now want to focus on the statistics of the current.
Understanding how microscopic dynamics determine the long-time averages of the current and its fluctuations is one of the main objectives of nonequilibrium statistical physics \cite{GC}-\cite{Jona2}, as this is a central observable characterizing macroscopic behavior out of equilibrium. Therefore we focus now on the probability $P_{\tau}(\vJJ)$ of observing a space\&time-averaged current
\beq
\vJJ=\frac{1}{\tau}\int_{0}^{\tau}dt\int_{\Lambda}d\vrr \vj(\vrr,t).
\label{ch1:currint}
\eeq
This probability can be written as
\beq
\text{P}_{\tau}(\vJJ)= \int^* {\cal D}\rho{\cal D}\vj \, \, \text{P}\left(\{\rho,\vj\}_0^{\tau} \right) \, \delta\left(\vJJ-\frac{1}{\tau}\int_0^{\tau} dt \int_{\Lambda} d\vrr \,\vj(\vrr,t) \right) \nonumber \, ,
\eeq
where the asterisk means that this path integral is restricted to histories $\{\rho,\vj\}_0^{\tau}$ coupled via Eq. (\ref{ch1:cont2}). As the exponent 
of $\text{P}\left(\{\rho,\vj\}_0^{\tau} \right)$ is extensive in both $\tau$ and $N^d$ \cite{PabloAIP}, see Eq. (\ref{ch1:LDpple}), for long times and large system sizes the above path integral is dominated by the associated saddle point, resulting in the following large deviation principle
\beq
\text{P}_{\tau}(\vJJ)\sim \exp[+\tau N^d G(\vJJ)],
\label{ch1:pj}
\eeq
where the rate functional $G(\vJJ)$ defines the current large deviation function (LDF)
\beq
G(\vJJ)=-\lim_{\tau\rightarrow \infty}\frac{1}{\tau}
\min_{\{\rho,\vj\}_0^{\tau}}\left\{\mint_{0}^{\tau}dt \mint_{\Lambda} d\vrr\dfrac{(\vj (\vrr,t)-\vQ_{\vE}[\rho])^2}{2 \sigma[\rho]}\right\}.
\label{ch1:ldf0}
\eeq
subject to the constraints (\ref{ch1:cont2}) and (\ref{ch1:currint}). The LDF $G(\vJJ)$ measures the (exponential) rate at which $\vJJ \to \vJJ_{st}$ as $\tau$ increases (notice that $G(\vJJ)\leq 0$, with $G(\vJJ_{st})=0$).
The optimal density and current fields solution of the (complex) variational problem Eq. (\ref{ch1:ldf0}), denoted here as $\rho_{\vJJ}(\vrr,t)$ and $\vj_{\vJJ}(\vrr,t)$, can be interpreted as the optimal path the system follows in mesoscopic phase space in order to sustain a long-time current fluctuation $\vJJ$. It is worth emphasizing here that the existence of an optimal path rests on the presence of a selection principle at play, namely a long time, large size limit  which selects, among all possible paths compatible with a given fluctuation, an optimal one via a saddle point mechanism. Despite its inherent complexity, the current LDF $G(\vJJ)$ obeys a symmetry property which stems from the reversibility of microscopic dynamics. This is the Gallavotti-Cohen fluctuation theorem \cite{GC}, which relates the probability of observing a long-time current fluctuation $\vJJ$ with the probability of the reverse event, $-\vJJ$,
\be
 \lim_{\tau\to\infty} \frac{1}{\tau N^d} \ln \left[ \frac{\textrm{P}_{\tau}(\vJJ)}{\textrm{P}_{\tau}(-\vJJ)}\right] = 2\vecep\cdot \vJJ \, ,
\label{GCtheo}
\ee
where $\vecep=\vvep+\vE$ is the driving force, a constant vector which depends on the boundary baths via $\vvep$ (see below) and on the external field $\vE$, and is directly related to the rate of entropy production in the nonequilibrium system of interest.

Finally, it is remarkable that although the MFT here described is in general applied to conservative systems, it can be generalized to dissipative systems characterized by a continuous loss of energy to the environment \cite{prados1}. In these cases the macroscopic evolution equation is given by
\beq
\partial_t \rho=-\vnabla \cdot (\vQ_{\vE}+\vxi)-\nu A[\rho],
\nonumber
\eeq
where the first term in the r.h.s describes the diffusive energy propagation, whereas the second term defines the energy dissipation rate through the functional $A[\rho]$ and the macroscopic dissipation coefficient $\nu$. In this case, the essential macroscopic observables which characterize the non-equilibrium behaviour are the current and the dissipated energy. Using the path integral formalism above described, it is possible to define the large deviation function of these observables and the optimal fields associated with their fluctuations. The extension of the MFT to dissipative systems has been recently developed and tested in Refs. \cite{BL1,BL2,prados1,prados2,prados3}.

MFT hence provides in general clear-cut variational formulae to understand current statistics in diffusive systems arbitrarily far from equilibrium, together with the optimal paths that, in order to facilitate a given current fluctuation, the system of interest traverses in phase space. The complexity of the problem is however humongous, and many difficult questions arise: How are the solutions to this complex variational problem? Can we classify them according to some hierarchical scheme? Are the optimal paths solution of this mathematical problem physically observable? How is the statistics of rare current fluctuations as compared to the Gaussian statistics naively expected from the central limit theorem? Is there nontrivial structure at the fluctuating level? Can we confirm the Gallavotti-Cohen symmetry, and even more, can we uncover hidden symmetries at the fluctuating level? Are there phase transitions in the fluctuating behavior of complex diffusive systems? The solution to these and many other fundamental and exciting questions calls for a detailed analysis of MFT and its predictions, together with a deep investigation of sound hypotheses and conjectures that may simplify the inherent complexities of the theory. Moreover, this work must be accompanied at every step by \emph{in silico} experiments, i.e. extensive numerical simulations of simplified models of transport using the novel techniques to simulate rare events. This numerical work will allow us to test and guide new theoretical ideas and to aid the formulation of bold conjectures, leading eventually to the discovery of entirely unexpected phenomena. We provide below a review of our work in these directions.

\section{Models of transport out of equilibrium}
\label{smodels}

MFT and its generalizations offer an unique opportunity to obtain general results for a large class of systems arbitrarily far from equilibrium, a possibility that we could only 
dream of some years ago. Therefore it is essential to test and validate the hypotheses underlying its formulation, as well as its detailed predictions. The investigation of rare event statistics in realistic systems with many degrees of freedom poses still today formidable challenges. It is therefore necessary to work with simplified models of reality which, while
capturing the essential ingredients characterizing more realistic systems,
maximally simplify the microscopic details irrelevant for the phenomenon being studied. Universality arguments then allow us to connect the results obtained for these simplified models with the physics of more realistic, albeit more complex, natural systems. For the particular problem of nonequilibrium fluctuations here studied, the ideal laboratory where to test these ideas is provided by stochastic lattice gases \cite{jona}, for which the local equilibrium hypothesis and the hydrodynamic evolution equations which form the basis of MFT can be rigorously derived in some cases \cite{Bertini,Spohn}. Although the microscopic random dynamics of these lattice models is different from the Hamiltonian evolution of more realistic systems, the relevant symmetries and conservation laws are the same, and hence we expect that the resulting macroscopic nonequilibrium behavior will be qualitatively independent of these details \cite{jona}.

Many different stochastic lattice gases exist in the literature, but we will focus in this paper in two paradigmatic diffusive models which have guided the advances in the field during the last two decades: the Kipnis-Marchioro-Presutti model of energy transport and the weakly-asymmetric simple exclusion process.

\subsection{Kipnis-Marchioro-Presutti model and generalizations}
\label{sKMP}

In 1982, C. Kipnis, C. Marchioro and E. Presutti \cite{kmp} proposed a simple lattice model in order to understand in a mathematically rigorous way energy transport in systems with many degrees of freedom. Since its original formulation, this model, dubbed KMP model in the literature, has become a paradigm in nonequilibrium statistical physics, where new theoretical ideas have been tested and novel breakthroughs have been developed.
In particular, KMP were able to show rigorously from first principles (i.e. starting from its microscopic Markovian dynamics) that this model obeys Fourier's law in 1D, a relation formulated in 1822 by Joseph Fourier
which states that the heat current flowing through a material in contact with two reservoirs at different temperatures
is proportional to the temperature gradient. The special features of this model turn it into the ideal ground where to test MFT and its extensions, and this has triggered a surge of interest among specialists in nonequilibrium physics which has resulted in a number of new and surprising results (see below).

\begin{figure}
\centerline{
\includegraphics[width=15cm]{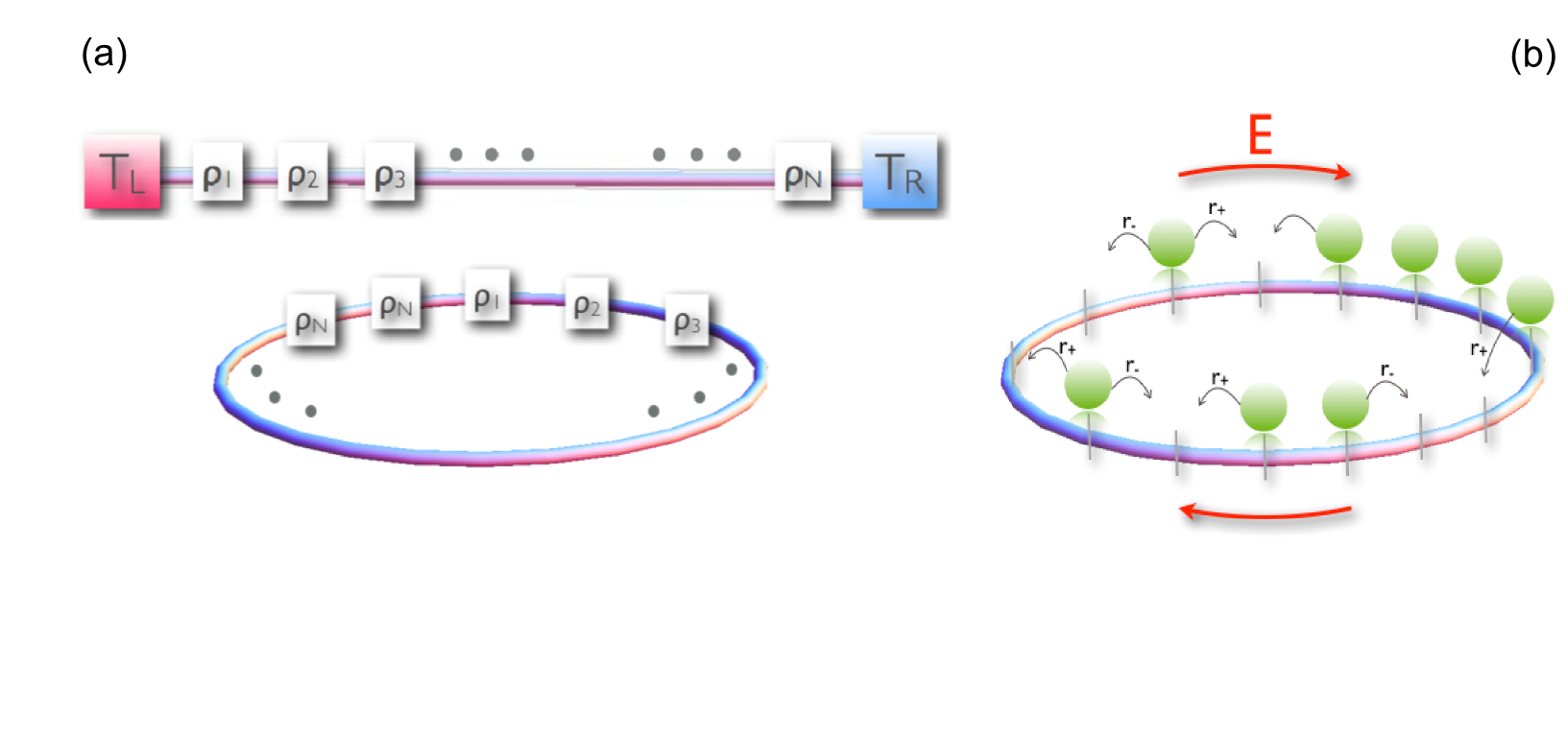}
}
\vspace{-2cm}
\caption{(Color online) (a) Kipnis-Marchioro-Presutti (KMP) model of transport in 1D lattices under different boundary conditions. Top: System coupled to boundary heat baths at different temperatures, $T_L\neq T_R$. Bottom: Periodic boundary conditions. Each lattice site is characterized by an energy $\rho_i\geq 0$, $i\in[1,N]$, and dynamics proceeds via stochastic collisions between nearest neighbors and involves a random redistribution of energy among the colliding pair. (b) Sketch of the weakly-asymmetric exclusion process (WASEP) defined on a 1D lattice subject to periodic boundary conditions. Particles jump stochastically to a right (left) empty nearest neighbor at a rate $r_+$ ($r_-$), which implies for $r_+\neq r_-$ that particles feel an external driving field $E=\frac{N}{2}\ln(\frac{r_+}{r_-})$.}
\label{sketchmodels}
\end{figure}

The KMP model is defined in a one-dimensional (1D) lattice with $N$ sites, see Fig. \ref{sketchmodels}.a, although it can be easily generalized to any type of lattice in arbitrary dimension. Each lattice site models a harmonic oscillator which is  mechanically uncoupled from its nearest neighbors but interacts with them through a random process which redistributes energy locally.
The system microscopic configuration is thus defined at any time by a set $\mathbf{\rho}\equiv \{\rho_i, i=1,...N \}$, where $\rho_i\in \mathbb{R}_+$
is the energy of the site $i\in [1,N]$. Dynamics is stochastics and proceeds
through random energy exchanges between randomly chosen nearest-neighbors according to a microcanonical procedure where the pair energy is kept constant. Hence,
$(\rho_i,\rho_{i+1})\rightarrow (\rho'_i,\rho'_{i+1})~~\forall i$ such that,
\begin{eqnarray}
\rho'_i&=&p(\rho_i+\rho_{i+1}) \label{ch2:kmpdyncont} \\
\rho'_{i+1}&=&(1-p)(\rho_i+\rho_{i+1}) \nonumber
\end{eqnarray}
where $p\in [0,1]$ is an uniform random number, and $\rho_i+\rho_{i+1}=\rho'_i+\rho'_{i+1}$. To complete the model definition, we must specify appropriate boundary conditions.
In the original paper \cite{kmp}, and in order to study energy transport, KMP considered open boundary conditions with extremal ($i= 1,N$) sites of the 1D chain connected to thermal baths at different temperatures, see top panel in Fig. \ref{sketchmodels}.a. In this case, extremal sites may interchange energy with thermal baths at temperatures $T_L$ for $i=1$ and $T_R$ for $i=N$, i.e., $\rho_{1,N} \rightarrow \rho'_{1,N}$ such that
\beq
\rho'_{1,N}=p({\tilde \rho}_{L,R}+\rho_{1,N})
\label{ch2:kmpbound}
\eeq
where $p\in [0,1]$ is again an uniform random number and ${\tilde \rho}_{L,R}$ is a random number drawn at each interaction from a Gibbs distribution at the corresponding temperature, $P(\rho_{k})=\beta_k \exp(-\rho_k\beta_k)$, $k=L,R$, with $\beta_k=T_k^{-1}$ so Boltzmann constant is set to one. For $T_L\neq T_R$ KMP proved rigorously \cite{kmp} that the system reaches a nonequilibrium steady state in the hydrodynamic scaling limit $N\to\infty$ described by Fourier's law, with a nonzero average energy current
\beq
J_{st}=-D[\rho]\frac{d\rho_{st}(x)}{dx} \, , \, x\in[0,1] \, ,
\label{ch2:currmed}
\eeq
where $D[\rho]=\frac{1}{2}$ is the conductivity (or diffusivity) for the KMP model, and a linear steady density profile
\beq
\rho_{st}(x)=T_L+x(T_R-T_L).
\label{ch2:statprof}
\eeq
In addition, convergence to the local Gibbs measure was proven in this limit \cite{kmp}, meaning that $\rho_i$, $i\in[1,N]$, has an exponential distribution with local temperature $\rho_{st}(\frac{i}{N+1})$ in the thermodynamic limit. However, corrections to Local Equilibrium (LE), though
vanishing in the $N\to \infty$ limit, become apparent at the fluctuation level \cite{Sp,BGL}. The mesoscopic evolution equation for this model is
\beq
\partial_t \rho + \partial_x\left(-\frac{1}{2}\partial_x \rho + \xi \right) = 0 \, ,
\label{ch2}
\eeq
which is the dynamical expression of the (fluctuating) Fourier's law, compare with Eq. (\ref{ch1:langevin}) above. The amplitude of the conserved noise term is given by the mobility $\sigma[\rho]$, see Eq. (\ref{ch1:noise}), which is the second transport coefficient needed by MFT to complete the macroscopic description of the system of interest. The mobility, which measures the variance of local energy current fluctuations in equilibrium ($\rho_L=\rho_R$), can be written for the KMP model as $\sigma[\rho]=\rho^2$. It is also worth noting that the microscopic dynamics in the KMP model obeys the local detailed balance condition, thus being time-reversible.

  KMP model is an optimal candidate to test and study MFT and its detailed predictions because: (a) MFT equations for this model are simple enough to admit full analytical solutions, and (b) its simple dynamical rules allow for a detailed numerical study of current fluctuations, both typical and rare, taking advantage of the novel computational methods to study rare event statistics \cite{sim,sim2,sim3}.

Another reason for the recent surge of interest in the KMP model is that it can be easily generalized to describe at a coarse-grained level many different \emph{nonlinear} diffusive microscopic processes \cite{Pablowave,prados1,prados2,prados3}. These type of processes abound in nature, with important examples in fields as diverse as fluid dynamics, heat transfer, mathematical biology, population dynamics, etc. For the standard KMP model, the mesoscopic evolution equation (\ref{ch2}) is linear in the density field, as results from a constant diffusivity, $D[\rho]=\frac{1}{2}$. However, it can be shown \cite{Pablowave,prados2} that a more realistic generalization of the KMP model, with collision rates between neighboring lattice sites depending explicitly on the energy of the colliding pair, gives rise to a mesoscopic description based on \emph{nonlinear} diffusion-type equations, a class of equations that characterize the physics of many natural complex systems. In particular, if the collision rate for pair $(i,i+1)$ is proportional to a power of its total energy, $(\rho_i+\rho_{i+1})^a$, it can be shown \cite{Pablowave,prados2} that the mesoscopic evolution equation is now
\be
\partial_t \rho + \partial_x\left(-\rho^a \partial_x \rho + \xi \right) = 0 \, ,
\label{langevinKMPnolin}
\ee
so the diffusivity is now a function of the density field, $D[\rho]=\rho^a$. Moreover, the mobility coefficient is also modified by the nonlinearity, $\sigma[\rho]\propto \rho^{a+2}$. The simplicity of this generalization of the KMP model allows us to investigate the nonequilibrium fluctuating behavior of strongly nonlinear systems and study in detail in this nonlinear regime the predictions of macroscopic fluctuation theory.

The nonlinear KMP model can be further generalized to include dissipative processes in competition with the main diffusive mechanism \cite{prados1,prados2,prados3}. Microscopically this is achieved by allowing the dissipation to the environment of a fraction of the pair energy in collisions, before the random redistribution of the remaining energy between colliding neighbors. It is easy to show that this apparently innocent modification of the KMP microscopic dynamics dramatically affects the system macroscopic evolution, which now follows a reaction-diffusion type of equation of the form
\be
\partial_t \rho + \partial_x\left(-\rho^a \partial_x \rho + \xi \right) + \nu\rho^{a+1} = 0 \, ,
\label{langevinKMPdisip}
\ee
where $\nu$ is a macroscopic dissipation coefficient. This generalization of KMP model contains the essential ingredients characterizing most dissipative media, namely: (i) nonlinear diffusive dynamics, (ii) bulk dissipation, and (iii) boundary injection. Moreover, it can be regarded as a  toy model for dense granular media: particles cannot freely move but may collide with their nearest neighbors, losing a fraction of the pair energy and exchanging the rest thereof randomly. The dissipation coefficient can be thus considered as the analogue to the restitution coefficient in granular systems \cite{PyL01}.

\subsection{Diffusive simple exclusion processes}
\label{sWASEP}

A second class of diffusive models widely studied in literature are simple exclusion processes (SEP)\footnote{We explicitly exclude in this description the asymmetric and totally-asymmetric exclusion processes (ASEP and TASEP, respectively), as these models are not diffusive. See \cite{sep} for a review on these interesting models of nonequilibrium behavior.}, in its symmetric (SSEP) and weakly-asymmetric (WASEP) versions \cite{sep,derrida_leb}.
As for the KMP model, these exclusion models can be defined on arbitrary lattices in any dimension,
and subject either to open or periodic boundary conditions.
In what follows we focus on 1D for simplicity, and start by considering the SSEP with open boundaries. This model is defined on a 1D lattice of size $N$, where each site $i\in[1,N]$ may contain at most one particle, so the state of the system is defined at any time by a set of occupation numbers, $\mathbf{n}\equiv \{n_i=0,1, \, i\in[1,N]\}$. The dynamics is stochastic and proceeds via sequential particle jumps to nearest neighbor sites, provided these are empty, at unit rate. At the two boundaries dynamics is modified to mimic the coupling with particle reservoirs, possibly at different densities $\rho_L\neq \rho_R$: at the left boundary ($i=1$) particles are injected at rate $\alpha$ (if this site is empty) and removed at rate $\gamma$ (if this site is occupied). Similarly, on site $N$ particles are injected at rate $\delta$ and removed at rate $\beta$. These injection and removal rates fix the densities of the left and right reservoirs to $\rho_L=\alpha/(\alpha+\gamma)$ and $\rho_R=\delta/(\beta+\delta)$, respectively.
For $\rho_L=\rho_R\equiv\rho$ the system is in equilibrium and the probability measure has a product form: $\text{prob}_{\text{eq}}(\mathbf{n})=\prod_{i=1}^{N}\rho^{n_i}(1-\rho)^{1-n_i}=e^{\sum_{i=1}^{N}\eta n_i}/(1+e^{\eta})^N$,
where $\eta=\log(\rho/(1-\rho))$ is the chemical potential. As soon as $\rho_L \neq \rho_R$, the system is out of equilibrium, a current is established, and the problem becomes nontrivial, with long range correlations. In particular, the SSEP reaches an steady state with an average density profile given, in the large $N$ limit, by \cite{derrida_leb}
\beq
\la n_i \ra=\rho_{st} (x)=\rho_L+x(\rho_R-\rho_L) \, ,
\label{ch2:steprofssep}
\eeq
which is equivalent to the linear profile of the KMP model, see Eq. (\ref{ch2:statprof}) above, and we have introduced a macroscopic coordinate $x=i/N$.
The average current in the steady state is then proportional to the density gradient, obeying Fick's law
\beq
J_{st}=-D[\rho]\frac{d\rho_{st}(x)}{dx} \, , \, x\in[0,1],
\label{ch2:statcurrssep}
\eeq
with $D[\rho]=1/2$. The SSEP thus obeys the following mesoscopic evolution equation \cite{Spohn}
\beq
\partial_t \rho + \partial_x\left(-\partial_x \rho + \xi \right) = 0 \, ,
\label{ch2:ssepmacro}
\eeq
which corresponds to the dynamical expression of Fick's law. Moreover, the mobility coefficient characterizing the equilibrium fluctuations of the current is $\sigma[\rho]=\rho(1-\rho)$ for the SSEP \cite{Spohn}. Note that, as compared to the KMP mobility coefficient, $\sigma_{\text{SSEP}}[\rho]$ is bounded and shows a maximum, a property that will have implications for the current statistics of this model (in particular for the existence of phase transitions at the fluctuating level \cite{BD2,PabloSSB}).

To end this section, we now consider the weakly asymmetric exclusion process (WASEP) on a 1D lattice with periodic boundary conditions. 
This model is analogous to SSEP except for the presence of a weak external field, $E$, which bias particle jumps in a preferential direction, 
and the periodic boundary conditions used. Therefore we have a 1D lattice with $N$ sites, where a fixed number of $Z=\sum_{i=1}^N n_i \le N$ 
particles live, see Fig. \ref{sketchmodels}.b, so the total density, $\rho=Z/N$, is fixed.
As before, particles perform stochastic sequential jumps to neighboring sites, provided these are empty, but now the jump 
rates are defined as $r_{\pm}\equiv \frac{1}{2} \exp(\pm E/N)$ for jumps along the $\pm\hat{x}$-direction\footnote{Note that these rates 
converge for large $N$ to the standard ones found in literature, namely $\frac{1}{2}(1\pm\frac{E}{N})$, but avoid problems with 
negative rates for small $N$. In any case, the hydrodynamic descriptions of both variants of the model are identical.}. Here $E$ plays the 
role of a weak external field which drives the system to a nonequilibrium steady state characterized by a homogeneous 
average density profile $\la n_i \ra=\rho_{st} (x)=\rho$ and a nonzero net average current $J_{st} = \rho_(1-\rho) E$. The 
mesoscopic evolution equation for WASEP now reads \cite{Spohn}
\be
\partial_t \rho + \partial_x\left(-D[\rho] \partial_x \rho + \sigma[\rho]E + \xi \right) = 0 \, ,
\label{ch2:easepmacro}
\ee
where $D[\rho]=1/2$ and $\sigma[\rho]=\rho(1-\rho)$ are the SSEP transport coefficients, as otherwise expected.

\section{Monte Carlo evaluation of current large-deviation functions}
\label{ssim}

As described above, these and other stochastic lattice gases provide the ideal ground where to investigate the large deviation statistics of currents out of equilibrium. The main reason is that their simple dynamical rules allow for an extensive analysis of fluctuations, both typical and rare, taking advantage of novel computational methods to study rare event statistics \cite{sim,sim2,sim3}. It is important to notice that, in general, large deviation functions are very hard to measure in experiments or simulations because they involve by definition exponentially-unlikely events, see e.g.  Eq. (\ref{ch1:pj}). Recently, Giardin\`a, Kurchan and Peliti \cite{sim}, and Tailleur and Lecomte \cite{sim2}, have introduced efficient algorithms to measure the probability of a large deviation for time-extensive observables such as the current or the activity in discrete- and continuous-time stochastic many-particle systems, see \cite{sim3} for a review. The main idea consists in modifying in a mathematically controlled way the underlying stochastic dynamics so that the rare events responsible of a given large deviation are no longer rare.

Let $U_{C' C}$ be the transition rate from configuration $C$ to $C'$ for the stochastic model of  interest, and define $\vqq_{C' C}$ as the elementary 
current involved in this microscopic transition. The probability of measuring a total time-integrated current $\vQ_t$ after a time $t$ starting from 
a configuration $C_0$ can be thus written as 
\be
\pp(\vQ_t,t;C_0) = \sum_{C_{t}..C_1} U_{C_t C_{t-1}}..U_{C_1 C_0} \, \delta (\vQ_t - \sum_{k=0}^{t-1}\vqq_{C_{k+1} C_k}) \, ,
\label{recurr1}
\ee
where $U_{C_t C_{t-1}}..U_{C_1 C_0}$ is nothing but the probability of a path $C_0\to C_1\to\ldots\to C_t$ in phase space. For long times we 
expect the information on the initial state $C_0$ to be lost, so $\pp(\vQ_t,t;C_0) \to \pp(\vQ_t,t)$. In this limit $\pp(\vQ_t,t)$ obeys the usual large 
deviation principle\footnote{Note that the macroscopic current $\vJJ$ is related to this microscopic current $\vqq$ through $\vJJ=\vqq N$.
Hence the microscopic large deviation function ${\cal F}(\vqq)$ scales with 
the system size as ${\cal F}(\vqq)=N^{d-2} G(\vJJ=\vqq N)$, where $G(\vJJ)$ is now the current LDF appearing in the diffusively-scaled 
macroscopic fluctuation theory of \S\ref{sMFT}, see Eq. (\ref{ch1:pj}). 
This can be proved by noting that in the microscopic case 
we have $P(\vqq)=\pp(\vQ_t,t)\sim \exp[+t {\cal F}(\vqq)]$ while in macroscopic limit we have 
$P(\vJJ)\sim \exp[+\tau N^d G(\vJJ)]$.
Thus, by writing the latter probability in terms of the diffusive-scaled time variable $\tau=t/N^2$, we get
that $P(\vJJ=\vqq N)\sim \exp[+t N^{d-2} G(\vJJ=\vqq N)]$ which compared to the microscopic probabilty gives us the scaling mentioned above.
In a similar manner, 
if $\theta(\vla)=\max_{\vqq}[{\cal F}(\vqq) + \vla\cdot\vqq]$ and $\mu(\vla)=\max_{\vJJ}[G(\vJJ) + \vla\cdot\vJJ]$ are the Legendre 
transforms of ${\cal F}(\vqq)$ and $G(\vJJ)$, respectively, they are related via the simple scaling 
relation $\mu(\vla)=N^{2-d}\theta(\vla N^{d-1})$. 
} 
$\pp(\vQ_t,t)\sim \exp[+t {\cal F}(\vqq=\vQ_t/t)]$. In most cases it 
is convenient to work with the moment-generating function of the above distribution
\be
\Pi(\vla,t) \equiv \sum_{\vQ_t} \text{e}^{\vla \cdot \vQ_t} P(\vQ_t,t)  = \sum_{C_{t}..C_1} U_{C_t C_{t-1}}..U_{C_1 C_0} \, \text{e}^{\vla \cdot \sum_{k=0}^{t-1}\vqq_{C_{k+1} C_k}} \, .
\label{pi1}
\ee
For long $t$, we have {$\Pi(\vla,t) \to \exp[+t \theta(\vla)]$, where the new LDF $\theta(\vla)$ is connected to the current LDF via Legendre transform,
\be
\theta(\vla)= \max_\vqq [{\cal F}(\vqq) + \vla \cdot \vqq] = {\cal F}[\vqq^*(\vla)] + \vla \cdot \vqq^*(\vla) \, ,
\label{LegT}
\ee
with $\vqq^*(\vla)$ the current conjugated to parameter $\vla$, which is solution of the equation ${\cal F}'(\vqq^*) + \vla = 0$. We can now define a modified dynamics,
\be
\tilde{U}_{C' C}\equiv \text{e}^{\vla \cdot \vqq_{C' C}}\, U_{C' C} \, ,
\label{moddyn}
\ee
and therefore
\begin{equation}
\Pi(\vla,t)  = \sum_{C_{t}\ldots C_1} \tilde{U}_{C_t C_{t-1}} \ldots \tilde{U}_{C_1 C_0}  \, .
\label{pi2}
\end{equation}
Note however that this dynamics is not normalized, $\sum_{C'} \tilde{U}_{C' C}\neq 1$. We now introduce Dirac's bra and ket notation, useful in the context of the quantum Hamiltonian formalism for the master equation \cite{schutz,schutz2}, see also \cite{sim,Rakos}. The idea is to assign to each system configuration $C$ a vector $|C\ra$ in phase space, which together with its transposed vector $\la C |$, form an orthogonal basis of a complex space and its dual \cite{schutz,schutz2}. For instance, for systems with a finite number of available configurations, one could write $|C\ra^T = \la C|=(0 \ldots 0, 1, 0 \ldots 0)$, i.e. all components equal to zero except for the component corresponding to configuration $C$, which is $1$. In this notation, $\tilde{U}_{C' C}= \la C' | \tilde{U} | C \ra$, and a probability distribution can be written as a probability vector
\begin{equation}
| \pp(t) \ra = \sum_C \pp(C,t) |C\ra \nonumber \, ,
\end{equation}
where $\pp(C,t)=\la C| \pp(t) \ra$ with the scalar product $\la C' | C\ra = \delta_{C'C}$. If $\la s|=(1\ldots 1)$, normalization then implies $\la s|\pp(t)\ra =1$. With the previous notation, we can now write the spectral decomposition of \emph{operator} $\tilde{U}(\vla)$ as
\be
\tilde{U}(\vla)=\sum_j \text{e}^{\Lambda_j(\vla)} |\Lambda_j^R(\vla) \ra \la \Lambda_j^L(\vla) | \, ,
\label{specdecomp}
\ee
where we assume that a complete biorthogonal basis of right and left eigenvectors for matrix $\tilde{U}$ exists,
\be
\tilde{U} |\Lambda_j^R(\vla) \ra = \text{e}^{\Lambda_j(\vla)} |\Lambda_j^R(\vla) \ra \qquad \text{and} \qquad \la \Lambda_j^L(\vla)| \tilde{U} = \text{e}^{\Lambda_j(\vla)} \la \Lambda_j^L(\vla)| \, .
\label{eigU}
\ee
Denoting as $\text{e}^{\Lambda(\vla)}$ the largest eigenvalue of $\tilde{U}(\vla)$, with associated right and left eigenvectors $|\Lambda^R(\vla) \ra$ and $\la \Lambda^L(\vla)|$, respectively, and writing $\Pi(\vla,t) = \sum_{C_t} \la C_t | \tilde{U}^t | C_0 \ra$, see Eq. (\ref{pi2}), we find for long times
\begin{equation}
\Pi(\vla,t) \xrightarrow{t \gg 1} \text{e}^{+t \Lambda(\vla)}  \langle \Lambda^L(\vla) | C_0 \rangle
\left( \sum_{C_t} \langle C_t | \Lambda^R(\vla) \rangle \right)\, ,
\label{muasymp}
\end{equation}
where we have used the spectral decomposition (\ref{specdecomp}). In this way we have $\theta(\vla)=\Lambda(\vla)$, so the Legendre transform of the current LDF is given by the natural logarithm of the largest eigenvalue of $\tilde{U}(\vla)$.

\begin{figure}
\centerline{
\includegraphics[width=6cm]{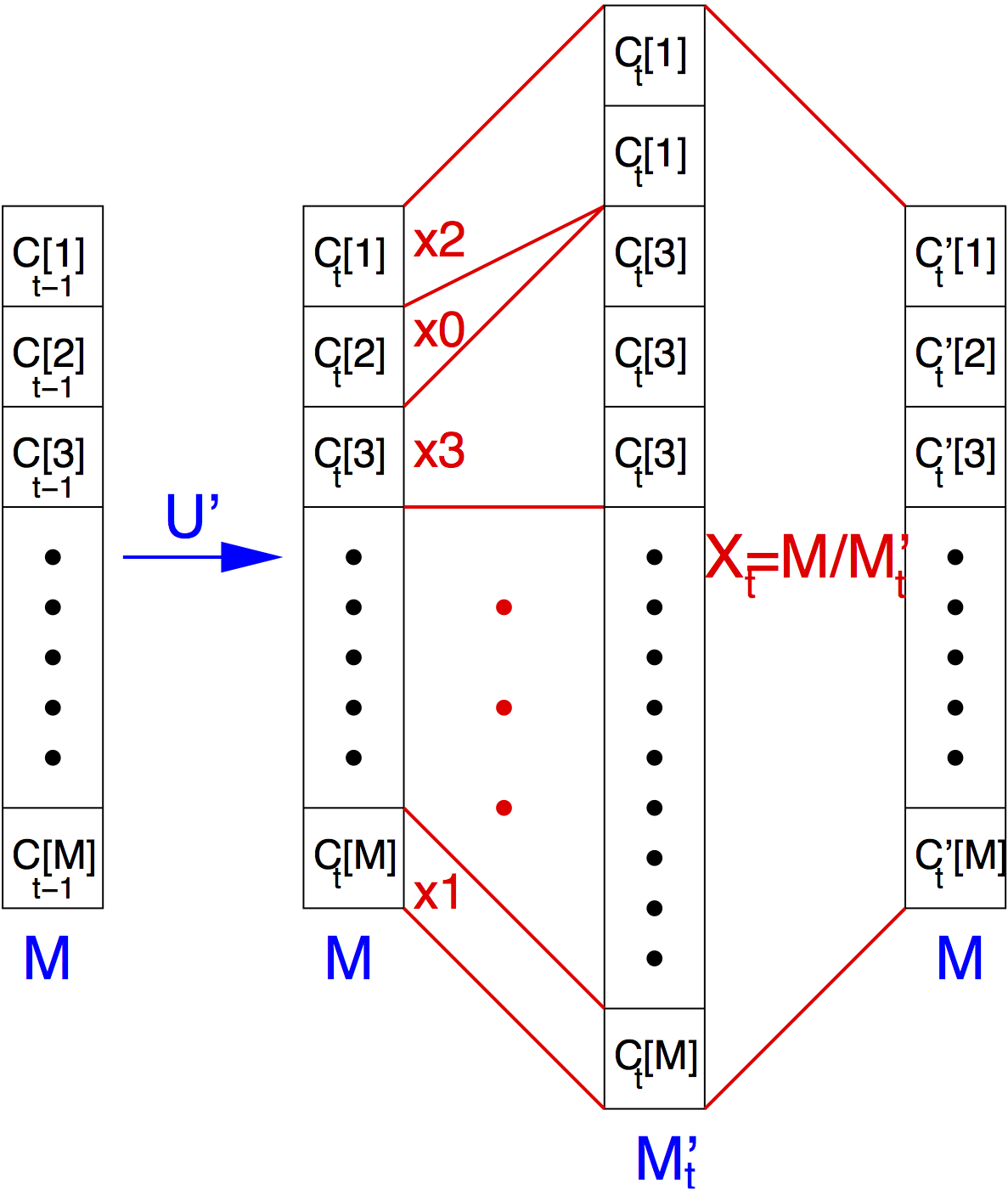}
}
\caption{(Color online) Sketch of the evolution and cloning of the copies during the evaluation of the large deviation function.}
\label{sketchclon}
\end{figure}

In order to measure this eigenvalue in Monte Carlo simulations, and given that dynamics $\tilde{U}$ is not normalized, we introduce the exit rates $Y_C=\sum_{C'} \tilde{U}_{C' C}$, and define the normalized dynamics $U'_{C' C}\equiv Y_C^{-1} \tilde{U}_{C' C}$. Now
\begin{equation}
{\Pi (\vla,t)=  \sum_{C_{t}\ldots C_1} Y_{C_{t-1}} U'_{C_t C_{t-1}} \ldots Y_{C_0} U'_{C_1 C_0} }
\label{pilambda}
\end{equation}
This sum over paths can be realized by considering an ensemble of $M \gg 1$ copies (or \emph{clones}) of the system, evolving sequentially according to the following Monte Carlo scheme\footnote{This simulation scheme is well-suited for discrete-time Markov chains. A slightly different though equivalent version of the algorithm exists for continous-time stochastic lattice gases \cite{sim2}.} \cite{sim}:
\begin{enumerate}
\item[(I)] Each copy evolves independently according to modified normalized dynamics $U'_{C' C}$.
\item[(II)] Each copy $m\in [1,M]$ (in configuration $C_t[m]$ at time $t$) is cloned with rate $Y_{C_t[m]}$. This means that, for each copy $m\in [1,M]$,
we generate a number $K_{C_t[m]}=\lfloor Y_{C_t[m]} \rfloor +1$ of identical clones with probability $Y_{C_t[m]} - \lfloor Y_{C_t[m]} \rfloor$, or
$K_{C_t[m]}=\lfloor Y_{C_t[m]} \rfloor$ otherwise (here $\lfloor x \rfloor$ represents the integer part of $x$). Note that if $K_{C_t[m]}=0$ the copy
may be killed and leave no offspring. This procedure gives rise to a total of $M'_t=\sum_{m=1}^M K_{C_t[m]}$ copies after cloning all of the original
$M$ copies.
\item[(III)] Once all copies evolve and clone, the total number of copies $M'_t$ is sent back to $M$ by an uniform cloning probability $X_t=M/M'_t$.
\end{enumerate}
Fig. \ref{sketchclon} sketches this procedure. It then can be shown that, for long times, we recover $\theta(\vla)$ via
\begin{equation}
\theta(\vla)  = -\frac{1}{t} \ln \left(X_t \cdots X_0 \right)   \qquad \text{for } t\gg 1
\label{musim}
\end{equation}
In order to derive this expression, first consider the cloning dynamics above, but without keeping the total number of clones constant, i.e. forgetting about step (III). In this case, for a given history $\{C_t,C_{t-1}\ldots C_1,C_0 \}$, the number ${\cal N}(C_t\ldots C_0,t)$ of copies in configuration $C_t$ at time $t$ obeys ${\cal N}(C_t\ldots C_0,t)=Y_{C_{t-1}} U'_{C_t C_{t-1}}  {\cal N}(C_{t-1}\ldots C_0,t-1)$, so that
\begin{equation}
{\cal N}(C_t\ldots C_0,t)=Y_{C_{t-1}} U'_{C_t C_{t-1}} \ldots Y_{C_0} U'_{C_1 C_0} {\cal N}(C_0,0) \, .
\label{fraccop}
\end{equation}
Summing over all histories of duration $t$, see Eq. (\ref{pilambda}), we find that the average of the total number of clones at long times shows
exponential behavior, $\la {\cal N} (t)\ra = \sum_{C_t\ldots C_1} {\cal N}(C_t\ldots C_0,t) \sim {\cal N}(C_0,0) \exp[+t \theta(\vla)]$. Now, going back to
step (III) above, when the fixed number of copies $M$ is large enough, we have $X_t = \la {\cal N} (t-1)\ra/\la {\cal N} (t)\ra$ for the global
cloning factors, so $X_t \cdots X_1 = {\cal N} (C_0,0)/\la {\cal N} (t)\ra$ and we recover expression (\ref{musim}) for $\theta(\vla)$.

In the following sections we apply this Monte Carlo method to measure in detail both the statistics of current fluctuations in some of the stochastic lattice gases described in \S\ref{smodels} as well as the optimal paths in phase space responsible of these rare events. These \emph{in silico} experiments are then confronted with the predictions derived within macroscopic fluctuation theory.

\section{Additivity of current fluctuations}
\label{saddit}

We now go back to macroscopic fluctuation theory and its predictions for the statistics of the space\&time-averaged current $\vJJ$, see \S\ref{sMFT}.A. Let us write again the current LDF as obtained within MFT
\beq
G(\vJJ)=-\lim_{\tau\rightarrow \infty}\frac{1}{\tau}
\min_{\{\rho,\vj\}_0^{\tau}}\left\{\mint_{0}^{\tau}dt \mint_{\Lambda} d\vrr\dfrac{(\vj (\vrr,t)-\vQ_{\vE}[\rho])^2}{2 \sigma[\rho]}\right\} \, .
\label{ch1:ldf00}
\eeq
This defines a highly complex variational problem in space and time for the optimal density and current fields, whose solution remains challenging in most cases of interest \cite{Bertini,Derrida}. However, the following hypothesis, well supported on physical grounds as we will see below, greatly simplify the complexity of the associated problem:
\begin{enumerate}
\item[(H1)] The optimal density and current fields responsible of a given current fluctuation are assumed to be time-independent, $\rho_{\vJJ}(\vrr)$ and $\vj_{\vJJ}(\vrr)$. This, together with the continuity equation (\ref{ch1:cont2}) which couples both fields, implies that the optimal current vector field is also divergence-free, $\vnabla\cdot \vj_{\vJJ}(\vrr)=0$.
\item[(H2)] A further simplification consists in assuming that this optimal current field has no spatial structure, i.e. is constant across space, which implies together with constraint (\ref{ch1:currint}) on the current that $\vj_{\vJJ}(\vrr)=\vJJ$.
\end{enumerate}
The physical picture behind these hypotheses corresponds to a system that, after a short transient  time at the beginning of the large deviation event (microscopic in the diffusive timescale $\tau$), settles into a time independent state with an structured density field (which can be different from the stationary one) and a spatially uniform current field equal to $\vJJ$.
This behavior is expected to minimize the cost of a fluctuation at least for small and moderate deviations from the average behavior. Hypotheses (H1)-(H2) are the straightforward generalization to high-dimensional systems ($d\ge 1$) of the additivity principle introduced by Bodineau and Derrida for one-dimensional (1D) diffusive media \cite{BD}. As we shall see below, the validity of this conjecture has been checked numerically to a high degree of accuracy for some stochastic transport models in a wide interval of current fluctuations, though it is known that additivity may be violated in some particular cases for large enough fluctuations, where time-dependent optimal paths in the form of traveling waves emerge as dominant solution to the variational problem (\ref{ch1:ldf0}). We will analyze this additivity breakdown below.
Provided that hypotheses (H1)-(H2) hold, the current LDF (\ref{ch1:ldf00}) can be written as  \cite{PabloIFR,Derrida,BD,PabloAIP}
\beq
G(\vJJ) =-\min_{\rho(\vrr)} \int_{\Lambda} \frac{\displaystyle \left(\vJJ -\vQ_{\vE}[\rho]\right)^2} {\displaystyle 2 \sigma[\rho(\vrr)] } d \vrr \, ,
\label{ch1:ldf2}
\eeq
In this way the probability $\text{P}_{\tau}(\vJJ)$ is simply the Gaussian weight associated with the optimal density field responsible for such fluctuation. Note however that the minimization procedure gives rise to a nonlinear problem which results in general in a current distribution with non-Gaussian tails \cite{Derrida,Bertini,PabloPRL,PabloPRE}. As opposed to the general problem in Eq. (\ref{ch1:ldf00}), its simplified version, Eq. (\ref{ch1:ldf2}), can be readily used to obtain quantitative predictions for the current statistics in a large variety of non-equilibrium systems.
The optimal density profile $\rho_{\vJJ}(\vrr)$  is now solution of the following equation
\beq
\frac{\delta \pi_2[\rho(\vrr)]}{\delta \rho(\vrr')} - 2\vJJ\cdot \frac{\delta \vpi_1[\rho(\vrr)]}{\delta \rho(\vrr')} +
\vJJ^2 \frac{\delta \pi_0[\rho(\vrr)]}{\delta \rho(\vrr')}  = 0 \, ,
\label{ch1:optprof}
\eeq
which must be supplemented with appropriate boundary conditions. In the above equation, $\frac{\delta}{\delta \rho(\vrr')}$ stands for functional derivative, and
\beq
\vpi_n[\rho(\vrr)] \equiv  \int_{\Lambda} d\vrr \, \textbf{W}_n[\rho(\vrr)] \qquad \text{with} \qquad \textbf{W}_n[\rho(\vrr)]\equiv \frac{\vQ_{\vE}^n[\rho(\vrr)]}{\sigma[\rho(\vrr)]} \, .
\label{ch1:defs1}
\eeq
We will be interested below in diffusive systems without external field. In this case
$\vQ_{\vE=0}[\rho]=-D[\rho]\vnabla\rho$, and the resulting differential equation (\ref{ch1:optprof}) for the optimal profile takes the form
\beq
\vJJ^2 a'[\rho_{\vJJ}]-c'[\rho_{\vJJ}](\vnabla \rho_{\vJJ})^2-2c[\rho_{\vJJ}]\vnabla^2 \rho_{\vJJ}=0,
\label{ch1:optprofdiff}
\eeq
where $a[\rho_{\vJJ}]=(2\sigma[\rho_{\vJJ}])^{-1}$, $c[\rho_{\vJJ}]=D^2[\rho_{\vJJ}]a[\rho_{\vJJ}]$, and $'$ denotes the derivative with respect to the argument. Multiplying the above equation by $\vnabla \rho_{\vJJ}$, we obtain after one integration step
\beq
D[\rho_{\vJJ}]^2 (\vnabla \rho_{\vJJ})^2=\vJJ^2\left(1+2\sigma[\rho_{\vJJ}]K(\vJJ^2)\right)
\label{ch1:optprofdiffcomp}
\eeq
where $K(\vJJ^2)$ is a constant of integration which guarantees the correct boundary conditions for $\rho_{\vJJ}(\vrr)$. Eqs. (\ref{ch1:ldf2}) and (\ref{ch1:optprofdiffcomp}) then completely determine the current distribution $\text{P}_{\tau}(\vJJ)$ in diffusive media, which is in general non-Gaussian (except for small current fluctuations). Explicit solutions to these equations for particular models of diffusive transport in varying dimensions can be obtained \cite{PabloPRE,PabloAIP}. Appendix \ref{sadditKMP} summarizes the calculation for the KMP model of energy transport, for which we explore below its current statistics using the advanced Monte Carlo methods of \S\ref{ssim}.

Before turning to numerics notice that, as described above, hypotheses (H1)-(H2) have been shown \cite{Bertini} to be equivalent to the additivity principle 
for 1D diffusive systems \cite{BD}. To understand its original formulation, let $\text{P}_N(J,\rho_L,\rho_R,\tau)$ be the probability of observing a 
time-averaged current $J$ during a long time $\tau$ in a 1D system of size $N$ in contact with boundary reservoirs at densities $\rho_L$ and $\rho_R$. 
The additivity principle relates this probability with the probabilities of sustaining the same current in subsystems of 
lengths $N-\ell$ and $\ell$, i.e, \footnote{Note that this is the original formulation of the additivity principle for the integrated current stated by Bodineau and 
Derrida in \cite{BD}. In \cite{BGL}, Bertini et al. state an additivity principle for the density field which involves either a maximization 
or a minimization depending on the convexity of the rate functional for the considered model.}
\be
P_{N}(J,\rho_L,\rho_R,\tau) = \max_{\rho}[P_{N-\ell}(J,\rho_L,\rho,\tau)\times P_{\ell}(J,\rho,\rho_R,\tau)] \, .
\label{addit1}
\ee
The maximization over the contact density $\rho$ can be rationalized by writing this probability as an integral over $\rho$ of the product of probabilities for subsystems and noticing that these should obey also a large deviation principle. Hence a saddle-point calculation in the long-$\tau$ limit leads to the above expression. The additivity principle can be rewritten for the current LDF as $N G(J,\rho_L,\rho_R) = \max_{\rho}[(N-\ell) G(J,\rho_L,\rho) + \ell G(J,\rho,\rho_R)$. Slicing iteratively the 1D system of length $N$ into smaller and smaller segments, and assuming locally-Gaussian current fluctuations, it is easy to show that in the continuum limit a variational form for $G(J,\rho_L,\rho_R)$ is obtained which is just the 1D counterpart of Eq. (\ref{ch1:ldf2}). Interestingly, for 1D systems the conjecture of time-independent optimal profiles implies that the optimal current profile must be constant. This is no longer true in higher dimensions, as any divergence-free current field with spatial integral equal to $\vJJ$ is compatible with the equations. This gives rise to a variational problem with respect to the (time-independent) density and current fields which still poses many technical difficulties. Therefore an additional assumption is needed, namely the constancy of the optimal current vector field across space. These two hypotheses are equivalent to the iterative procedure of the additivity principle in higher dimensions.

\subsection{Testing the additivity conjecture in one and two dimensions}
\label{saddittest}

The additivity principle previously described provides a relatively simple and straightforward recipe to compute the statistics of typical and rare current fluctuations,
opening the door to the systematic calculation of large deviation statistics in general nonequilibrium systems. It is a very general conjecture of broad applicability, expected to hold for a large family of systems of classical interacting particles, both deterministic or stochastic, in arbitrary dimension and independently of the details of the interactions between the particles or the coupling to the thermal reservoirs or external fields. Furthermore, equivalent results to those obtained with the additivity principle have been derived for interacting quantum systems \cite{Derrida}. 
The only requirement is that the system at hand must be diffusive, i.e. described by a mesoscopic evolution equation of the form of 
Eq. (\ref{ch1:langevin})\footnote{Note however that the additivity hypothesis has been recently confirmed in Hamiltonian models with anomalous, 
non-diffusive transport properties \cite{additHam}. These results considerably broaden the range of applicability of the additivity conjecture.}, and that the 
prior hypotheses H1 and H2 hold.
If this is the case, the additivity principle predicts the full current distribution in terms of its first two cumulants.
Moreover, the additivity conjecture can be applied to a multitude of different situations once appropriately generalized \cite{prados1,prados3}.

\begin{figure}
\centerline{
\includegraphics[width=16cm]{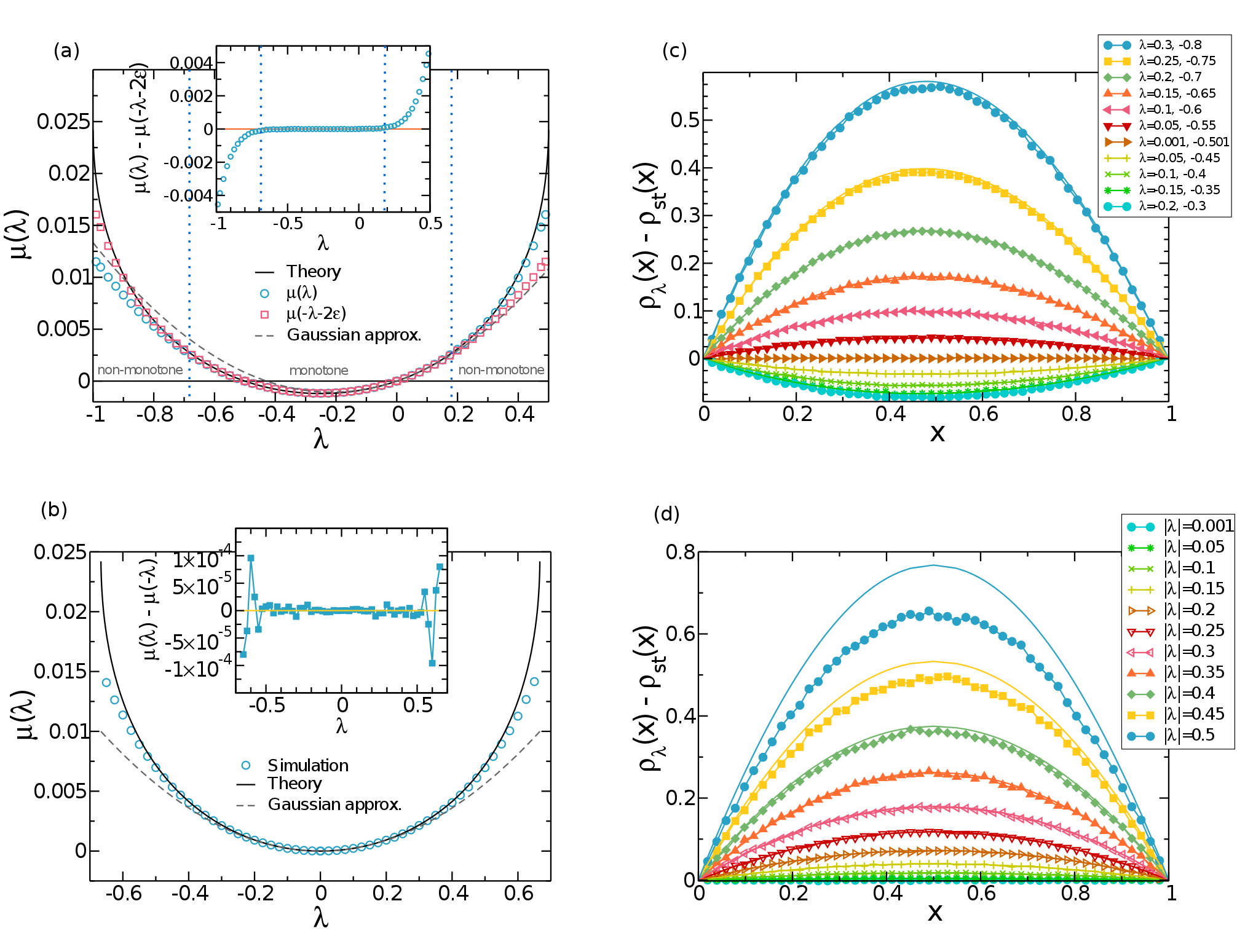}
}
\caption{(Color online) (a) Legendre transform of the current LDF for the KMP model in one dimension with a temperature gradient ($\rho_L=2$, $\rho_R=1$) and (b) in equilibrium ($\rho_L=1.5=\rho_R$). Symbols correspond to numerical simulations, full lines to theoretical predictions based on the additivity conjecture, and dashed lines to Gaussian approximations (see text). Errorbars (with 5 standard deviations) are always smaller than symbol sizes. The vertical dotted lines in panel (a) signal the transition between monotone (inner region) or non-monotone (outer region) optimal profiles.
Note that in equilibrium profiles are non-monotone for all current fluctuations.
The inset in the panels (a) and (b) tests the Gallavotti-Cohen relation by plotting the difference $\mu(\lambda)-\mu(-\lambda-2{\cal E})$ in case (a) and $\mu(\lambda)-\mu(-\lambda)$ in (b).
Right panels correspond to excess temperature profiles for different current fluctuations, (c) for a system subject to a temperature gradient, ($\rho_L=2$, $\rho_R=1$) and (d) in equilibrium ($\rho_L=1.5=\rho_R$). In all cases, agreement with theoretical predictions based on the additivity hypothesis (lines) is very good within the range of validity of the computational method.}
\label{figadit1}
\end{figure}

It is therefore essential to test the emerging picture in detailed numerical experiments to confirm the validity of this hypothesis and asses its range of applicability. With this aim in mind, we have recently performed extensive numerical simulations of the one- and two-dimensional KMP models of energy transport in open lattices subject to a boundary-induced temperature gradient, $T_L\neq T_R$ \cite{PabloPRL,PabloPRE,Carlostesis}.
In particular, we applied the cloning Monte Carlo method of \S\ref{ssim} \cite{sim,sim2,sim3} to measure the Legendre transform of the current LDF, defined as
\be
\mu(\vla)=\max_{\vJJ}[G(\vJJ) + \vla \cdot \vJJ] \, .
\label{mulamb}
\ee
The method of \S\ref{ssim} yields the \emph{macroscopic} LDF $\mu(\vla)$ in terms of the logarithm of the largest eigenvalue of the modified dynamics $\tilde{U}(\vla)$ via the simple scaling relation $\mu(\vla)=N^{2-d}\theta(\vla N^{d-1})$ (see footnote of page 11) for a system of linear size $N$ in dimension $d$, where $\theta(\vla)$ is the \emph{microscopic} LDF, see Eq. (\ref{LegT}) above.

We first measured $\mu(\lambda)$ for the 1D KMP model with $N=50$, $T_L=2$ and $T_R=1$, see Fig. \ref{figadit1}.a, where we compare simulation data with predictions derived from macroscopic fluctuation theory once supplemented with the additivity conjecture (theory denoted as MFT$_{\text{Ad}}$ hereafter). Explicit details of the theory can be found in Appendix \ref{sadditKMP}, where it is shown that, for the particular case of the KMP model, the optimal profiles solution of Eq. (\ref{ch1:optprofdiffcomp}) can be either monotone for small current fluctuations or non-monotone with a single maximum for larger fluctuations. The agreement between the measured $\mu(\lambda) $ and MFT$_{\text{Ad}}$, see 
Fig. \ref{figadit1}.a, is excellent for a wide $\lambda$-interval, say $-0.8<\lambda<0.45$, which corresponds to a large range of current 
fluctuations, say $-1.5<J<4$ (see\footnote{See inset of figure 11 of Ref. \cite{PabloPRE} where J versus $\lambda$ is displayed.}) (note that $\lambda$-space is bounded, $\lambda\in[-T_R^{-1},T_L^{-1}]$ while the current-space is not, see Appendix \ref{sadditKMP}). Moreover, the deviations observed for extreme current fluctuations (i.e. large values of $|\lambda|$) are due to known limitations of the algorithm \cite{PabloPRL,sim,sim2,PabloJSTAT,sim3}, so no violations of additivity are observed in this context. Notice that these spurious differences seem to occur earlier for currents \emph{against the gradient}, i.e. $\lambda <0$. These deviations can be traced back to sampling biases introduced by the cloning Monte Carlos scheme for finite number of clones, and extreme value statistics can be used to derive bounds in $\lambda$-space as a function of the clone population for the applicability of the cloning method \cite{PabloJSTAT}. Interestingly, we can use the Gallavotti-Cohen symmetry, which in $\lambda$-space now reads $\mu(\lambda)=\mu(-\lambda-2\epsilon)$ with the driving force $\epsilon= (T_R^{-1}-T_L^{-1})/2$, to offer a blind bound for the range of validity of the algorithm: Violations of the fluctuation relation signal the onset of the systematic bias in the estimations provided by the method of Ref. \cite{sim}. Fig. \ref{figadit1}.a and its inset show that the Gallavotti-Cohen symmetry holds in the large current interval for which the additivity principle predictions agree with measurements, thus confirming its validity in this range. However, we cannot discard the possibility of an additivity breakdown for extreme current fluctuations due to the onset of time-dependent optimal profiles expected in general in MFT \cite{Bertini,BD2,PabloSSB,CarlosSSB}, although we stress that such scenario is not observed here. We will explore this interesting possibility in \S\ref{sSSB} below for the same model with periodic boundary conditions.

We also measured the current LDF in canonical equilibrium, i.e. for $T_L=T_R=1.5$, see Fig. \ref{figadit1}.b. The agreement with MFT$_{\text{Ad}}$ is again excellent within the range of validity of our measurements, which expands a wide current interval, see inset to Fig. \ref{figadit1}.b, where we show that the fluctuation relation is verified except for extreme currents deviations, for which the algorithm fails to provide reliable results. Notice that, both in the presence of a temperature gradient and in canonical equilibrium, $\mu(\lambda)$ is parabolic around $\lambda=0$ meaning that current fluctuations are approximately Gaussian for $J\approx J_{st}$, as demanded by the central limit theorem, see eqs. (\ref{Gsmallq})-(\ref{musmalllambda}) in Appendix \ref{sadditKMP}. This observation is particularly interesting in equilibrium, where large fluctuations in canonical and microcanonical ensembles behave differently (see below).

The additivity principle leads to the minimization of a functional of the density field, $\rho_J(x)$, see  eqs. (\ref{ch1:ldf2}) and (\ref{ch1:optprofdiffcomp}). A relevant question is whether this optimal field is actually observable. We naturally define $\rho_J(x)$ in simulations as the average \emph{energy} profile adopted by the system during a large deviation event of (long) duration $t=\tau N^2$ and time-integrated current $J t$, measured at an \emph{intermediate time} $1\ll t' \ll t$ \cite{PabloPRL,PabloPRE}. Figs. \ref{figadit1}.c-d show the measured $\rho_{\lambda}(x)$ for both the nonequilibrium (c) and equilibrium (d) settings, and the agreement with MFT$_{\text{Ad}}$ predictions is again very good in all cases, with discrepancies appearing only for extreme  current fluctuations, as otherwise expected. Notice that Fig. \ref{figadit1}.c include data both in the monotone and non-monotone profile regimes, see Appendix \ref{sadditKMP}. These observations confirm the idea that the system indeed modifies its density profile to facilitate the deviation of the current, validating the additivity principle as a powerful conjecture to compute both the current LDF and the associated optimal profiles.

Notice that in the canonical equilibrium case ($T_L=T_R$) optimal density profiles are always non-monotone, see Appendix \ref{sadditKMP}, with a single maximum for any current fluctuation $J\neq J_{st}$ (the stationary profile is obviously flat). This is in stark contrast to the behavior predicted for current fluctuations in \emph{microcanonical} equilibrium, i.e. for a one-dimensional closed diffusive system on a ring \cite{Bertini,PabloSSB}, see \S\ref{sSSB} below. In this case the optimal profiles remain flat and current fluctuations are Gaussian up to a critical current value, at which profiles become time-dependent (traveling waves) \cite{PabloSSB}. Hence current statistics can differ considerably depending on the particular equilibrium ensemble at hand, despite their equivalence for average quantities in the thermodynamic limit.

Remarkably, our numerical results show that optimal density fields in equilibrium and nonequilibrium are indeed independent of the sign of the current, $\rho_{\lambda}(x)=\rho_{-\lambda-2\epsilon}(x)$ or equivalently $\rho_J(x)=\rho_{-J}(x)$, a counter-intuitive symmetry resulting (as the Gallavotti-Cohen fluctuation theorem) from the reversibility of microscopic dynamics\footnote{For equilibrium dynamics, $T_L=T_R$, this symmetry is obvious as the system has $x\leftrightarrow 1-x$ symmetry.}. We will show in \S\ref{sGC} below from a microscopic point of view that not only the optimal density field, but the whole statistics during a current large deviation event, remain invariant under  current sign reversal \cite{PabloPRE,PabloJSTAT}.

As a final note, we just mention that our numerical simulations can be used to explore the physics beyond the additivity conjecture by studying the fluctuations of the total energy in the system, which exhibit the trace left by corrections to local equilibrium resulting from the presence of weak long-range correlations in the nonequilibrium steady state \cite{PabloPRL,PabloPRE}. In addition, one can extend the additivity hypothesis to study the joint, coupled fluctuations of the current and the density profile which appear for long but finite times, when the density profile associated with a given current fluctuation is subject to fluctuations itself \cite{PabloPRL,PabloPRE}.

\begin{figure}
\centerline{
\includegraphics[width=16cm]{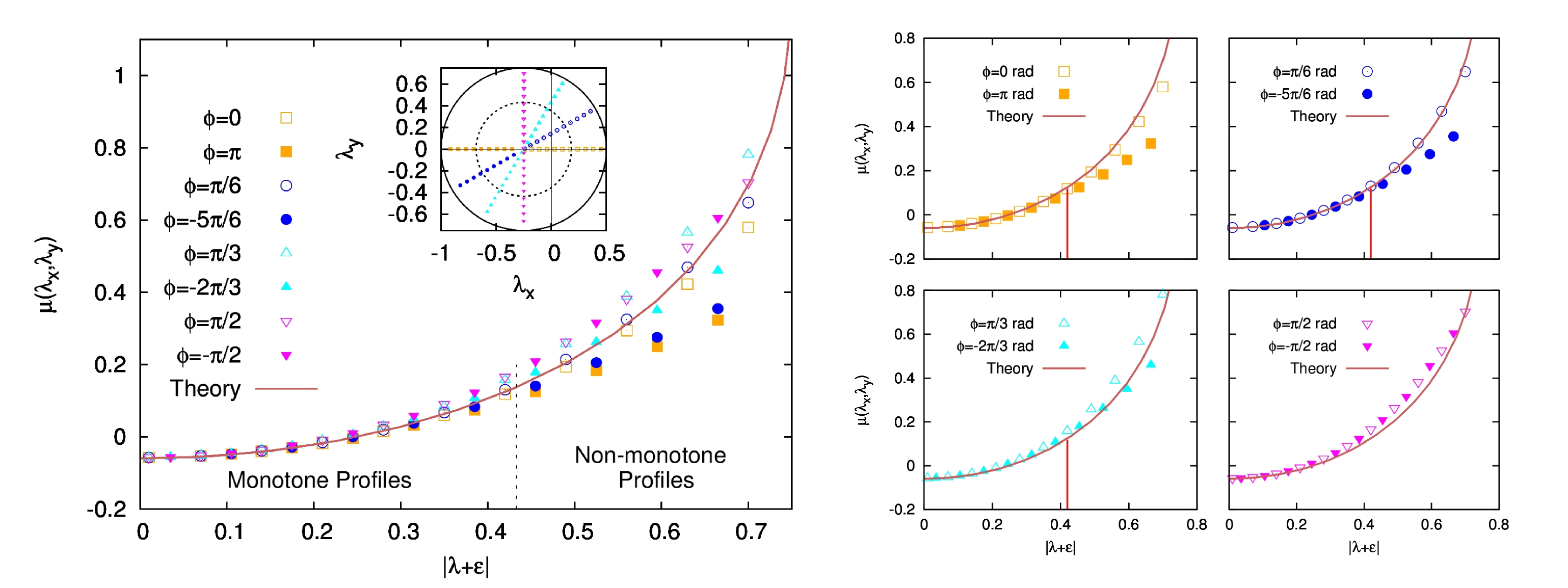}
}
\caption{(Color online) (a) Measured $\mu(\vla)$ for the 2D KMP model with $\rho_L=2$, $\rho_R=1$ and $N=20$ for different angles versus $|\vla+\vecep|$. The solid line corresponds to the theoretical prediction. Inset: Points in $\vla$-space for different angles where measurements are taken. The dashed circle is where the non-monotone regime starts. (b) Measured $\mu(\vla)$ with $\rho_L=2$, $\rho_R=1$, $N=20$ and $10^3$ clones for different pairs of angles $(\varphi,\pi-\varphi)$ corresponding to opposite currents, which are coupled by time reversibility (see the inset of panel (a)). The red vertical line indicates the threshold value of $|\vla+\vecep|$ up to which the GC symmetry holds.
}
\label{figadit2}
\end{figure}

One-dimensional systems are oversimplified models of reality. In order to establish the generality and usefulness of the additivity conjecture to compute large deviation statistics in general nonequilibrium systems,
it is mandatory to test additivity in more complex, higher-dimensional systems. In order to do so,
we have measured the statistics of the space\&time-averaged current $\vJJ$ for the 2D KMP model with linear size $N=20$, $T_L=2$ and $T_R=1$ using both standard simulations and the advanced Monte Carlo technique of \S\ref{ssim} \cite{Carlostesis}.
In particular, we measured the current statistics as a function of the magnitude of the current vector for different current orientations, i.e. for different angles $\varphi=\tan^{-1}\left({J_y}/{J_x}\right)$, where $J_{\alpha}$ is the $\alpha$-component of the current vector $\vJJ$. Note that in the conjugated $\vla$-space, this angle can be written as $\varphi=\tan^{-1}\left({\lambda_y}/(\lambda_x +\epsilon_x)\right)$ If $\epsilon_y=0$. MFT$_{\text{Ad}}$ predicts that the Legendre transform of the current LDF, $\mu(\vla)$, depends exclusively on $|\vla+\vecep|$ but not on $\varphi$. As we will see in \S\ref{sIFR}, this is not a mathematical curiosity but a deep result related with hidden symmetries of the current LDF. Fig. \ref{figadit2}.a shows the measured $\mu(\vla)$ as a function of $|\vla+\vecep|$, for different constant angles $\varphi$ (see inset to Fig. \ref{figadit2}.a), together with the MFT$_{\text{Ad}}$ prediction. We observe that there is a good agreement for a broad interval of current fluctuations such that $|\vla+\vecep|\leq 0.35$.
From this value on clear deviations from the theoretical predictions are observed which depend on $\varphi$. The origin of such disagreement is twofold: (i) standard finite size effects, as MFT is a macroscopic theory but we could only simulate reliably systems of small size ($N\leq 32$), and (ii) a different class of finite size effects related to the finite number of clones used to sample the large-deviation statistics \cite{PabloJSTAT}. As for the 1D case above, we can use the Gallavotti-Cohen (GC) symmetry to detect the regime where the finite population of clones introduces a bias in simulation results.
Consequently, in figure \ref{figadit2}.b we compare the LDF for current fluctuations coupled by time reversibility by plotting $\mu(\vla)$ versus $|\vla+\vecep|$, for 
a fixed pairs of angles $\varphi$ and $\varphi-\pi$ (see the inset of figure \ref{figadit2}.b). This analysis shows that GC holds to a good degree of accuracy 
for $|\vla+\vecep|\lesssim 0.42$, a critical value above which Monte Carlo results are biased due to the finite population of clones. In this way, the disagreement 
with the theory for $0.35\leq |\vla+\vecep|\leq 0.42$ can be then traced back to standard finite size effects (small $N$). This can be corroborated by studying the 
dependence of $\mu(\vla)$ on the system size (see Fig. \ref{figIFR}.d below): a slow but clear convergence toward the theoretical prediction is observed for all 
angles $\varphi$ as $N$ grows \cite{Carlostesis}. Therefore, as for the 1D case, and excluding the different finite size effects discussed, no violations of 
additivity are observed in 2D, confirming the validity of this hypothesis in higher dimensions.

\subsection{Invariance of rare event statistics under current reversal}
\label{sGC}

We have shown numerically in the previous section that the measured optimal density field associated with a current fluctuation does not depend on the current sign, i.e. $\rho_{\vJJ}(\vrr)=\rho_{-\vJJ}(\vrr)$, in agreement with the MFT$_{\text{Ad}}$ prediction. In fact,
Eq. (\ref{ch1:optprofdiffcomp}) for the optimal profile clearly shows that this object is independent of the sign of the current, $\vJJ \to -\vJJ$. Such counter-intuitive symmetry results from the time reversibility of microscopic (stochastic) dynamics, and goes hand by hand with the Gallavotti-Cohen fluctuation theorem \cite{GC}, see Eq. (\ref{GCtheo}). In fact, it can be shown  starting from the microscopic, Markov-chain description of the large deviation problem \cite{PabloPRE,PabloJSTAT} that not only the optimal profile remains invariant under current reversal,
but also the whole statistics during the large deviation event.

To show this remarkable invariance of rare-event statistics under time reversal, we must first define time-reversibility in stochastic dynamics. 
The condition that plays the role of the time-reversal invariance of deterministic dynamics in stochastic systems is known 
as \emph{local detailed balance} \cite{LS}, and reads
\be
p_{\text{eff}}(C) \, U_{C' C}  = p_{\text{eff}}(C') \, U_{C C'}  \text{e}^{2\vecep \cdot \vqq_{C' C}} \, ,
\label{ldb}
\ee
where $\vecep$ is the driving force, $p_{\text{eff}}(C)$ is an effective statistical weight for configuration $C$ different 
from the steady-state measure which for the 1D KMP model 
takes the form $p_{\text{eff}}(C)=\exp[-\sum_{i=1}^N \beta_i e_i]$ with $e_i$ being the energy of each site and $\beta_i=T_L^{-1}+2\vecep \frac{i-1}{N-1}$ \cite{PabloPRE}. 
Recall that $U_{C' C}$ is the transition 
rate for the jump $C\to C'$, which involves an elementary current $\vqq_{C' C}$.
We may now use the modified dynamics $\tilde{U}(\vla)$ defined in \S\ref{ssim} to write 
condition (\ref{ldb}) as $\tilde{U}_{C C'}(\vla)=p_{\text{eff}}^{-1}(C') \tilde{U}_{C' C}(-\vla-2\vecep)p_{\text{eff}}(C)$,
or in matrix form
\begin{equation}
\tilde{U}^{\text{T}} (\vla) = \pp_{\text{eff}}^{-1} \tilde{U}(-\vla-2\vecep) \pp_{\text{eff}} \, ,
\label{transpose}
\end{equation}
where $\pp_{\text{eff}}$ is a diagonal \emph{matrix} with entries $\la C|\pp_{\text{eff}}|C'\ra=p_{\text{eff}}(C) \delta_{C,C'}$. Eq. (\ref{transpose}) implies a 
symmetry between the modified dynamics for a current fluctuation and the modified dynamics for the negative current fluctuation. In particular, the similarity 
relation (\ref{transpose}) implies that all eigenvalues of $\tilde{U}(\vla)$ and $\tilde{U}(-\vla-2\vecep)$ are equal, and in particular the largest, so
\be
\theta(\vla)=\theta(-\vla-2\vecep) \, .
\label{GCtheta}
\ee
This is just the Gallavotti-Cohen fluctuation theorem, ${\cal F}(\vqq)-{\cal F}(-\vqq)=2\vecep\cdot\vqq$, see also Eq. (\ref{GCtheo}), written in terms of the the Legendre transform of the current LDF.
The similarity relation (\ref{transpose}) can be further exploited to show that the statistics during a current large deviation event remains invariant under time reversal. Let $\bar{\pp}(C_{t'},\vQ_t,t',t)$ be the probability that the system was in configuration $C_{t'}$ at time $t'$ when at time $t$ the total integrated current is $\vQ_t$. Timescales are such that $1 \ll t' \ll t$, so all times involved are long enough for the memory of the initial state $C_0$ to be lost. We can write now
\begin{equation}
\bar{\pp}(C_{t'},\vQ_t,t',t) = \sum_{C_{t}\ldots C_{t'+1} C_{t'-1} \ldots  C_1} U_{C_t C_{t-1}} \cdots U_{C_{t'+1} C_{t'}}
U_{C_{t'} C_{t'-1}} \cdots U_{C_1 C_0}\, \delta \Big(\vQ_t - \sum_{k=0}^{t-1}\vqq_{C_{k+1} C_k}\Big) \, ,
\label{probCmid}
\end{equation}
where we do not sum over $C_{t'}$. Defining the moment-generating function of the above distribution,
\bea
\bar{\Pi}(C_{t'},\vla,t',t)&=&\sum_{\vQ_t} \text{e}^{\vla\cdot \vQ_t} \bar{\pp}(C_{t'},\vQ_t,t',t) = \sum_{C_{t}\ldots C_{t'+1} C_{t'-1} \ldots  C_1} \tilde{U}_{C_t C_{t-1}} \cdots \tilde{U}_{C_{t'+1} C_{t'}} \tilde{U}_{C_{t'} C_{t'-1}} \cdots \tilde{U}_{C_1 C_0} \nonumber \\
&=& \sum_{C_{t}} \la C_t |  \tilde{U}^{t-t'} |C_{t'}\ra \la C_{t'} |  \tilde{U}^{t'} |C_0\ra   \, ,
\label{piC}
\eea
it is easy to show that for long times such that $1 \ll t' \ll t$ the probability weight of configuration $C_{t'}$ at intermediate time $t'$ in a 
large deviation event of current $\vqq=\vQ_t/t$ can be written as
\be
\pp_{\vqq}(C_{t'}) \equiv \frac{\bar{\pp}(C_{t'},\vQ_t,t',t)}{\pp(\vQ_t,t)}=  \frac{\bar{\Pi}(C_{t'},\vla,t',t)}{\Pi(\vla,t)} \equiv P_{\vla}(C_{t'})
\label{probCq}
\ee
where $\vqq$ and $\vla$ are conjugated parameters univocally related via Legendre transform, $\vqq=\vqq^*(\vla)$, 
see Eq. (\ref{LegT}) above.
This relation is easily demonstrated by noting that 
\bea
\pp_{\vla}(C_{t'})=\frac{\bar{\Pi}(C_{t'},\vla,t',t)}{\Pi(\vla,t)}=\frac{1}{\Pi(\vla,t)}\sum_{\vQ_t} \text{e}^{\vla\cdot \vQ_t} \bar{\pp}(C_{t'},\vQ_t,t',t)=
\frac{\pp_{\vqq}(C_{t'})}{\Pi(\vla,t)}\sum_{\vQ_t} \text{e}^{\vla\cdot \vQ_t} \pp(\vQ_t,t)\xrightarrow{t \gg 1} \nonumber\\
\xrightarrow{t \gg 1}\pp_{\vqq}(C_{t'})\frac{\text{e}^{t\max_{\vqq}[F(\vqq)+\vla\cdot\vqq]}}{\Pi(\vla,t)}=\pp_{\vqq}(C_{t'}),
\nonumber
\eea
where we have used in the last step that in the long time limit $\pp(\vQ_t,t)\sim \text{e}^{tF(\vqq)}$ and $\Pi(\vla,t)\sim \text{e}^{t\mu(\vla)}$ with 
$\mu(\vla)=\max_{\vqq}[F(\vqq)+\vla\cdot\vqq]$.
Using the spectral decomposition of the operator $\tilde{U}(\vla)$, Eq. (\ref{specdecomp}), one thus finds
\begin{equation}
\pp_{\vla}(C_{t'}) = \dfrac{\la C_{t'} |  
\tilde{U}^{t'} |C_0\ra\sum_{C_{t}} \la C_t |  \tilde{U}^{t-t'} |C_{t'}\ra}{\sum_{C_{t}} \la C_t |  \tilde{U}^{t} |C_{0}\ra},
\nonumber
\end{equation}
so in the long time limit we arrive at
\begin{equation}
\pp_{\vla}(C_{t'}) \xrightarrow{t \gg 1} \la \Lambda^L(\vla) | C_{t'} \ra \la C_{t'} | \Lambda^R(\vla) \ra \, .
\label{profmid1}
\end{equation}
Here $|\Lambda^R(\vla)\ra$ and $\la \Lambda^L(\vla)|$ are the right and left eigenvectors associated with the largest eigenvalue $\text{e}^{\Lambda(\vla)}$
of modified transition rate $\tilde{U}(\vla)$, respectively. They are different because $\tilde{U}$ is not symmetric. Therefore the probability of a configuration during a large fluctuation of the current is proportional to the projection of this configuration on the left and right eigenvectors associated with the largest eigenvalue of $\tilde{U}(\vla)$.

Interestingly, $|\Lambda^L(\vla) \ra$ is the \emph{right} eigenvector of the transpose matrix $\tilde{U}^{\text{T}}(\vla)$ with 
eigenvalue $\text{e}^{\Lambda(\vla)}$. This right eigenvector of $\tilde{U}^{\text{T}}(\vla)$ can be in turn related to the corresponding right 
eigenvector $|\Lambda^R(-\vla-2\vecep)\ra$ of matrix $\tilde{U}(-\vla-2\vecep)$ using the similarity relation (\ref{transpose})
which stems from the local detailed balance condition (\ref{ldb}).
Using the basis expansion $|\Lambda^R(-\vla-2\vecep)\ra = \sum_C \la C | \Lambda^R(-\vla-2\vecep)\ra \, |C\ra$, it is easy to show that
\begin{equation}
|\Lambda^L(\vla)\ra = \sum_C  \frac{\la C | \Lambda^R(-\vla-2\vecep)\ra}{p_{\text{eff}}(C)} |C\ra \, ,
\label{eigenleft}
\end{equation}
is the right eigenvector of $\tilde{U}^{\text{T}}(\vla)$ with eigenvalue $\text{e}^{\Lambda(\vla)}$.  In fact,
\bea
\tilde{U}^{\text{T}}(\vla) \, |\Lambda^L(\vla)\ra &=& \pp_{\text{eff}}^{-1} \tilde{U}(-\vla-2\vecep) \pp_{\text{eff}} \sum_C  \frac{\la C | \Lambda^R(-\vla-2\vecep)\ra}{p_{\text{eff}}(C)} |C\ra = \pp_{\text{eff}}^{-1} \tilde{U}(-\vla-2\vecep) \, |\Lambda^R(-\vla-2\vecep)\ra \nonumber\\
&=&  \text{e}^{\Lambda(\vla)} \sum_C  \frac{\la C | \Lambda^R(-\vla-2\vecep)\ra}{p_{\text{eff}}(C)} |C\ra = \text{e}^{\Lambda(\vla)} \, |\Lambda^L(\vla)\ra
\label{demo}
\eea
In this way, by using Eq. (\ref{eigenleft}) into Eq. (\ref{profmid1}) we find
\begin{equation}
\pp_{\vla}(C) \propto \frac{\la  \Lambda^R(-\vla-2\vecep) | C \ra \, \la C | \Lambda^R(\vla) \ra}{p_{\text{eff}}(C)} \, , \nonumber
\label{profmid01}
\end{equation}
where we assumed real components for the eigenvectors associated with the largest eigenvalue. Remarkably, this equation 
implies that $P_{\vla}(C)=P_{-\vla-2\vecep}(C)$ or equivalently $P_{\vqq}(C)=P_{-\vqq}(C)$, so the statistics during an arbitrary current fluctuation does 
not depend on the current sign, i.e. it remains invariant under time reversal. This implies in particular that the average density profile during a current 
large deviation event is invariant under $\vqq\to -\vqq$, but also that all higher-order moments of the density field as well as all $n$-body spatial 
correlations during a given current fluctuation exhibit this remarkable symmetry.

Starting from equations similar to eqs. (\ref{probCmid})-(\ref{profmid1}) above, it is easy to show that the probability of observing a given configuration $C$ \emph{at the end} of a current large deviation event parameterized by $\vla$ (i.e., for time $t'=t$) is just $\pp_{\vla}^{\text{end}}(C)\propto \la C | \Lambda^R(\vla) \ra$ \cite{PabloPRE}, allowing us to relate midtime and endtime current large-deviation statistics in a simple manner,
\begin{equation}
\pp_{\vla}(C) =A \, \frac{\pp_{\vla}^{\text{end}}(C) \pp_{-\vla-2\vecep}^{\text{end}}(C)}{p_{\text{eff}}(C)} \, ,
\label{profmid2}
\end{equation}
with $A$ a normalization constant. This relation implies that configurations with a significant contribution to the current large deviation statistics are those 
with an important probabilistic weight at the end of both the large deviation event and its time-reversed process \cite{PabloPRE}.

\section{The isometric fluctuation relation}
\label{sIFR}

When discussing the additivity of current fluctuations for the 2D KMP model in \S\ref{saddittest}, we have noticed a remarkable invariance of the Legendre 
transform of the current LDF with the angle of the current vector. As we discuss in this section, this observation is not a quirk of the 2D KMP model but 
a deep result related with hidden symmetries of the current LDF. 
To show this, notice  that
the optimal profile $\rho_{\vJJ}(\vrr)$ solution of Eq. (\ref{ch1:optprof}) depends exclusively on $\vJJ$ and $\vJJ^2$. Such a simple quadratic dependence, inherited from the locally-Gaussian nature of fluctuations, has important consequences at the level of symmetries of the current distribution. In fact, it is clear from Eq. (\ref{ch1:optprof}) that the condition
\beq
\frac{\delta \vpi_1[\rho(\vrr)]}{\delta \rho(\vrr')} = 0 \, ,
\label{ch4:cond1}
\eeq
implies that $\rho_{\vJJ}(\vrr)$ will depend exclusively on the \emph{magnitude} of the current vector, via $\vJJ^2$, not on its \emph{orientation}. In this way, 
all isometric current fluctuations characterized by a constant $|\vJJ|$ will have the same associated optimal profile, $\rho_{\vJJ}(\vrr)=\rho_{|\vJJ|}(\vrr)$, 
independently of whether the current vector $\vJJ$ points along the gradient direction, against it, or along any arbitrary direction. In other words, 
the optimal profile is invariant under current rotations if condition (\ref{ch4:cond1}) holds.  It turns out that condition (\ref{ch4:cond1}) follows from the 
time-reversibility of the dynamics, in the sense that the evolution operator in the Fokker-Planck formulation of the Langevin equation (\ref{ch1:langevin}) obeys 
a local detailed balance condition \cite{K,LS}. In this case, we can write in general
\beq
\textbf{W}_1[\rho(\vrr)] \equiv \frac{\vQ_{\vE}[\rho(\vrr)]}{\sigma[\rho(\vrr)]} = -\vnabla \frac{\delta {\cal H}[\rho]}{\delta\rho} \, ,
\label{ch4:detbal}
\eeq
where ${\cal H}[\rho(\vrr)]$ is the effective Hamiltonian for the system of interest. Now, if this condition holds, it is easy to show by using vector integration by parts that
\beq
\frac{\delta}{\delta \rho(\vrr')} \int_{\Lambda} d\vrr \textbf{W}_1[\rho(\vrr)] \cdot \cvA(\vrr) = - \frac{\delta}{\delta \rho(\vrr')} \int_{\partial\Lambda} d\Gamma \frac{\delta {\cal H}[\rho]}{\delta \rho} \cvA(\vrr) \cdot \vnn =0 \, ,
\label{ch4:Acond1}
\eeq
for any divergence-free vector field $\cvA(\vrr)$. The second integral is taken over the boundary $\partial \Lambda$ of the domain $\Lambda$ where the system 
is defined, and $\vnn$ is the unit vector normal to the boundary at each point. In particular, by taking $\cvA(\vrr)=\vJJ$ constant, Eq. (\ref{ch4:Acond1}) implies 
that $\delta \vpi_1[\rho(\vrr)]/\delta \rho(\vrr') = 0$. Hence for time-reversible systems --in the Fokker-Planck sense 
of Eq. (\ref{ch4:detbal})-- the optimal profile $\rho_{\vJJ}(\vrr)$ remains invariant under rotations of the current $\vJJ$, see Eq. (\ref{ch1:optprof}). We can 
now use this invariance to relate in a simple way the probability of any pair of isometric current fluctuations $\vJJ$ and $\vJJ'$, with $|\vJJ|=|\vJJ'|$, 
see eqs. (\ref{ch1:pj}) and (\ref{ch1:ldf2}). This allows us to write the following Isometric Fluctuation Relation (IFR) \cite{PabloIFR}
\be
 \lim_{\tau\to\infty} \frac{1}{\tau N^d} \ln \left[ \frac{\textrm{P}_{\tau}(\vJJ)}{\textrm{P}_{\tau}(\vJJ')}\right]  = \vecep\cdot (\vJJ-\vJJ') \, .
\label{IFT}
\ee
where $\vecep$ is the driving force.
The previous statement, which includes as a particular case the Gallavotti-Cohen (GC) result for $\vJJ'=-\vJJ$, see Eq. (\ref{GCtheo}), relates in a strikingly simple manner the probability of a given fluctuation $\vJJ$ with the likelihood of any other current fluctuation on the $d$-dimensional hypersphere of radius $|\vJJ|$, see Fig. \ref{sketchIFR}, projecting a complex $d$-dimensional problem onto a much simpler one-dimensional theory\footnote{In fact, it suffices to determine the current distribution along an arbitrary direction, say e.g. $\vJJ=|\vJJ| \hat{\bf x}$, to reconstruct the whole distribution $\textrm{P}_{\tau}(\vJJ)$ using the IFR.}. By recalling the definition of the current LDF, see Eq. (\ref{ch1:ldf0}), we can write an alternative formulation of the IFR
\beq
G(\vJJ) - G(\vJJ') = \vecep\cdot (\vJJ-\vJJ') = |\vecep| |\vJJ| (\cos \theta - \cos \theta') \, ,
\label{IFTcos}
\eeq
where $\theta$ and $\theta'$ are the angles formed by vectors $\vJJ$ and $\vJJ'$, respectively, with the constant vector $\vecep=\vvep + \vE$. This last expression will be useful when checking in simulations the IFR, see section \ref{ssimul} below. By letting now $\vJJ$ and $\vJJ'$ differ by an infinitesimal angle, the IFR can be cast in a simple differential form
\be
\frac{\partial G(\vJJ)}{\partial \theta} = |\vecep| |\vJJ| \sin \theta \, ,
\label{IFTdiff}
\ee
which reflects the high level of symmetry imposed by time-reversibility on the current distribution. Unlike the GC relation which is a non-differentiable symmetry involving the inversion of the current sign, $\vJJ \to -\vJJ$, Eq. (\ref{IFT}) is valid for arbitrary changes in orientation of the current vector, as reflected by Eq. (\ref{IFTdiff}) above. This makes the experimental test of the above relation a feasible problem, as data for current fluctuations involving different orientations around the average can be gathered with enough statistics to ensure experimental accuracy. Finally, it is important to stress that the IFR is valid for arbitrarily large fluctuations, i.e. even for the non-Gaussian far tails of current distribution.

\begin{figure}
\vspace{-0.3cm}
\centerline{
\includegraphics[width=9cm]{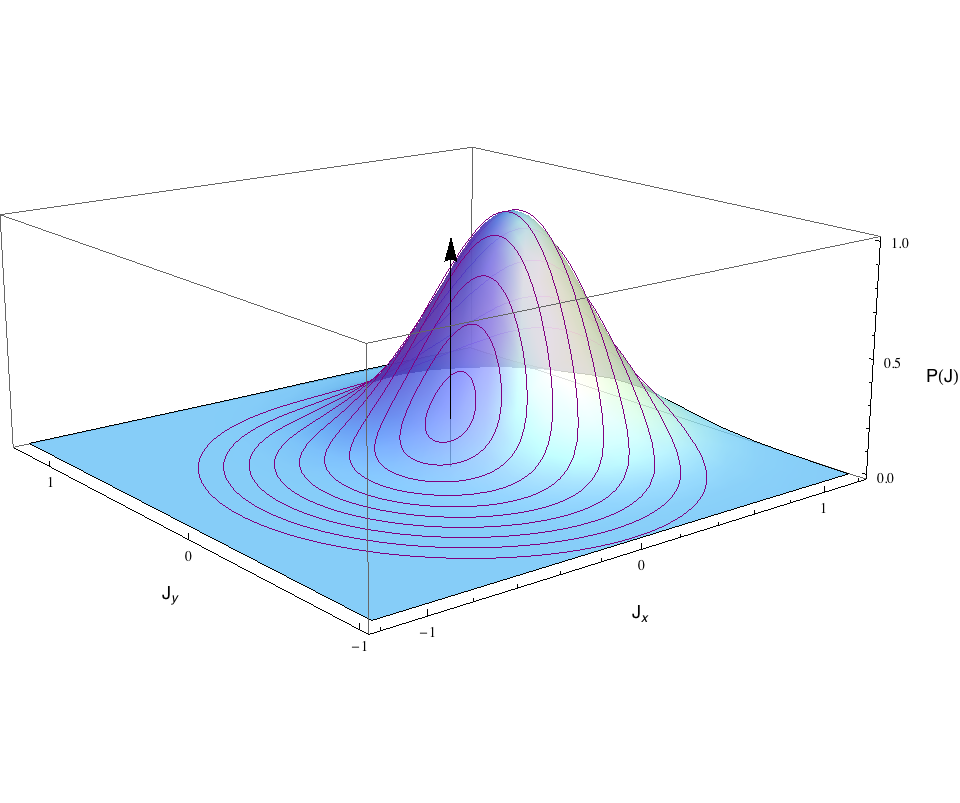}
\includegraphics[width=7cm]{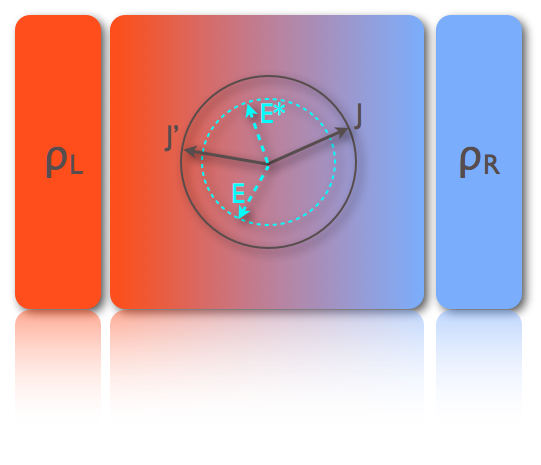}
}
\vspace{-1cm}
\caption{(Color online) {Left: The isometric fluctuation relation at a glance.} Sketch of the current distribution in two dimensions, peaked around its average $\la \vJJ\ra_{\epsilon}$, and isometric contour lines for different $|\vJJ|$'s. The isometric fluctuation relation, Eq. (\ref{IFT}), establishes a simple relation for the probability of current fluctuations along each of these contour lines. Right: The IFR relates in a strikingly simple manner the probability of a given fluctuation $\vJJ$ with the likelihood of any other current fluctuation on the $d$-dimensional hypersphere of radius $|\vJJ|$. Moreover, under the additional condition (\ref{ch4:constm1}), the generalized IFR (\ref{ch4:IFR2}) relates any current fluctuation $\vJJ$ in the presence of an external field $\vE$ with any other isometric current fluctuation $\vJJ'$ in the presence of an arbitrarily-rotated external field $\vE^*$}
\label{sketchIFR}
\end{figure}

The IFR and the GC theorem are deep statements on the subtle but enduring consequences of microscopic time reversibility at the macroscopic level. Particularly important here is the observation that microscopic symmetries are reflected at the fluctuating macroscopic level arbitrarily far from equilibrium.
This crucial observation suggest to apply and extend the ideas and tools associated with the concept of symmetry, so successful in other areas of theoretical physics, to study the statistics of fluctuations out of equilibrium. In physics, symmetry means invariance of an object under certain transformation rules. In this context the physically relevant and natural object whose invariance we are interested in is the optimal path that a system with many degrees of freedom transits in mesoscopic (or coarse-grained) phase space to facilitate a given fluctuation. As we have seen above, the invariance of this optimal path under rotations of the associated current vector has led to a new insight, the Isometric Fluctuation Relation. We anticipate that invariance principles of this kind can be applied with great generality in diverse fields where fluctuations play a fundamental role, opening an unexplored route toward a deeper understanding of nonequilibrium physics by bringing symmetry principles to the realm of fluctuations. Below we show that  the IFR can be extended to different situations by using this general idea.

Before developing these generalizations, it is important to notice that the condition $\delta \vpi_1[\rho(\vrr)]/\delta \rho(\vrr')=0$ can be seen as a conservation law. It implies that the observable $\vpi_1[\rho(\vrr)]$ is in fact a \emph{constant of motion}, $\vecep\equiv \vpi_1[\rho(\vrr)]$, independent of the profile $\rho(\vrr)$, which can be related with the rate of entropy production via the Gallavotti-Cohen theorem \cite{GC,ECM,K,LS}. In a way similar to Noether's theorem, the conservation law for $\vecep$ implies a symmetry for the optimal profiles under rotations of the current and an isometric fluctuation relation for the current LDF. This constant can be easily computed under very general assumptions, see appendix \ref{ap2}.

\subsection{Numerical investigation of isometric current fluctuations}
\label{ssimul}

\begin{figure}
\centerline{
\includegraphics[width=16cm]{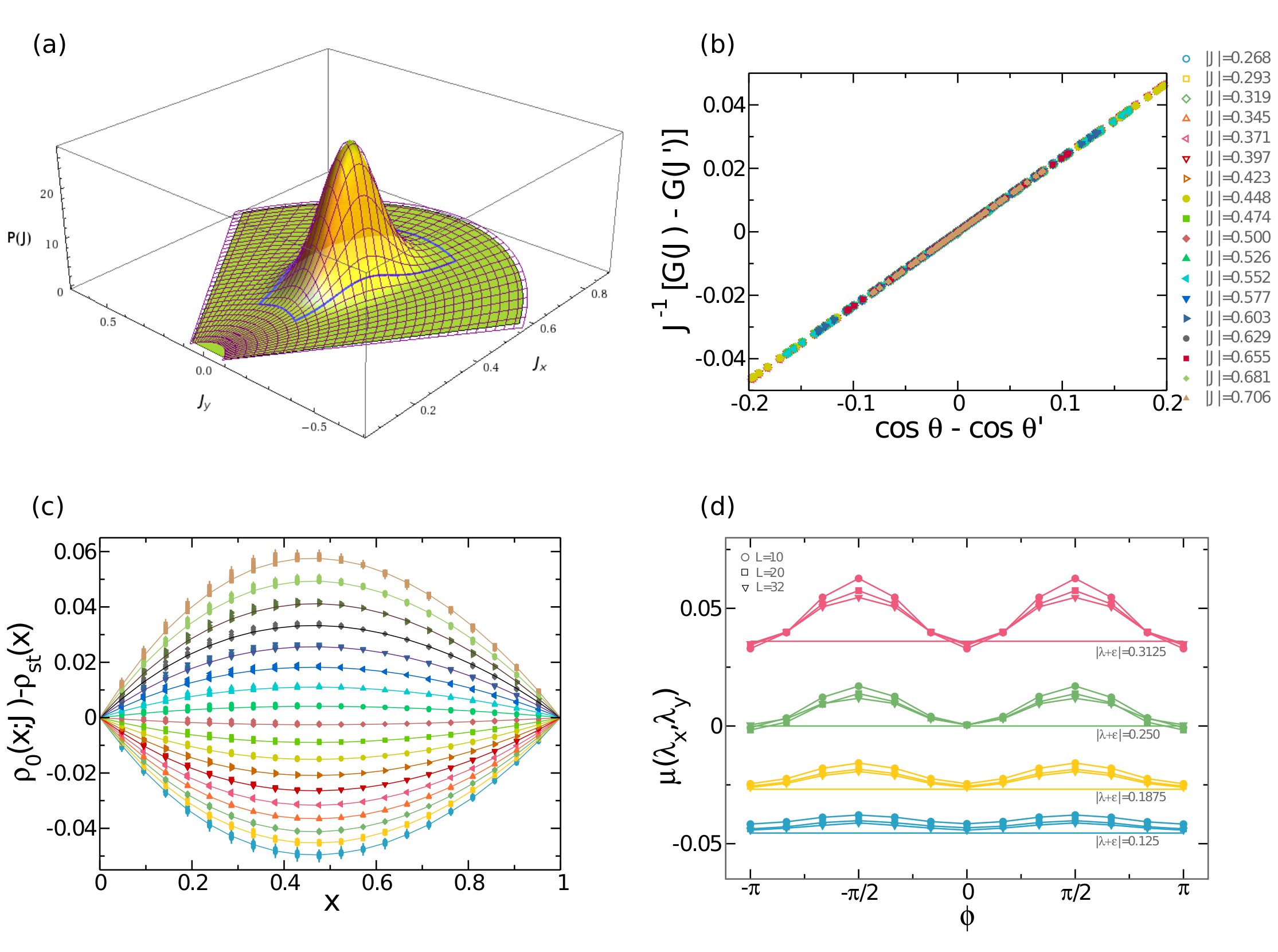}
}
\caption{(Color online) {Confirmation of IFR in a diffusive system.} (a) Current distribution as measured in standard simulations of the 2D KMP model subject to a temperature gradient, together with the polar binning used to test the IFR. (b) The IFR predicts that $|\vJJ|^{-1}[G(\vJJ)-G(\vJJ')]$ collapses onto a linear function of $\cos\theta-\cos\theta'$ for all values of $|\vJJ|$, see Eq. (\ref{IFTcos}). This collapse is confirmed here in the KMP energy diffusion model for a wide range of values for $|\vJJ|$.
(c) Average profiles for different but isometric current fluctuations all collapse onto single curves, confirming the invariance of optimal profiles under current rotations. Angle range is $|\theta|\le 16.6^{\circ}$, see marked region in the histogram, and lines are MFT$_{\text{Ad}}$ predictions. (d) Legendre transform of the current LDF for the 2D KMP model, for different values of $|\vla+\vecep|$ corresponding to very large current fluctuations, different rotation angles $\phi$ such that $\vla'={\cal R}_{\phi}(\vla+\vecep)-\vecep$, and increasing system sizes. Lines are theoretical predictions. The IFR predicts that $\mu(\vla)=\mu[{\cal R}_{\phi}(\vla+\vecep)-\vecep]$ $\forall \phi\in[0,2\pi]$. The isometric fluctuation symmetry emerges in the macroscopic limit as the effects associated with the underlying lattice fade away.
}
\label{figIFR}
\end{figure}

We now set out to validate the IFR in extensive simulations of the 2D KMP model, using both standard Monte Carlo simulations and the advanced cloning technique discussed in \S\ref{ssim}. For the former case, we performed a large number of steady-state simulations of long duration $\tau>N^2$ (the unit of time is the Monte Carlo step) for $N=20$, $T_L=2$ and $T_R=1$, accumulating statistics for the space\&time-averaged current vector $\vJJ$. The measured current distribution $\pp_{\tau}(\vJJ)$ is shown in Fig. \ref{figIFR}.a, together with a fine polar binning which allows us to compare the probabilities of isometric current fluctuations along each polar corona, see Eq. (\ref{IFT}). 
Taking $G(\vJJ)=(\tau N^d)^{-1} \ln \text{P}_{\tau}(\vJJ)$, Fig. \ref{figIFR}.b confirms the IFR prediction that $G(\vJJ)-G(\vJJ')$, once scaled by $|\vJJ|^{-1}$, collapses onto a linear function of $\cos\theta-\cos\theta'$ for all values of $|\vJJ|$, see Eq. (\ref{IFTcos}). Here $\theta$, $\theta'$ are the angles formed by the isometric current vectors $\vJJ$, $\vJJ'$ with the $x$-axis ($\vE=0$ in this case). We also measured in standard simulations the average energy profile associated with each current fluctuation, $\rho_{\vJJ}(\vrr)$, see Fig. \ref{figIFR}.c.  As predicted above, profiles for different but isometric current fluctuations all collapse onto a single curve well predicted by MFT$_{\text{Ad}}$, confirming the invariance of optimal profiles under current rotations.

Standard simulations allow us to explore moderate fluctuations of the current around the average. In order to test the IFR in the far tails of the current distribution, corresponding to exponentially unlikely rare events, we measured the current statistics using the cloning algorithm of \S\ref{ssim}. As discussed above, this method yields the Legendre transform of the current LDF, $\mu(\vla)$, so we must first write the IFR in terms of $\mu(\vla)$.
To do so, we now write $\vJJ'={\cal R}\vJJ$, with ${\cal R}$ any $d$-dimensional rotation matrix, and use the IFR, see Eq. (\ref{IFTcos}), in the definition of $\mu(\vla)$, Eq. (\ref{mulamb}), to obtain
\beq
\mu(\vla)=\max_{\vJJ}[G(\vJJ)+\vla\cdot \vJJ]=\max_{\vJJ'}[G(\vJJ')+({\cal R}(\vla+\vecep)-\vecep) \cdot \vJJ']=\mu[{\cal R}(\vla+\vecep)-\vecep],
\nonumber
\eeq
where we have used that projections remain invariant under equal rotation of the vectors involved, i.e. $(\vla+\vecep)\cdot {\cal R}^{-1} \vJJ' = {\cal R}(\vla+\vecep)\cdot \vJJ'$. Hence, the IFR can be stated for $\mu(\vla)$ as
\beq
\mu(\vla)=\mu[{\cal R}(\vla+\vecep)-\vecep] \quad \forall {\cal R} \, .
\label{ch4:IFRmu}
\eeq
Therefore, in order to test the IFR we measured $\mu(\vla)$ in increasing manifolds of constant $|\vla+\vecep|$, see Fig. \ref{figIFR}.d, as the IFR (\ref{ch4:IFRmu}) 
implies that $\mu(\vla)$ is constant along each of these manifolds.
${\cal R}_{\phi}$ a rotation in 2D of angle $\phi$.
Fig. \ref{figIFR}.d shows the measured $\mu(\vla)$ for different values of $|\vla+\vecep|$ corresponding to very large current fluctuations, different 
rotation angles $\phi$ and increasing system sizes, together with the theoretical predictions (see Appendix \ref{sadditKMP}). As a result of the finite, 
discrete character of the lattice system studied here, we observe violations of IFR in the far tails of the current distribution, specially for 
currents orthogonal to the driving force $\vecep$. These violations, already encountered when investigating additivity, are expected since a prerequisite 
for the IFR to hold is the existence of a macroscopic limit, i.e. Eq. (\ref{ch1:langevin}) should hold strictly, which is not the case for the relatively 
small values of $N$ studied here. However, as $N$ increases, a slow but clear convergence toward the IFR prediction is observed as the effects associated with 
the underlying lattice fade away, strongly supporting the validity of IFR in the macroscopic limit.

In order to investigate the universality of the IFR, we also measured current fluctuations in a Hamiltonian hard-disk fluid subject to a temperature gradient (not shown) \cite{PabloIFR}. This model is a paradigm in liquid state theory, condensed matter and statistical physics, and has been widely studied during last decades. Our results unambiguously confirm the IFR and the associated invariance of optimal profiles under current rotations in this model for a wide range of  fluctuations. Interestingly, the hard-disk fluid is a fully hydrodynamic system, with 4 different locally-conserved coupled fields possibly subject to memory effects, defining a far more complex situation than the one studied here within MFT$_{\text{Ad}}$, see Eq. (\ref{ch1:langevin}). Therefore the validity of IFR in this context suggests that this fluctuation relation, based on the invariance of optimal profiles under symmetry transformations, is in fact a rather general result valid for arbitrary fluctuating hydrodynamic systems.
We also mention that recent work \cite{Rodrigo} has shown that the IFR can be suitably generalized to anisotropic systems, and this idea has been confirmed in great detail for the anisotropic zero-range process.

\subsection{Hierarchies for the cumulants and nonlinear response coefficients}
\label{sIFRhier}

The isometric fluctuation relation, Eq. (\ref{IFT}), has far-reaching and nontrivial consequences. For instance, the IFR implies remarkable hierarchies of 
equations for the current cumulants and the nonlinear response coefficients, which go far beyond Onsager's reciprocity relations and Green-Kubo formulas. 
These hierarchies can be derived starting from the moment-generating function associated with $\text{P}_{\tau}(\vJJ)$,
\beq
\Pi(\vla,\tau)=\int \text{P}_{\tau}(\vJJ)\text{e}^{\tau N^d \vla \cdot \vJJ}d\vJJ \, ,
\label{ch4:momentgen}
\eeq
which scales for long times as $\Pi(\vla,\tau)\sim \exp[+\tau N^d \mu(\vla)]$, where $\mu(\vla)=\max_{\vJJ}[G(\vJJ)+\vla\cdot \vJJ]$
works now as the cumulant generating function.
The cumulants of the current distribution can be obtained from the derivatives of $\mu(\vla)$ evaluated at $\vla=0$, i.e.
\beq
\mu_{(n_1 ... n_d)}^{(n)}\equiv \left[\frac{\partial^n \mu(\vla)}{\partial \lambda_1^{n_1} ... \partial\lambda_d^{n_d}}\right]_{\vla=0},
\label{ch4:muderiv}
\eeq
where $\lambda_i$ is the $i$-th component of vector $\vla$ and $\sum_{i=1}^{d}n_i=n$. Note that for $n\leq 3$ we have that 
$\mu_{(n_1 ... n_d)}^{(n)}=(\tau L^d)^{n-1}\la \Delta J_1^{n_1}...\Delta J_d^{n_d} \ra$ 
where $\Delta J_k\equiv J_k-(1-\delta_{n,1})\la J_k\ra_{\epsilon}$, with $\la J_k\ra_{\epsilon}$ being the average current 
along the $k$-direction for external driving $\vecep$.
We now may use the IFR as expressed for $\mu(\vla)$, see Eq. (\ref{ch4:IFRmu}),
in the definition of the $n$-th order cumulant. 
As we shall show below, in the limit of 
infinitesimal rotations, ${\cal R}=\mathbb{I}+\Delta \phi {\cal L}$, with $\mathbb{I}$ the identity matrix,
the IFR implies that
\beq
n_{\alpha} {\cal L}_{\beta \alpha} \mu_{(n_1 ... n_{\alpha}-1 ... n_{\beta}+1 ... n_d)}^{(n)} +   \epsilon_{\nu} {\cal L}_{\gamma\nu} \mu_{(n_1 ... n_{\gamma}+1 ... n_d)}^{(n+1)}= 0 \, ,
\label{ch4:cumul}
\eeq
where ${\cal L}$ is any generator of $d$-dimensional rotations, and summation over repeated Greek indices ($\in[1,d]$) is assumed. The above hierarchy 
relates in a simple way current cumulants of order $n$ and $n+1$ $\forall n \ge 1$, and is valid arbitrarily far from equilibrium.
\\
We now set out to derive Eq. (\ref{ch4:cumul}). To do so, we start by slightly modifying our notation by 
replacing in Eq. (\ref{ch4:muderiv}) the $n_i$-th order derivative with respect to $\lambda_i$ by 
$n_i$ first order derivatives with respect to $\lambda_i$, i.e.,
\beq
\mu_{(n_1 ... n_d)}^{(n)}\equiv \left[\frac{\partial^n \mu(\vla)}{\partial \lambda_1^{n_1} ... \partial\lambda_d^{n_d}}\right]_{\vla=0}=
\left[\frac{\partial^n \mu(\vla)}{\partial \lambda_{i_1} \partial \lambda_{i_2}... \partial\lambda_{i_n}}\right]_{\vla=0},
\label{ch4:muderiv2}
\eeq
with 
\beq
i_k=\left\{
\begin{array}{llll}
1~~k\in[1,n_1]  \\
2~~k\in[n_1+1,n_1+n_2] \\
\vdots \\
d~~k\in[\sum_{i=1}^{d-1}n_i+1,\sum_{i=1}^dn_i]
\end{array}
\right.
\label{notation}
\eeq
We now may use the IFR as expressed for $\mu(\vla)$, see Eq. (\ref{ch4:IFRmu}), to write
\beq
\mu_{(n_1 ... n_d)}^{(n)}=\left[\frac{\partial^n \mu(\vla)}{\partial \lambda_{i_1} \partial \lambda_{i_2}... \partial\lambda_{i_n}}\right]_{\vla=0}=
\left[\frac{\partial^n \mu({\cal R}(\vla+\vecep)-\vecep)}{\partial \lambda_{i_1} \partial \lambda_{i_2}... \partial\lambda_{i_n}}\right]_{\vla=0}
\label{eqcum1}
\eeq
Recall that all througout this section summation over repeated Greek indices ($\in[1,d]$) is assumed. By using the chain rule Eq. (\ref{eqcum1}) reads
\beq
\left[\frac{\partial^n \mu({\cal R}(\vla+\vecep)-\vecep)}{\partial \lambda_{i_1}... \partial\lambda_{i_n}}\right]_{\vla=0}=
R_{\omega_1 i_1}...R_{\omega_n i_n}
\left[\frac{\partial^n \mu(\vla')}{\partial \lambda_{\omega_1}^{'} ... \partial \lambda_{\omega_n}^{'}}\right]_{\vla'={\cal R}\vecep-\vecep},
\label{eqcum2}
\eeq
where we have used that the $k$-th component of vector ${\cal R}(\vla+\vecep)-\vecep$ is just $R_{k\alpha}(\lambda_{\alpha}+\epsilon_{\alpha})-\epsilon_{k}$.

In the limit of 
infinitesimal rotations, ${\cal R}=\mathbb{I}+\Delta \phi {\cal L}$, with $\mathbb{I}$ the identity matrix, we have that 
$\vla'={\cal R}\vecep-\vecep\simeq \Delta\phi {\cal L}\vecep$ and we can write
\beq
\left[\frac{\partial^n \mu(\vla')}{\partial \lambda_{\omega_1}^{'} ... \partial \lambda_{\omega_n}^{'}}\right]_{\vla'=\Delta\phi {\cal L}\vecep}\simeq
\left[\frac{\partial^n \mu(\vla')}{\partial \lambda_{\omega_1}^{'} ... \partial \lambda_{\omega_n}^{'}}\right]_{\vla'=0}+
\Delta \phi \epsilon_{\nu}{\cal L}_{\gamma \nu}\left[\frac{\partial^{n+1} \mu(\vla')}{\partial \lambda_{\omega_1}^{'} ...\partial \lambda_{\gamma}^{'}... \partial \lambda_{\omega_n}^{'}}\right]_{\vla'=0}.
\label{eqcum3}
\eeq
As we are considering infinitesimal rotations, we can also expand the prefactors $R_{\omega_1 i_1}...R_{\omega_n i_n}$ to first order in $\Delta \phi$ as
\beq
R_{\omega_1 i_1}...R_{\omega_n i_n}=\delta_{\omega_1 i_1}...\delta_{\omega_n i_n}+
\Delta\phi \sum_{k=1}^n\delta_{\omega_1 i_1}...\delta_{\omega_{k-1} i_{k-1}}{\cal L}_{\omega_{k} i_{k}}\delta_{\omega_{k+1} i_{k+1}}...\delta_{\omega_n i_n}.
\label{eqcum4}
\eeq
Thus, by using eqs. (\ref{eqcum3})-(\ref{eqcum4}) in the r.h.s of Eq. (\ref{eqcum2}) and retaining up to linear terms in $\Delta \phi$, we can write 
Eq. (\ref{eqcum1}) as
\begin{eqnarray}
\mu_{(n_1 ... n_d)}^{(n)}=\delta_{\omega_1 i_1}...\delta_{\omega_n i_n}\left[\frac{\partial^n \mu(\vla')}{\partial \lambda_{\omega_1}^{'} ... \partial \lambda_{\omega_n}^{'}}\right]_{\vla'=0}+
\Delta \phi\left[\delta_{\omega_1 i_1}...\delta_{\omega_n i_n} \epsilon_{\nu}{\cal L}_{\gamma \nu}\left[\frac{\partial^{n+1} \mu(\vla')}{\partial \lambda_{\omega_1}^{'} ...\partial \lambda_{\gamma}^{'}... \partial \lambda_{\omega_n}^{'}}\right]_{\vla'=0}\right.+ \nonumber   \\
\left.+\sum_{k=1}^n\delta_{\omega_1 i_1}...\delta_{\omega_{k-1} i_{k-1}}{\cal L}_{\omega_{k} i_{k}}\delta_{\omega_{k+1} i_{k+1}}...\delta_{\omega_n i_n}\left[\frac{\partial^n \mu(\vla')}{\partial \lambda_{\omega_1}^{'} ... \partial \lambda_{\omega_n}^{'}}\right]_{\vla'=0}\right]
\nonumber
\end{eqnarray}
It is easy to see that the first summand of the r.h.s of the above equation is just the original $\mu_{(n_1 ... n_d)}^{(n)}$, thus the sum in brackets must be zero, i.e.,
\beq
\sum_{k=1}^n {\cal L}_{\omega_{k} i_{k}}\left[\frac{\partial^n \mu(\vla')}{\partial \lambda_{i_1}^{'} ...\partial\lambda_{\omega_k}^{'}... \partial \lambda_{i_n}^{'}}\right]_{\vla'=0}
+\epsilon_{\nu}{\cal L}_{\gamma \nu}\left[\frac{\partial^{n+1} \mu(\vla')}{\partial \lambda_{i_1}^{'} ...\partial \lambda_{\gamma}^{'}... \partial \lambda_{i_n}^{'}}\right]_{\vla'=0}=0.
\nonumber
\eeq
To end the calculation and get Eq. (\ref{ch4:cumul}) we now go back to the original notation used in Eq. (\ref{notation}). We can thus inmmediately identify 
the second summand of the l.h.s of the above equation with the second summand of Eq. (\ref{ch4:cumul}). In order to derive the first summand of 
Eq. (\ref{ch4:cumul}), we take for instance $k\in[1,n_1]$ in the first summand of the above equation. In that case we have $i_k=1$ meaning that 
there are $n_1$ equal summands involving $n_1-1$ derivatives with respect to $\lambda_1$ and an additional derivative with respect to $\lambda_{\omega_1}$, 
so the first $n_1$ summands in the first term of the l.h.s of the above equation are just $n_1{\cal L}_{\omega_11}\mu_{n_1-1,n_2...n_{\omega_1}+1...n_d}$. 
In general, we can write this summation as $n_{\alpha} {\cal L}_{\beta \alpha} \mu_{(n_1 ... n_{\alpha}-1 ... n_{\beta}+1 ... n_d)}^{(n)}$, recovering 
the hierarchy in Eq. (\ref{ch4:cumul}).
\\

As an example, the first two sets of relations ($n=1,2$) of the hierarchy given by Eq. (\ref{ch4:cumul}) in two dimensions are
\begin{eqnarray}
 \la J_x\ra _{\epsilon}& = & \tau N^2 \left[\epsilon_x \la\Delta J_y^2\ra _{\epsilon}- \epsilon_y\la\Delta J_x \Delta J_y\ra_{\epsilon}  \right] \, \label{ch4:cumul2dn1} \\
 \la J_y\ra_{\epsilon} & = & \tau N^2 \left[\epsilon_y \la\Delta J_x^2\ra_{\epsilon} - \epsilon_x\la\Delta J_x \Delta J_y\ra_{\epsilon}  \right] \, \nonumber \\
\phantom{aaa} \nonumber \\
2\la \Delta J_x \Delta J_y\ra_{\epsilon} & = & \tau N^2 \left[\epsilon_y \la\Delta J_x^3\ra_{\epsilon} - \epsilon_x\la\Delta J_x^2 \Delta J_y\ra_{\epsilon}  \right]  \, \label{ch4:cumul2dn2} \\
& = & \tau N^2 \left[\epsilon_x \la\Delta J_y^3\ra_{\epsilon} - \epsilon_y\la\Delta J_x \Delta J_y^2\ra_{\epsilon}  \right] \, \nonumber \\
\la \Delta J_x^2\ra_{\epsilon} - \la \Delta J_y^2\ra_{\epsilon} & = & \tau N^2 \left[\epsilon_x \la\Delta J_x \Delta J_y^2\ra_{\epsilon} - \epsilon_y\la\Delta J_x^2 \Delta J_y\ra_{\epsilon}  \right] \, . \nonumber
\end{eqnarray}

In a similar way, we can explore the consequences of the IFR on the linear and nonlinear response coefficients. For that, we now expand 
the cumulants of the current in powers of the driving force $\vecep$
\beq
\mu_{(n_1 ... n_d)}^{(n)} (\vecep) = \sum_{k=0}^{\infty} \frac{1}{k!} \sum_{\substack{k_1... k_d=0 \\ \sum_i k_i=k}}^k \phantom{,} _{(n)}^{(k)}\chi_{(n_1 ... n_d)}^{(k_1 ... k_d)} \, \epsilon_1^{k_1} ... \epsilon_d^{k_d} \, ,
\label{ch4:expand}
\eeq
where the nonlinear response coefficients are defined as
\beq
_{(n)}^{(k)}\chi_{(n_1 ... n_d)}^{(k_1 ... k_d)} \equiv \left[\frac{\partial^{n+k} \mu(\vla)}{\partial \lambda_1^{n_1}...\partial \lambda_d^{n_d} \partial \epsilon_1^{k_1}...\partial\epsilon_d^{k_d}}\right]_{\lambda=0=\ecep} \, ,
\label{ch4:respcoeff}
\eeq
and measure the $k$-th order response of the $n$-th order current cumulant to the external driving.
Inserting expansion (\ref{ch4:expand}) into the cumulant hierarchy, Eq. (\ref{ch4:cumul}), we have
\bea
n_{\alpha} {\cal L}_{\beta \alpha} \sum_{k=0}^{\infty} \frac{1}{k!} \sum_{\substack{k_1... k_d=0 \\ \sum_i k_i=k}}^k \phantom{,} _{(n)}^{(k)}\chi_{(n_1 ... n_{\alpha}-1 ... n_{\beta}+1 ... n_d)}^{(k_1 ... k_d)} \, \epsilon_1^{k_1} ... \epsilon_d^{k_d} + \nonumber\\  
+{\cal L}_{\gamma\nu} \sum_{k^{'}=0}^{\infty} \frac{1}{k^{'}!} \sum_{\substack{k^{'}_1... k'_d=0 \\ \sum_i k^{'}_i=k^{'}}}^{k^{'}} \phantom{,} _{(n+1)}^{(k^{'})}\chi_{(n_1 ... n_{\gamma}+1 ... n_d)}^{(k^{'}_1 ... k^{'}_d)} \, \epsilon_1^{k^{'}_1} ... \epsilon_{\nu}^{k^{'}_{\nu}+1} ... \epsilon_d^{k^{'}_d}= 0 \, ,
\nonumber
\eea
To match order by order in $k$ the two terms in the above expression we have that $k^{'}_1=k_1$, ..., $k^{'}_{\nu}+1=k_{\nu}$ ,..., $k^{'}_d=k_d$. Therefore
$k'=\sum_i k^{'}_i=\sum_i k_i-1=k-1$ and we get
\beq
\sum_{k=0}^{\infty} \sum_{\substack{k_1... k_d=0 \\ \sum_i k_i=k}}^k\left[ n_{\alpha} {\cal L}_{\beta \alpha}  \frac{1}{k}\phantom{,} _{(n)}^{(k)}\chi_{(n_1 ... n_{\alpha}-1 ... n_{\beta}+1 ... n_d)}^{(k_1 ... k_d)} \,  +   
{\cal L}_{\gamma\nu} \phantom{,} _{(n+1)}^{(k-1)}\chi_{(n_1 ... n_{\gamma}+1 ... n_d)}^{(k_1 ... k_{\nu}-1 ... k_d)} \,\right]\frac{\epsilon_1^{k_1} ... \epsilon_d^{k_d}}{(k-1)!}=0
\nonumber
\eeq
This relation must be valid for arbitrary $\vecep$, so the expansion coefficient for each $k$ must vanish exactly. This leads to 
another interesting hierarchy for the response coefficients of the different cumulants. For $k=0$ we get
\beq
n_{\alpha} {\cal L}_{\beta \alpha} \phantom{,} _{(n)}^{(0)}\chi_{(n_1 ... n_{\alpha}-1 ... n_{\beta}+1 ... n_d)}^{(0 ... 0)} = 0 \, ,
\label{ch4:nonlincoefs1}
\eeq
which is a symmetry relation for the equilibrium ($\vecep=0$) current cumulants. For $k\ge 1$ we obtain
\begin{eqnarray}
\sum_{\substack{k_1... k_d=0 \\ \sum_i k_i=k\ge 1}}^k \Bigg[ \frac{n_{\alpha}}{k} {\cal L}_{\beta \alpha} \phantom{,}
_{(n)}^{(k)}\chi_{(n_1 ... n_{\alpha}-1 ... n_{\beta}+1 ... n_d)}^{(k_1 ... k_d)}
+ {\cal L}_{\gamma\nu} \, _{(n+1)}^{(k-1)}\chi_{(n_1 ... n_{\gamma}+1 ... n_d)}^{(k_1... k_{\nu}-1... k_d)}\Bigg] = 0 \, ,
\label{ch4:nonlincoefs2}
\end{eqnarray}
which relates $k$-order response coefficients of $n$-order cumulants with $(k-1)$-order coefficients of $(n+1)$-order cumulants. 
Relations (\ref{ch4:nonlincoefs1})-(\ref{ch4:nonlincoefs2}) for the response coefficients result from the IFR in the limit of infinitesimal rotations. For a 
finite rotation ${\cal R}=-\mathbb{I}$, which is equivalent to a current inversion, we have $\mu(\vla)=\mu(-\vla-2\vecep)$, an alternative formulation of the 
GC fluctuation theorem expressed for the Legendre transform of the current LDF, and we may use this in the definition of response coefficients, Eq. (\ref{ch4:respcoeff}), to obtain a complementary relation for the response coefficients
\beq
_{(n)}^{(k)}\chi_{(n_1 ... n_d)}^{(k_1 ... k_d)} = k! \sum_{p_1=0}^{k_1}... \sum_{p_d=0}^{k_d} \frac{(-1)^{n+p} 2^p}{p!(k-p)!} \phantom{,} _{(n+p)}^{(k-p)}\chi_{(n_1+p_1 ... n_d+p_d)}^{(k_1-p_1 ... k_d-p_d)} \, ,
\label{ch4:nonlincoefs3}
\eeq
where $p=\sum_i p_i$. A similar equation was derived in \cite{Gaspard} from the standard GC fluctuation theorem, although the IFR (which includes the GC fluctuation theorem for currents as a particular case) adds a whole set of new relations, eqs. (\ref{ch4:nonlincoefs1})-(\ref{ch4:nonlincoefs2}). All together, eqs. (\ref{ch4:nonlincoefs1})-(\ref{ch4:nonlincoefs3}) imply deep relations between the response coefficients at arbitrary orders which go far beyond Onsager's reciprocity relations and Green-Kubo formulae.

To better grasp the meaning of these hierarchies, we now go back to our two-dimensional example, and let $_{(n)}^{(k)}\chi_{(n_x,n_y)}^{(k_x,k_y)}$ be the response 
coefficient of the cumulant $\mu^{(n)}_{(n_x,n_y)}(\vecep)$ to order $\ecep_x^{k_x}\ecep_y^{k_y}$, with $n=n_x+n_y$ and $k=k_x+k_y$. To the 
lowest order these hierarchies imply Onsager's reciprocity symmetries and Green-Kubo relations for the linear response coefficients of the current, with the 
additional prediction that the linear response matrix is in fact proportional to the identity. In our notation, this translates to
\beq
_{(1)}^{(1)}\chi_{(1,0)}^{(1,0)}={_{(1)}^{(1)}\chi_{(0,1)}^{(0,1)}}={_{(2)}^{(0)}\chi_{(2,0)}^{(0,0)}}={_{(2)}^{(0)}\chi_{(0,2)}^{(0,0)}} \, ,
\nonumber
\eeq
while
\beq
_{(1)}^{(1)}\chi_{(1,0)}^{(0,1)}=0={_{(1)}^{(1)}\chi_{(0,1)}^{(1,0)}} \, .
\nonumber
\eeq
The first nonlinear coefficients of the current can be simply written in terms of the linear coefficients of the second cumulants as
\beq
_{(1)}^{(2)}\chi_{(1,0)}^{(2,0)} = 2 {_{(2)}^{(1)}\chi_{(2,0)}^{(1,0)}}~~~\text{and}~~~_{(1)}^{(2)}\chi_{(1,0)}^{(0,2)} = -2 {_{(2)}^{(1)}\chi_{(1,1)}^{(1,0)}} \, ,
\nonumber
\eeq
while the cross-coefficient reads
\beq
_{(1)}^{(2)}\chi_{(1,0)}^{(1,1)} = 2 \left[{_{(2)}^{(1)}\chi_{(2,0)}^{(0,1)}} + {_{(2)}^{(1)}\chi_{(1,1)}^{(0,1)}}\right]
\nonumber
\eeq
(symmetric results hold for $n_x=0$, $n_y=1$). Linear response coefficients for the second-order cumulants also obey simple relations,
e.g.
\beq
_{(2)}^{(1)}\chi_{(1,1)}^{(1,0)} = - {_{(2)}^{(1)}\chi_{(1,1)}^{(0,1)}}~~~\text{and}~~~
_{(2)}^{(1)}\chi_{(2,0)}^{(1,0)} + {_{(2)}^{(1)}\chi_{(2,0)}^{(0,1)}} = {_{(2)}^{(1)}\chi_{(0,2)}^{(1,0)}} + {_{(2)}^{(1)}\chi_{(0,2)}^{(0,1)}},
\nonumber
\eeq
and
the set of relations continues to arbitrary high orders. In this way hierarchies (\ref{ch4:nonlincoefs1})-(\ref{ch4:nonlincoefs3}),
which derive from microreversibility as reflected in the IFR, provide new deep insights into nonlinear response theory for nonequilibrium systems \cite{Gaspard}.

\subsection{Generalizations of the IFR}
\label{sIFRgen}

The IFR and the above hierarchies all follow from the invariance of optimal profiles under certain transformations. This idea can be further exploited in more general settings. In fact, by writing explicitly the dependence on the external field $\vE$ in Eq. (\ref{ch1:optprof}) for the optimal profile, one realizes that if the following condition holds,
\beq
\frac{\delta}{\delta \rho(\vrr')} \int_{\Lambda} \vQ[\rho(\vrr)] d\vrr = 0 \, ,
\label{ch4:constm1}
\eeq
together with the time-reversibility condition, Eq. (\ref{ch4:cond1}), the resulting optimal profiles are invariant under \emph{independent} rotations of the current and the external field. It thus follows that the current LDFs for pairs $(\vJJ,\vE)$ and $(\vJJ'={\cal R} \vJJ,\vE^*={\cal S} \vE)$, with ${\cal R}$, ${\cal S}$ independent $d$-dimensional rotations, obey a generalized Isometric Fluctuation Relation
\beq
G_{\vE}(\vJJ) - G_{\vE^*}(\vJJ') = \vvep\cdot(\vJJ-\vJJ') - \vnu\cdot(\vE-\vE^*) + \vJJ\cdot\vE-\vJJ'\cdot\vE^* \, ,
\label{ch4:IFR2}
\eeq
where we write explicitly the dependence of the current LDF on the external field. The vector $\vnu\equiv \int_{\Lambda} \vQ[\rho(\vrr)] d\vrr$ is now another constant
of motion, independent of $\rho(\vrr)$, which can be easily computed (see Appendix \ref{ap2}). For a fixed boundary gradient, the above equation relates any current fluctuation $\vJJ$ in the presence of an external field $\vE$ with any other isometric current fluctuation $\vJJ'$ in the presence of an arbitrarily-rotated external field $\vE^*$, see right panel in Fig. \ref{sketchIFR}, and reduces to the standard IFR for $\vE=\vE^*$. Interestingly, condition (\ref{ch4:constm1})
is rather general, as most time-reversible systems with a local mobility $\sigma[\rho]$ do fulfill this condition (as, e.g., diffusive systems as those studied here).

The IFR can be further generalized to cases where the current profile is not constant, relaxing hypothesis (H2) of the additivity conjecture, see \S\ref{saddit} above. Let $\text{P}_{\tau}[\cvJ(\vrr)]$ be the probability of observing a time-averaged current field $\cvJ(\vrr)=\tau^{-1}\int_0^{\tau} dt \, \vj (\vrr,t)$. Notice that this vector field must be divergence-free because it is coupled via the continuity equation (\ref{ch1:cont2}) to an optimal density profile which is assumed to be time-independent, see hypothesis (H1) in \S\ref{saddit}. This probability also obeys a large deviation principle,
\beq
\text{P}_{\tau}[\cvJ(\vrr)]\sim \exp\left(+\tau N^d G[\cvJ(\vrr)] \right),
\label{ch4:ldpplext}
\eeq
with a current LDF equivalent to that in Eq. (\ref{ch1:ldf2}) but with a space-dependent current field $\cvJ(\vrr)$. The optimal density profile $\rho_{\cvJ(\vrr)}[\vrr]$ is now solution of
\beq
\frac{\delta}{\delta \rho(\vrr')}\int_{\Lambda} d\vrr \, \Big(W_2[\rho(\vrr)] - 2 \cvJ(\vrr) \cdot \textbf{W}_1[\rho(\vrr)] +  \cvJ^2(\vrr)W_0[\rho(\vrr)] \Big)   = 0 \, ,
\label{ch4:optprofJr}
\eeq
which is the equivalent to Eq. (\ref{ch1:optprof}) in this case. For time-reversible systems condition (\ref{ch4:Acond1}) holds and $\rho_{\cvJ(\vrr)}[\vrr]$ remains invariant under (local or global) rotations of $\cvJ(\vrr)$. In this way we can simply relate $\text{P}_{\tau}[\cvJ(\vrr)]$ with the probability of any other divergence-free current field $\cvJ'(\vrr)$ locally-isometric to $\cvJ(\vrr)$, i.e. $\cvJ'(\vrr)^2 = \cvJ(\vrr)^2$ $\forall \vrr$, via a generalized Isometric Fluctuation Relation,
\beq
\lim_{\tau\to \infty}\, \frac{1}{\tau} \, \ln \left[\frac{\displaystyle \text{P}_{\tau}[\cvJ(\vrr)]}{\displaystyle \text{P}_{\tau}[\cvJ'(\vrr)]}\right] =
\int_{\partial\Lambda} d\Gamma \, \frac{\delta {\cal H}[\rho]}{\delta \rho} \vnn \cdot [\cvJ'(\vrr)-\cvJ(\vrr)] \, ,
\label{ch4:gIFR}
\eeq
where the integral (whose result is independent of $\rho(\vrr)$) is taken over the boundary $\partial \Lambda$ of the domain $\Lambda$ where the
system is defined, and $\vnn$ is the unit vector normal to the boundary at each point.
Notice that in general an arbitrary local or global rotation of a divergence-free vector field does not conserve the zero-divergence property, so this constrains
the current fields and/or local rotations for which this generalized IFR applies.
Note that the probability of observing a time averaged integrated current, $\text{P}_{\tau}(\vJJ)$, is given by the following path integral
\beq
\text{P}_{\tau}(\vJJ)=\int {\cal D}{\cvJ}\text{P}_{\tau}[\cvJ(\vrr)]\delta \left(\vJJ-\small\int_{\Lambda} d\vrr \, \cvJ(\vrr)\right).
\label{ch4:pjj}
\eeq
Hence, taking into account the above equation and that for long times Eq. (\ref{ch4:ldpplext}) holds and $\text{P}_{\tau}(\vJJ)\sim\exp\left(+\tau N^d G(\vJJ) \right)$, we can relate the large deviation function for the space- and time-averaged current, $G(\vJJ)$, to $G[\cvJ(\vrr)]$ via a contraction principle \cite{Ellis,Touchette}
\beq
G(\vJJ)=\max_{\substack{\tiny \cvJ(\vrr) : \vnabla\cdot\cvJ(\vrr)=0 \\ \tiny \vJJ=\int_{\Lambda} d\vrr \, \cvJ(\vrr)}} G[\cvJ(\vrr)] \, .
\label{ch4:contrac}
\eeq
The optimal, divergence-free current field $\cvJ_{\vJJ}(\vrr)$ solution of this variational problem may have spatial structure in general. Eq. (\ref{ch4:gIFR}) 
generalizes the IFR to situations where hypothesis (H2) of \S\ref{saddit} is violated, opening the door to isometries based on local 
(in addition to global) rotations. However, numerical results and phenomenological arguments strongly suggest that the constant 
solution, $\cvJ_{\vJJ}(\vrr)=\vJJ$, is the optimizer at least for a wide interval of current fluctuations, showing that hypothesis (H2) above is 
not only plausible but also well justified on physical grounds. In any case, the range of validity of this hypothesis can be explored by studying 
the limit of local stability of the constant current solution using tools similar to those in Ref. \cite{BD2}.

We could also try to generalize the IFR to the full space- and time-dependent variational problem of Eq. (\ref{ch1:ldf0}), i.e. to situations where the additivity conjecture breaks down. In this case, we would study the probability $\text{P}\left(\{\vj(\vrr,t)\}_0^{\tau} \right)$ of observing a particular history for the current field, which can be written as the path integral of the probability in Eq. (\ref{ch1:LDpple}) over histories of the density field $\{\rho(\vrr,t)\}_0^{\tau}$ coupled to the desired current field via the continuity Eq. (\ref{ch1:cont2}) at every point in space and time. This probability obeys another large deviation principle, with an optimal history of the density field $\{\rho_{\vJJ}(\vrr,t)\}_0^{\tau}$ which is solution of an equation similar to Eq. (\ref{ch4:optprofJr}) but with time-dependent profiles. However, as opposed to the cases above, the current field $\vj(\vrr,t)$ is not necessarily divergence-free because of the time-dependence of the associated $\rho_{\vJJ}(\vrr,t)$, resulting in a violation of condition (\ref{ch4:Acond1}). In this way the optimal $\rho_{\vJJ}(\vrr,t)$ depends on both $\vj(\vrr,t)$ and $\vj(\vrr,t)^2$ so it does not remain invariant under (local or global) instantaneous rotations of the current field, and hence no generalization of the IFR to the \emph{general} time-dependent regime can be done. However, as we shall see below, it is quite remarkable that additivity violations seem to happen in a controled way: once additivity breaks down, optimal paths depend on space and time via simple scaling laws (as for instance in a travelling wave form, $\vrr-\mathbf{v}t$, see below), and this can be used in turn to extend the validity of the IFR to time-dependent regimes \cite{CarlosPRL}.

\section{Additivity violation and spontaneous symmetry-breaking at the fluctuating level}
\label{sSSB}

Macroscopic fluctuation theory allows us to study dynamic fluctuations in diffusive media \cite{Bertini}, offering predictions for both the LDFs of interest and the optimal path in mesoscopic (or coarse-grained) phase space responsible for a given fluctuation. This optimal path is a dynamical object, which can be in general time-dependent \cite{Bertini}. However, we have argued above that, for a broad spectrum of fluctuations,
we expect the optimal path that minimizes the cost of a fluctuation to be time-independent, an idea captured by the additivity conjecture of \S\ref{saddit}. The physical picture behind this hypothesis corresponds to a system that, after a short transient of time at the beginning of the large deviation event (microscopic in the diffusive timescale $\tau$), settles into a time independent state with an structured density field (which can be different from the stationary one) and a spatially uniform current field equal to $\vJJ$. As we show in this section, this additivity scenario eventually breaks down for large current fluctuations via a dynamic phase transition at the fluctuating level involving a symmetry breaking event \cite{Bertini,BD2}, where time-dependent optimal paths in the form of traveling waves emerge as dominant solution to the variational problem (\ref{ch1:ldf0}).

We now study this phenomenon in detail. In order to simplify the discussion below, we focus on a 1D diffusive system on the unit interval, $\Lambda=[0,1]$, subject to periodic boundary conditions and possibly to a conservative external field $E$. In this case, and according to MFT, the LDF of the space\&time-averaged current
\be
J=\frac{1}{\tau}\int_0^{\tau}dt\int_0^1 j(x,t)dx \, ,
\label{curr1D}
\ee
is given by Eq. (\ref{ch1:ldf0}), which once particularized for our 1D driven diffusive system reads
\beq
G(J)=-\lim_{\tau\rightarrow \infty}\frac{1}{\tau} \min_{\{\rho,j\}_0^{\tau}}\left\{ \mint_{0}^{\tau}dt \mint_{0}^1 dx\dfrac{(j+D[\rho]\partial_x\rho-E\sigma[\rho])^2}{2 \sigma[\rho]} \right\},
\label{ch5:ldf0p}
\eeq
subject to the constraints imposed by the continuity equation $\partial_t \rho+\partial_x j=0$ and Eq. (\ref{curr1D}).
Periodic boundary conditions then imply that $\rho(0,t)=\rho(1,t)$ and $j(0,t)=j(1,t)$, and the total density is a conserved quantity
\be
\int_0^1 dx\rho(x,t)=\rho_0 \, .
\label{Arho0}
\ee
The stationary profile is the flat one (uniformly equal to $\rho_0$) and the average current is $J_{st}=\sigma[\rho_0]E$.
Finding the optimal fields solution of (\ref{ch5:ldf0p}) is in general a complex spatiotemporal problem whose solution remains challenging in most cases. 
The problem becomes much simpler however in different limiting cases. For instance, one expects that small current fluctuations around the 
average, $J\simeq J_{st}$, result from the random superposition of weakly-correlated local fluctuations of the microscopic jump process. In this case it is reasonable to assume the optimal density field to be just the flat, steady-state one,  $\rho_J(x,t)=\rho_0$, and hence $j_J(x,t)=J$, resulting in a simple quadratic form for the current LDF
\be
G_{\text{flat}}(J) = - \frac{(J-\sigma(\rho_0) E)^2}{2\sigma(\rho_0)} \, ,
\label{gflat0}
\ee
or an equivalent quadratic form for its Legendre transform
\beq
\mu_{\text{flat}}(\lambda)=\frac{\lambda (\lambda+2E)\sigma(\rho_0)}{2}.
\label{ch5:muflat}
\eeq
Therefore Gaussian statistics is obtained for small (i.e. typical) current fluctuations, in agreement with the central limit theorem. The previous argument, revolving around small fluctuations, breaks down however for moderate current deviations where correlations may play a relevant role. In fact, Bodineau and Derrida have shown recently \cite{BD2} that the flat profile indeed becomes unstable, in the sense that $G(J)$ \emph{increases} (compared to the flat solution) by adding a small time-dependent periodic perturbation to the otherwise constant profile, whenever
\be
8 \pi^2D^2(\rho_0)\sigma(\rho_0)+(E^2\sigma^2(\rho_0)-J^2)\sigma''(\rho_0)<0 \, ,
\label{Acond}
\ee
where $\sigma''$ denotes second derivative. This condition implies a well-defined critical current
\be
|J_c|=\sqrt{\frac{8\pi^2D^2(\rho_0)\sigma(\rho_0)}{\sigma''(\rho_0)}+E^2\sigma^2(\rho_0)} \, ,
\label{Aqc}
\ee
where the instability emerges.
Notice that, for the instability to exist, the strength of the driving field, $|E|$, must be large enough to guarantee a positive discriminant in Eq. (\ref{Aqc}), namely
\be
|E|\geq |E_c| \equiv \text{Re}\left[\sqrt{-\frac{8\pi^2D(\rho_0)^2}{\sigma(\rho_0)\sigma''(\rho_0)}}\right] \, .
\label{AEc}
\ee
Therefore, since the mobility $\sigma(\rho)$ is positive definite, a non-zero threshold field only exists for models such that $\sigma''(\rho)<0$, as for instance the WASEP, where $\sigma(\rho)=\rho(1-\rho)$ \cite{CarlosSSB}. For the other transport model of interest here, the KMP model \cite{kmp,PabloSSB}, we have $\sigma(\rho)=\rho^2$ so $\sigma''(\rho)>0$ and hence $|E_c|=0$, thus exhibiting the aforementioned instability even in the absence of external fields \cite{PabloSSB}. In fact, for $\sigma''(\rho)>0$ the instability dominates whenever $|J|>|J_c|$, while for $\sigma''(\rho)<0$ the instability controls the $|J|<|J_c|$ regime.

When the instability kicks in, an analysis of the resulting perturbation \cite{BD2} suggests that the dominant form of the optimal profile is a traveling wave moving at constant velocity $v$
\be
\rho_J(x,t)=\omega_J(x-vt) \, ,
\label{Aonda}
\ee
which implies via the continuity equation $\partial_t \rho + \partial_x j=0$ that
\be
j_J(x,t)=J-v\rho_0+v\omega_J(x-vt) \, .
\label{q1}
\ee
Provided that the traveling-wave form remains as the optimal solution for currents well-beyond the critical threshold, the current LDF can now be written as
\be
G(J)=-\min_{\omega_J(x),v}\int_0^1\frac{dx}{2\sigma[\omega_J]}[J-v\rho_0+v\omega_J(x)+D[\omega_J]\omega'_J(x)-\sigma[\omega_J]E]^2 \, ,
\label{Aldfwave}
\ee
where we have absorbed the time dependence after a change of variables due to the periodic boundary conditions,
and the minimum is now taken over the traveling wave profile $\omega_J(x)$ and its velocity $v$. Expanding now the square in Eq. (\ref{Aldfwave}), we notice that the terms linear in $\omega_J'$ give a null contribution due again to the system periodicity. Taking also into account the constraint $\int_0^1\omega_J(x) dx=\rho_0$, see Eq. (\ref{Arho0}), one gets
\be
G(J)=-\min_{\omega_J(x),v}\left[\int_0^1dx (X[\omega_J]+\omega'_J(x)^2Y[\omega_J]\right]+JE \, ,
\label{LDF3}
\ee
where, borrowing the notation of ref. \cite{BD2},
\beq
X[\omega_J]=\frac{[J-v(\rho_0-\omega_J)]^2}{2\sigma[\omega_J)]}+\frac{E^2\sigma[\omega_J)]}{2} \qquad \text{and} \qquad Y[\omega_J]=\frac{D[\omega_J)]^2}{2\sigma[\omega_J]}.
\label{XYw}
\eeq
The differential equation for the optimal profile solution of the variational problem Eq. (\ref{LDF3}) can be written as
\beq
X[\omega_J]-\omega'_J(x)^2Y[\omega_J]=C_1+C_2\omega_J \, ,
\label{optprofw}
\eeq
with $C_1$ and $C_2$ constants to be determined below. This equation generically yields a symmetric optimal profile with a $\omega_J(x)$ with a single minimum $\omega_1=\omega_J(x_1)$ and a 
single maximum $\omega_0=\omega_J(x_0)$ such that $|x_0-x_1|=1/2$ \footnote{The optimal traveling wave profile solution of Eq. (\ref{optprofw}) must be 
symmetric because this differential equation remains invariant under the transformation $x\to 1-x$. Periodicity implies that profile extrema, if any, come 
in pairs (maximum and minimum). Moreover, the profile has at most a single pair of extrema because, when we make $w'_J(x)=0$ in Eq. (\ref{optprofw}), the 
resulting equation is third order for the KMP model and WASEP. This, together with the symmetry of the profile, implies in turn that the 
extrema $\omega_1=\omega_J(x_1)$ and $\omega_0=\omega_J(x_0)$ are such that $|x_0-x_1|=1/2$.}.
The optimal velocity also follows from the above variational problem,
\beq
v=-J\frac{\displaystyle\int_0^1dx\frac{(\omega_J(x)-\rho_0)}{\sigma[\omega_J]}}{\displaystyle\int_0^1dx\frac{(\omega_J(x)-\rho_0)^2}{\sigma[\omega_J]}}.
\label{eq1}
\eeq
It is worth emphasizing that the optimal velocity is proportional to $J$. This implies that the optimal profile solution of Eq. (\ref{optprofw}) depends exclusively on $J^2$ and not on the current sign, reflecting the Gallavotti-Cohen time-reversal symmetry. This invariance of the optimal profile under the transformation $J\leftrightarrow -J$ can now be used in Eq. (\ref{LDF3}) to show explicitly the GC symmetry in this case, $G(J)-G(-J)=2EJ$, which as expected remains valid in this time-dependent regime.

The constants $C_1$ and $C_2$ appearing in Eq. (\ref{optprofw}) can be expressed in terms of the extrema $\omega_1$ and $\omega_0$ of the profile via
\begin{eqnarray}
X(\omega_1)=C_1+C_2\omega_1 \, ,
\label{eq2} \\
X(\omega_0)=C_1+C_2\omega_0 \, .
\label{eq3}
\end{eqnarray}
Moreover, the extrema locations are fixed by the constraints on the distance between them and the total density of the system,
\begin{equation}
\frac{1}{2}=\int_{x_1}^{x_0}dx
=\int_{\omega_1}^{\omega_0}\sqrt{\frac{Y(\omega_J)}{X(\omega_J)-C_1-C_2\omega_J}}d\omega_J \\
\label{eq4}
\end{equation}
and
\begin{equation}
\frac{\rho_0}{2}=\int_{x_1}^{x_0}\omega_J(x) dx=\int_{\omega_1}^{\omega_0}\sqrt{\frac{\omega_J^2 Y(\omega_J)}{X(\omega_J)-C_1-C_2\omega_J}}d\omega_J \, ,
\label{eq5}
\end{equation}
where we have used in the last equality of both expressions the differential equation (\ref{optprofw}). In this way, for fixed values of the current $J$ and the density $\rho_0$ (provided externally), we use eqs. (\ref{eq1})-(\ref{eq5}) in order to determine the five constants $\omega_1,\omega_0,C_1,C_2,v$ which can be used in turn to obtain the shape of the optimal density profile $\omega_J(x)$ from Eq. (\ref{optprofw}).
Notice that the unknown variables $\omega_0$, $\omega_1$ appear as the integration limits in eqs. (\ref{eq4}) and (\ref{eq5}), making this problem remarkably difficult to solve numerically. It can be shown that, by performing a suitable change of variables \cite{CarlosSSB}, the integrals involved in the calculation can be transformed into known functions, as e.g. elliptic integrals of the first kind, thus allowing to derive an explicit analytical expression for $\omega_J(x)$ as a function of the relevant constants for the particular model of interest.

\subsection{Observation of spontaneous symmetry breaking at the fluctuating level}

\begin{figure}
\centerline{
\includegraphics[width=14cm]{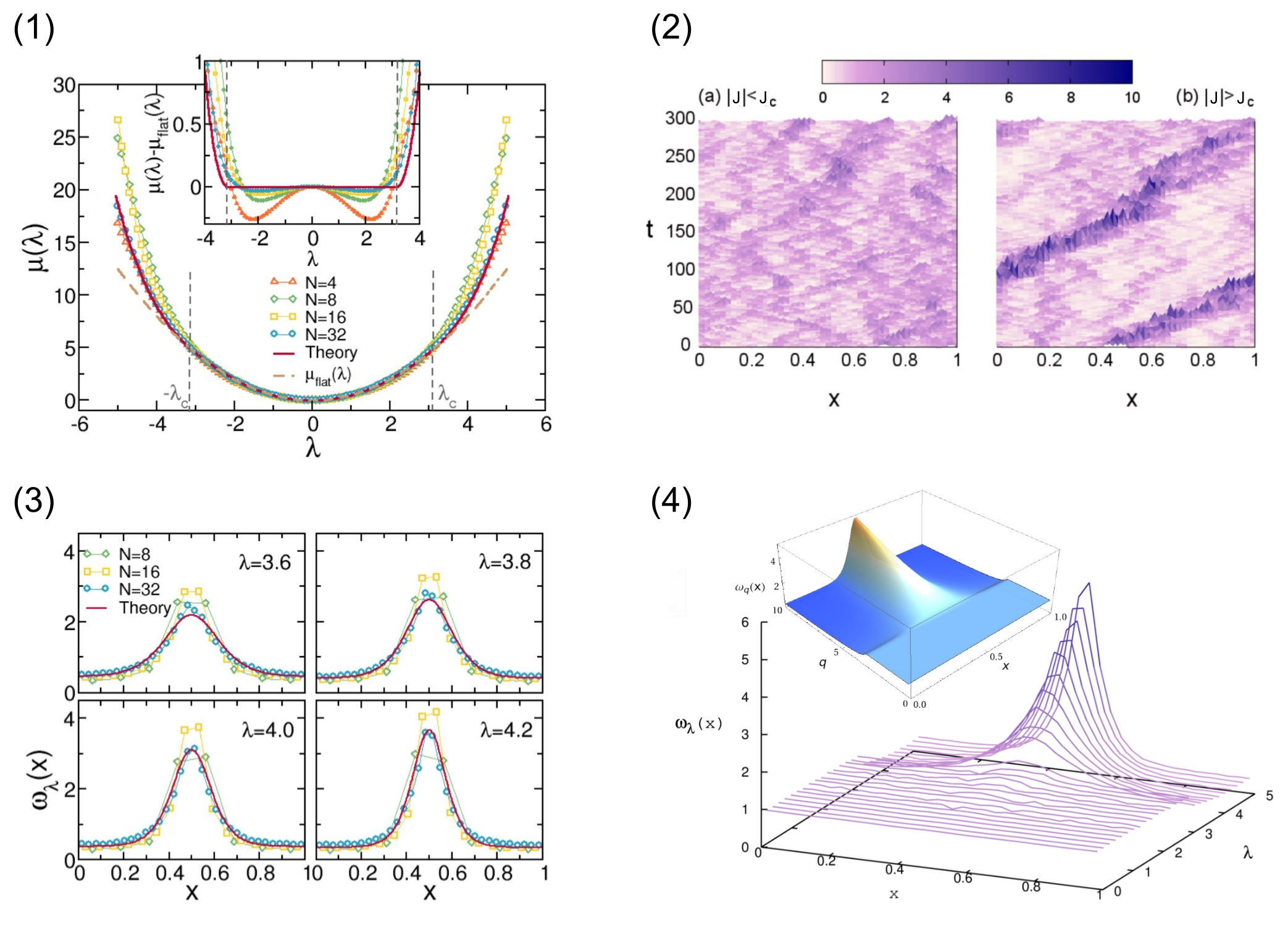}
}
\caption{(Color online) (1) Main: Measured $\mu(\lambda)$ for the 1D KMP model on the ring and increasing values of $N$, together with the MFT prediction and the Gaussian approximation. Inset: $\mu(\lambda)-\mu_{\text{flat}}(\lambda)$ for the same $N$. Data converge  to the MFT prediction as $N$ increases. (2) Typical evolution of the energy field for different current fluctuations in the 1D KMP model on a ring. (2.a) Small current fluctuations result from weakly-correlated local events. (2.b) However, for $|J|>J_c$ the system facilitates this unlikely deviation by forming a traveling wave. (3) Supercritical profiles for different $\lambda$ and varying $N$, and MFT predictions. (4) Measured profiles as a function of $\lambda$ for $N=32$. The inset shows the MFT prediction for the optimal profile $\omega_J(x)$. Profiles are flat up to the critical current, beyond which a nonlinear wave pattern develops.
}
\label{figSSB1}
\end{figure}

Can we observe the dynamic phase transition described in the previous section, related to the breakdown of additivity, in actual simulations of diffusive systems? To answer this question, we study now the fluctuating behavior of the 1D KMP and WASEP models under periodic boundary conditions. MFT predicts this dynamic phase transitions to happen for currents away from the average current, so we will certainly need the advanced Monte Carlo method of \S\ref{ssim} to sample the tails of the current distribution where signs of this instability may appear.

We first report compelling evidences of this phenomenon in the 1D KMP model of energy transport \cite{PabloSSB}, for which the relevant transport coefficients are $D[\rho]=1/2$ and $\sigma[\rho]=\rho^2$ (and hence $\sigma''[\rho]>0$, see discussion below Eq. (\ref{AEc})). In particular, we performed extensive simulations using the cloning algorithm with a large number $M\gtrsim 10^4$ of clones and fixed total density $\rho_0=1$. We are interested in the statistics of the total current $J$ flowing through the system, averaged over a long diffusive time $\tau$. For $\tau \to \infty$ this time average converges toward the 
ensemble average $J_{st}$, which is of course zero because the system is isolated and in equilibrium. However, for long but 
finite $\tau$ we may still observe fluctuations $J\ne J_{st}$, and their probability $\text{P}_{\tau}(J)$ is our primary object of interest.

To better grasp the physics behind this phenomenon, let us first examine the typical spacetime trajectories for small and large current fluctuations. 
As expected, , see Fig. \ref{figSSB1}.2.a, we find that small current fluctuations result indeed from the sum of weakly-correlated local random events in the density field, thus giving rise to a flat, structureless optimal density field in average and, as we will show below, Gaussian statistics as dictated by the central limit theorem. However, \emph{for large enough currents}, the KMP system self-organizes into a coherent traveling wave which facilitates this rare event by accumulating energy in a localized packet, 
see Fig. \ref{figSSB1}.2.b, with a critical current $J_c$ separating both regimes. For our particular case, it is easy to show that $J_c=\pi$, see Eq. (\ref{Aqc}). This phenomenon, predicted by MFT above \cite{Bertini,BD2}, is most striking for this model as it happens in an isolated equilibrium system in the absence of any external field, breaking spontaneously a symmetry (translation invariance) in 1D. This is an example of the general observation that symmetry-breaking instabilities forbidden in equilibrium steady states can however happen at the fluctuating level or in nonequilibrium settings \cite{Jona2}. Such instabilities may help explaining puzzling asymmetries in nature \cite{Jona2}, from the dominance of left-handed chiral molecules in biology to the matter-antimatter asymmetry in cosmology.

Fig. \ref{figSSB1}.1 shows simulation results for the LDF $\mu(\lambda)$ and increasing values of $N$. The MFT prediction of Gaussian current statistics for $|J|< J_c=\pi$ corresponds to quadratic behavior in $\mu(\lambda)= \mu_{\text{flat}}(\lambda)=\lambda^2/2$ up to a critical $|\lambda_c|=\pi$. 
This is fully confirmed in Fig. \ref{figSSB1}.1 as $N$ increases, meaning that small and intermediate current fluctuations have their origin in the superposition of weakly-correlated local events, giving rise to Gaussian statistics. However, for fluctuations above the critical threshold, $|\lambda|>\lambda_c$, deviations from this simple quadratic form are apparent, signaling the onset of a phase transition. In fact, as $N$ increases a clear convergence toward the MFT non-quadratic prediction is observed, with very good results already for $N=32$. Strong finite size effects associated with the finite population of clones $M$ prevent us from reaching larger system sizes \cite{PabloJSTAT}, but $N=32$
is already \emph{close} enough to the asymptotic hydrodynamic behavior. Still, small corrections to the MFT predictions are observed which quickly decrease with $N$, see inset to 
Fig. \ref{figSSB1}.1.

\begin{figure}
\centerline{
\includegraphics[width=16cm]{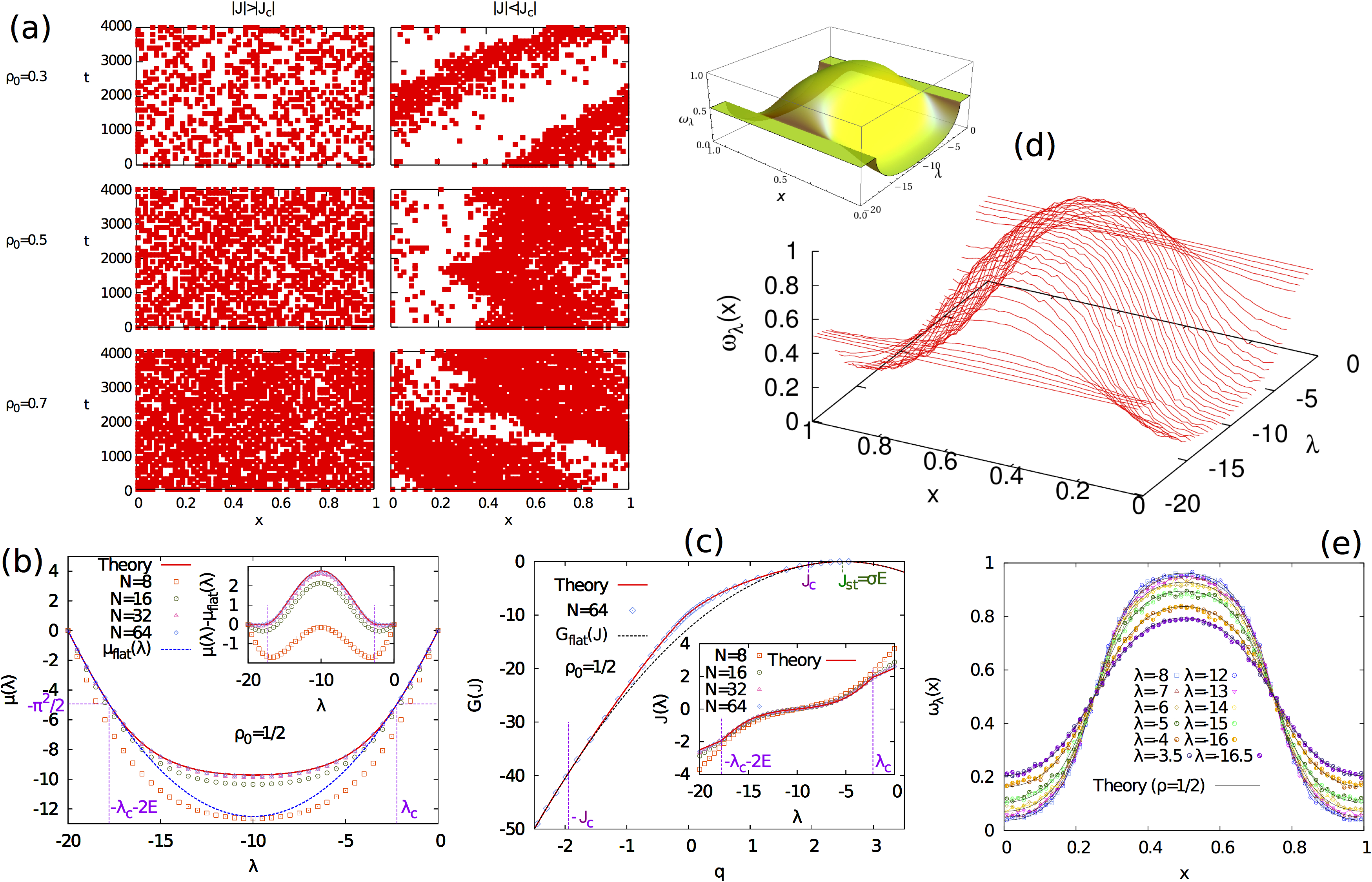}
}
\caption{(Color online) (a) Typical evolution of microscopic configurations for current fluctuations above and below the critical current for three 
different densities in the 1D WASEP (from top to bottom, $\rho_0=0.3,\, 0.5,\, 0.7$). Left panels correspond to currents above the critical one 
where the system remains homogeneous. Right panels correspond to subcritical current fluctuations where a traveling wave emerges. 
(b) Measured $\mu(\lambda)$ for $\rho_0=0.5$ and increasing $N$, together with the MFT result (solid red line) and the quadratic 
approximation (dashed blue line). The inset shows the difference $\mu(\lambda)-\mu_{\text{flat}}(\lambda)$. 
(c) Current LDF for $\rho_0=0.5$ and $N=64$. The traveling wave solution enhances the probability for fluctuations $|J|<|J_c|$ (solid red line) with 
respect to the flat profile associated with gaussian statistics (dashed black line). The inset shows the average current as a 
function of $\lambda$ as $N$ increases. (d) Measured density profiles as a function of $\lambda$ for $\rho_0=0.5$ and $N=64$, with the MFT 
prediction as inset. Optimal profiles are flat up to a critical current (equivalently $\lambda_c^{\pm}$) where a traveling wave emerges. 
(e) Collapse of measured profiles associated with different current fluctuations $\omega_{\lambda}(x)$ and their time-reversal 
partners $\omega_{-\lambda-2E}(x)$ for $N=64$ and $\rho_0=0.5$, together with theoretical predictions. Optimal profiles, both below and above 
the instability, remain invariant under change of sign of the current.
}
\label{figSSB2}
\end{figure}

The dynamical phase transition is most evident at the configurational level, as observed in Fig. \ref{figSSB1}.2, so we measured the average density profile associated with a given current fluctuation \cite{PabloPRL,CarlosSSB}, see Figs. \ref{figSSB1}.3-4. Because of the system periodicity, and in order not to blur away the possible structure present in microscopic configurations, we performed profile averages around the instantaneous center of mass. For that, we consider the system as a 1D ring embedded in two-dimensional space, 
see Fig. \ref{sketchmodels},  and compute the angular position of the center of mass, shifting it to the origin before averaging. In particular, we assign an angular position $\theta_i=2\pi i/N$ to each site $i\in[1,N]$ in the lattice. The angular position of the center of mass for a given microscopic configuration $\mathbf{\rho}=\{\rho_i,i=1,\ldots,N\}$, with $\rho_i\geq 0$ the on-site energies, is thus defined as
\be
\theta_{\text{CM}} \equiv  \tan^{-1}\left(\frac{Y_{\text{CM}}}{X_{\text{CM}}}\right) \qquad \text{with} \qquad X_{\text{CM}}= \frac{1}{N\rho_0} \sum_{i=1}^N \rho_i \cos \theta_i \quad ; \quad Y_{\text{CM}}= \frac{1}{N\rho_0} \sum_{i=1}^N \rho_i \sin \theta_i \, ,
\label{angCM}
\ee
and recall that $\rho_0=N^{-1}\sum_{i=1}^N \rho_i$ is the total energy per site. Notice that this center-of-mass averaging procedure yields a spurious weak structure in the Gaussian (homogeneous) fluctuation region, equivalent to averaging random density profiles around their (random) center of mass. Such a spurious profile is of course independent of the current $J$ and can be easily subtracted. On the other hand, once the instability is triggered average profiles exhibit a much more pronounced structure resulting from the appearance of a traveling wave. 
Fig. \ref{figSSB1}.3 shows the measured profile $\omega_{\lambda}(x)$ for different $\lambda>\lambda_c$ and varying $N$. Again, fast convergence toward the MFT result is observed, with excellent agreement for $N=32$ in all cases. 
Fig. \ref{figSSB1}.4 shows the measured profiles for $N=32$ and different $\lambda$, which closely resembles the MFT scenario, see inset. Finally, we mention that we also measured the average velocity associated with a given current fluctuation (not shown), and the comparison with theoretical predictions is again excellent, with a velocity linear in the current for subcritical fluctuations but following a strongly nonlinear relation above the threshold $J_c$ \cite{PabloSSB}.

We also studied this phenomenon in the weakly-asymmetric simple exclusion process (WASEP) described in \S\ref{smodels}, using the same tools as described above. Recall that in this case $D[\rho]=1/2$ and $\sigma[\rho]=\rho(1-\rho)$. In this way, the critical current and the threshold field for the WASEP, see eqs. (\ref{Aqc}) and (\ref{AEc}), are
\begin{eqnarray}
|J_c| & = & \rho_0(1-\rho_0)\sqrt{E^2 - E_c^2} \, \label{qc2} \\
|E_c| & = & \frac{\pi}{\sqrt{\rho_0(1-\rho_0)}} \, .
\label{Ec}
\end{eqnarray}
Equivalently, taking into account that $J_{st} = \rho_0(1-\rho_0) E$, we have that
\be
|J_c| = |J_{st} | \sqrt{1-\left(\frac{E_c}{E}\right)^2} \, ,
\label{qcWASEP}
\ee
so $|J_c| < |J_{st} |$ in WASEP for field strengths above the threshold, while no phase transition happens for $|E|<|E_c|$. This is in stark contrast with the phenomenology found for the KMP model, and stems from the fact that $\sigma''[\rho]<0$ for the WASEP. Therefore we expect Gaussian current fluctuations and a flat optimal density profile for $|J|>|J_c|$. Legendre-transforming the quadratic current LDF in Eq. (\ref{gflat0}) once particularized for WASEP, we have in this regime
\be
\mu_{\text{flat}}(\lambda)=\frac{\rho_0}{2}(1-\rho_0)\lambda (\lambda+2E) \, .
\label{muflat}
\ee
However, for currents $|J|<|J_c|$, a time-dependent regime with an optimal density profile in the form of a traveling wave is expected. This corresponds to values of the conjugate parameter $\lambda$ such that $|\lambda+E| < \tilde{\lambda}_c$, with
\be
\tilde{\lambda}_c \equiv \frac{q_c}{\rho_0(1-\rho_0)} = \sqrt{E^2 - \frac{\pi^2}{\rho_0(1-\rho_0)}} = \sqrt{E^2 - E_c^2}\, ,
\label{lambdac}
\ee
and where we have used eqs. (\ref{qc2})-(\ref{qcWASEP}). Therefore we expect traveling wave solutions for $\lambda_c^-<\lambda<\lambda_c^+$, where
\be
\lambda_c^{\pm}\equiv \pm\tilde{\lambda}_c - E \, .
\label{lambdacpm}
\ee
In this way the time-dependent fluctuation regime kicks in whenever $\mu(\lambda)<\mu(\lambda_c^{\pm})=-\pi^2/2$, see Eq. (\ref{muflat}).

We performed simulations of the 1d periodic WASEP for three different average densities, $\rho_0=0.3,\, 0.5$ and $0.7$, for increasing system sizes $N\in[8,64]$ and a fixed external field $E=+10$. This driving field is above the threshold $E_c$ for all three densities $\rho_0$, see Eq. (\ref{Ec}), so we expect the instability to appear on the basis of the previous analysis. 
Fig. \ref{figSSB2}.a shows the typical spacetime trajectories for these three densities and supercritical as well as subcritical current fluctuations. As for the KMP model, we find a dynamic phase transition in current fluctuations involving the spontaneous breaking of translation symmetry. However, for the WASEP the dynamic phase transition corresponds to the emergence of a macroscopic jammed state which hinders transport of particles to facilitate a current fluctuation well below the average as $|J_c|<|J_{st} |$.

Fig. \ref{figSSB2}.b shows simulation results for $\mu(\lambda)$ and increasing values of $N$ for a particular value of $\rho_0=0.5$ (similar results hold for other densities), together with the explicit MFT results. Gaussian current statistics corresponds to the quadratic behavior of Eq.
(\ref{muflat}), which is fully confirmed in the figure for $|\lambda+E|>\tilde{\lambda}_c$ and different values of $N$. This means that small current fluctuations, $J\simeq J_{st}$, originate from the superposition of uncorrelated (or at most weakly-correlated) local events of the stochastic jump process.
Interestingly, this observation also applies to the time-reversal partners of these small fluctuations, $-J$, which are far from typical. This is a direct consequence of the Gallavotti-Cohen symmetry \cite{GC,ECM,K,LS}, which implies that the statistics associated with a current fluctuation does not depend on the current sign \cite{PabloJSTAT}, see \S\ref{sGC}. On the other hand, for fluctuations below a critical threshold, $|J|<|J_c|$ or equivalently $\lambda_c^-<\lambda<\lambda_c^+$, deviations from this simple quadratic form announce the dynamical phase transition to a coherent traveling wave profile. As for the KMP model, finite size effects are strong but convergence toward the asymptotic MFT result is clearly observed.
On the other hand, the Gallavotti-Cohen fluctuation theorem for currents holds in the whole current range and irrespective of $N$ \footnote{This results from the microreversibility of the model at hand: while we need a large size limit in order to verify the predictions derived from MFT (which is a macroscopic theory), no finite-size corrections affect the fluctuation theorem, whose validity can be used to ascertain the range of applicability of the cloning algorithm used to sample large deviations \cite{PabloJSTAT}}.

We also measured the average current $\la J_{\lambda}\ra$ associated with a fixed value of the conjugate parameter $\lambda$, see the inset of Fig. \ref{figSSB2}.c, and
the agreement with MFT is again excellent (improving as $N$ increases).
We may use the measured $\la J_{\lambda}\ra$ to give a direct Monte Carlo estimate of the current LDF $G(J)$. In fact, $\la J_{\lambda}\ra$ is the current conjugated to a given $\lambda$ and hence we may write $G(J)=\mu(\lambda)+\lambda \la J_{\lambda}\ra$, where we combine the measured $\mu(\lambda)$ and $\la J_{\lambda}\ra$ in Figs. \ref{figSSB2}.b-c, respectively. The result for $G(J)$ for $\rho_0=0.5$ is plotted in 
Fig. \ref{figSSB2}.c, where we again find a good agreement between theory and Monte Carlo simulation. Notice in particular the deviation from quadratic behavior observed for currents $|J|<|J_c|$ resulting from the formation of macroscopic jammed states.

A most interesting observable consists in the optimal density profiles associated with different current fluctuations. Fig. \ref{figSSB2}.d shows
a three-dimensional plot of the measured profiles for $N=64$ and different $\lambda$, again for $\rho_0=0.5$, which closely resembles the MFT scenario plotted as inset to this figure. In general, the traveling wave profile grows from the flat form as $\lambda$ crosses the critical values $\lambda_c^{\pm}$ penetrating into the critical region, thus favoring a macroscopic jammed state that hinders transport of particles and thus facilitates a time-averaged current fluctuation well below the average current. This macroscopic jammed state reaches its maximum expression for $\lambda=-E$, or equivalently $J=0$ --see Fig. \ref{figSSB2}.c, so the system is maximally jammed for zero current irrespective of the driving field $E$.

Although the instabilities observed for the KMP and WASEP models are described equally well by MFT, the physical interpretation of the dynamic transition is quite different. In particular, in the KMP model the instability happens because the system optimizes the transport of a \emph{large} current by gathering energy in a localized packet (the wave) which then travels coherently, breaking spontaneously translation symmetry in the process. On the other hand, the WASEP instability happens in order to hinder the transport of particles via the formation of a macroscopic jammed states (again the wave), thus facilitating a current fluctuation well \emph{below} the average. Interestingly, both phenomena are sides of, essentially, the same instability.

\section{Discussion and outlook}
\label{sDO}

In this work we have studied the thermodynamics of currents in nonequilibrium diffusive systems, using both macroscopic fluctuation theory as theoretical framework and advanced Monte Carlo simulations of several stochastic lattice gases as a laboratory
to test the emerging picture.

Our starting point was a mesoscopic fluctuating hydrodynamic theory for the density field.
The validity of this hydrodynamic description can be rigorously demonstrated for a large family of stochastic microscopic models \cite{Spohn}, but it is expected to describe the mesoscopic, coarse-grained evolution of a broad class of real systems of theoretical and technological interest characterized by a locally conserved magnitude (e.g. energy, particles, momentum, charge, etc.).
From this fluctuating hydrodynamic description, and using a standard path integral formulation of the problem, we can write the probability of a path in mesoscopic phase space, that is, the space spanned by the slow hydrodynamic fields. The \emph{action} associated with this path can be then used to derive the large-deviation functions of the macroscopic observables of interest via contraction principles \cite{Ellis,Touchette}, and in particular the current LDF, which is expected to play in nonequilibrium physics a role equivalent to the density LDF in equilibrium.

This calculation yields a complex spatiotemporal variational problem whose solution remains challenging in most cases. However, a simple and powerful additivity conjecture simplifies the variational problem, allowing us to obtain explicit predictions for the current LDF and the optimal path that the system adopts to facilitate a given fluctuation. We have confirmed the validity of this additivity conjecture for a wide interval of current fluctuations in one and two dimensions for the KMP model of energy transport coupled to thermal baths at different temperatures \cite{PabloPRL,Carlostesis}. In particular, we found that the current distribution shows a Gaussian regime for small current fluctuations and non-Gaussian, exponential tails for large deviations of the current, such that in all cases the Gallavotti-Cohen fluctuation relation holds. Our results thus strongly support the additivity hypothesis as an useful tool to understand current statistics in diffusive systems, suggesting a general approach to attack a large class of nonequilibrium phenomena based on few simple principles.

As an important by-product of our numerical work, we have measured the precise optimal path that the system follows in phase space in order to sustain a given fluctuation, showing that this path
coincides with the solution to the variational problem posed by MFT. Interestingly, we have also shown that by demanding invariance of optimal paths under symmetry transformations, new and general fluctuation relations valid arbitrarily far from equilibrium are unveiled. This opens an unexplored route toward a deeper understanding of nonequilibrium physics by bringing symmetry principles to the realm of fluctuations. This general idea can be exploited to study hidden symmetries of the current distribution out of equilibrium. In particular we have derived an isometric fluctuation relation (FR) \cite{PabloIFR} which links in a strikingly simple manner the probabilities of any pair of isometric current fluctuations. This relation, which results from the time-reversibility of the dynamics, includes as a particular instance the Gallavotti-Cohen fluctuation theorem in this context but adds a completely new perspective on the high level of symmetry imposed by time-reversibility on the statistics of nonequilibrium fluctuations. The new symmetry implies remarkable hierarchies of equations for the current cumulants and the nonlinear response coefficients, going far beyond Onsager's reciprocity relations and Green-Kubo formulae. We have confirmed the validity of the new symmetry relation in extensive numerical simulations of the 2D KMP model, and our results suggest that the idea of symmetry in fluctuations as invariance of optimal paths has far-reaching consequences in diverse fields. In particular, generalizations of MFT to systems with several conserved, coupled fields (e.g.  fully hydrodynamic models) would allow to formulate an extended IFR for the joint distributions of the different hydrodynamic currents. We expect that this extended IFR would strongly constrain the shape of these joint current distributions, linking in a hierarchical manner the mixed nonlinear response coefficients and giving rise to unexpected cross-relations among the different coefficients. These relations may have direct application in complex nonequilibrium situations, e.g. from fluids and turbulence to thermo- and piezo-electricity, electrolytes, and chemical kinetics.

The IFR and its generalizations are fluctuation theorems which emerge from MFT. A fundamental issue consists in the demonstration of IFR starting from microscopic dynamics rather than MFT. Techniques similar to those used in \S\ref{sGC}, based on the spectral properties of the microscopic evolution operator, may prove useful but they must be supplemented with additional insights on its scaling properties as the macroscopic limit is approached. Also interesting is the possibility of an IFR for discrete isometries related with the underlying lattice in stochastic models. These open questions call for further study.

The optimal path solution of the MFT variational problem can be in general time dependent. The additivity of current fluctuations shows that this path is in fact time-independent for a broad range of fluctuations. We have shown however that this scenario eventually breaks down in isolated systems for large fluctuations via a dynamic phase transition at the fluctuating level involving a symmetry-breaking event: while small current fluctuations result from the superposition of weakly-correlated local events and thus obey Gaussian statistics, for large enough currents the system self-organizes in a coherent traveling wave which \emph{facilitates} the rare event by forming a localized density packet which breaks translation invariance. Moreover, the presence of these \emph{coherent structures} associated with large, rare fluctuations \cite{BD2,PabloSSB} imply in turn that these events are far more probable than previously anticipated. We have observed this phenomenon,
predicted by MFT \cite{BD2,PabloSSB,CarlosSSB}, in two different 1D diffusive models: the KMP model of energy transport and the WASEP.  Our results show unambiguously that the dynamical phase transitions observed in the KMP and WASEP models are continuous as conjectured in \cite{BD2}, excluding the possibility of a first-order scenario \cite{PabloSSB,CarlosSSB}. This suggests that a traveling wave is in fact the most favorable time-dependent profile once the instability is triggered. This observation may greatly simplify general time-dependent calculations, but the question remains of whether this is the whole story or if other, more complex solutions may play a dominant role for even larger fluctuations. An interesting, related question concerns the properties of time-dependent solutions for systems with open boundaries, where traveling-wave patterns are not appropriate. The time-independent profiles in these cases, from which a suitable perturbative analysis would hint at the form of the time-dependent solution, are far more complex than the trivial homogeneous profiles that appear for periodic systems, difficulting progress along this line. In fact, a recent study \cite{Gorissen} has found no evidence of dynamical phase transition in WASEP with open boundaries. In any case, it seems clear that extremely rare events call in general for coherent, self-organized patterns in order to be sustained \cite{rare}.

Another interesting direction to explore in a near future is the appearance of this phenomenon in higher-dimensional systems. In this case the solution of the associated MFT is far more complicated, with no guarantee of an unique solution and several patterns may appear \cite{CarlosPRL}. The role of numerical simulation will hence prove essential to explore rare current fluctuations in high-dimensional systems and to understand the appearance of dynamical phase transitions at the fluctuation level \cite{Carlostesis}. Furthermore, the simplicity and elegance of this phenomenon suggests that it might be a rather general property of any fluctuating field theory, with possible expressions in quantum field theory, hydrodynamics, etc.

Finally, although we have not focused here on this issue, let us mention that while the MFT is usually applied to conservative systems, it has been recently generalized to dissipative systems characterized by a continuous loss of energy to the environment \cite{prados1,prados2,prados3,BL1,BL2}. In this case, the essential macroscopic observables which characterize the nonequilibrium behavior are the current and the dissipated energy. Using the path integral formalism described in \S\ref{sMFT}, it is possible to define the large deviation function of these observables and the optimal fields associated with their fluctuations. This extension of MFT to dissipative media has been recently tested in detail in Refs. \cite{BL1,BL2,prados1,prados2,prados3}.

All these results are exciting and suggest that MFT and its generalizations provide a powerful theoretical framework to understand in detail the physics of many different nonequilibrium systems.The proposed scheme is very general, as MFT is based solely on (a) the knowledge of the conservation laws and symmetries governing a system, which allow to write down the balance equations for the fluctuating fields, and (b) a few characteristic transport coefficients appearing in the fluctuating balance equations. This opens the door to further general results in nonequilibrium physics that we plan to explore in a near future.

\emph{We thank T. Bodineau, B. Derrida, A. Lasanta, J.L. Lebowitz, V. Lecomte, A. Prados and J. Tailleur for illuminating discussions.}

\appendix

\section{Additivity predictions for current fluctuations in the 2D KMP model}
\label{sadditKMP}

In this appendix we solve the simplified variational problem posed by MFT after the additivity conjecture of \S\ref{saddit} for the particular case of the two-dimensional Kipnis-Marchioro-Presutti model defined in \S\ref{smodels}, when subject to a temperature gradient along the $x$-direction (i.e. in contact with thermal baths at the left and right boundaries at temperatures $T_L$ and $T_R$, respectively) and with periodic boundary conditions along the $y$-direction. Due to the symmetry of the problem, the calculations here described are also useful for the 1D KMP model subject to the same temperature gradient, for which predictions follow by simply making zero all $y$-components.

The equation for the optimal density field under the additivity conjecture reads
\beq
D[\rho_{\vJJ}]^2 (\vnabla \rho_{\vJJ})^2=\vJJ^2\left(1+2\sigma[\rho_{\vJJ}]K(\vJJ^2)\right)
\label{optprof}
\eeq
where $K(\vJJ^2)$ is a constant of integration which guarantees the correct boundary conditions for $\rho_{\vJJ}(\vrr)$, which in this particular case are
\be
\rho_{\vJJ}(x=0,y)=T_L \quad ; \quad \rho_{\vJJ}(x=1,y)=T_R \quad ; \quad \rho_{\vJJ}(x,y=0)=\rho_{\vJJ}(x,y=1) \, .
\label{bcond}
\ee
Moreover, for the KMP model recall that the diffusivity and mobility transport coefficients entering MFT are $D[\rho]=1/2$ and $\sigma[\rho]=\rho^2$, respectively. Interestingly, the optimal field solution of Eq. (\ref{optprof}) depends exclusively on the modulus of the current vector $J=|\vJJ|$, that is
$\rho_{\vJJ}(\vrr)=\rho_J(\vrr)$. In addition,
the symmetry of the problem suggests that $\rho_J(\vrr)$ depends exclusively on $x$, with no structure in the $y$-direction, compatible with the presence of an external gradient along the $x$-direction, i.e. $\rho_J(\vrr)\equiv \rho_J(x)$. In the last expression we have simplified our notation in order not to clutter formulas below. Under these considerations, the differential equation (\ref{optprof}) for the optimal profile in the 2D KMP model becomes
\beq
\left( \dfrac{d\rho_{J}(x)}{dx} \right)^2=4J^2\Big(1+2K(J) \rho_{J}^2(x)\Big)
\label{optprofKMP}
\eeq
Here two different scenarios appear. On one hand, for large enough $K(J)$ the r.h.s. of Eq. (\ref{optprofKMP}) does not vanish $\forall x\in[0,1]$ and the resulting profile is monotone. In this case, and assuming $T_L>T_R$ henceforth without loss of generality,
\beq
\frac{d \rho_{J}(x)}{dx} = - 2J \sqrt{1+2\rho_{J}^2(x) K(J)}  \, .
\label{optprofKMPmonot}
\eeq
On the other hand, for $K(J)<0$ the  r.h.s. of Eq. (\ref{optprofKMP}) may vanish at some points, resulting in a $\rho_{J}(x)$ that is non-monotone and takes an unique value $\rho_J^*\equiv \sqrt{-1/2K(J)}$ in the extrema. Notice that the r.h.s. of Eq. (\ref{optprofKMP}) may be written in this case as $4J^2[1-(\rho_{J}(x)/\rho_J^*)^2]$. It is then clear that, if non-monotone, the profile $\rho_{J}(x)$ can only have a single maximum because: (i) $\rho_{J}(x)\le \rho_J^*$ $\forall x\in[0,1]$ for the profile to be a real function, and (ii) several maxima are not possible because they should be separated by a minimum, which is not allowed because of (i). Hence for the non-monotone case (recall $T_L>T_R$)
\begin{eqnarray}
\frac{d \rho_{J}(x)}{dx} \!=\!
\left\{ \! \begin{array}{cc}
+2J \sqrt{1- \left(\frac{\displaystyle \rho_{J}(x)}{\displaystyle \rho_J^*}\right)^2} \, ,&
\, {\displaystyle x<x^* } \\ \\
- 2J \sqrt{1-\left(\frac{\displaystyle \rho_{J}(x)}{\displaystyle \rho_J^*}\right)^2} \, ,&
\, {\displaystyle x>x^* }
\end{array}
\right.
\label{optprofKMPnomonot}
\end{eqnarray}
where $x^*$ locates the profile maximum. This leaves us with two separated regimes for current fluctuations, with the crossover happening for $J_0\equiv \frac{T_L}{2}\left[\frac{\pi}{2} - \sin^{-1}\left(\frac{T_R}{T_L}\right) \right]$.
This crossover current can be obtained from Eq. (\ref{jnomon}) below by letting $\rho_J^*\rightarrow T_L$

\subsection{Region I: Monotonous Regime ($J<J_0$)}

In this case, using Eq. (\ref{optprofKMPmonot}) to change variables in Eq. (\ref{ch1:ldf2}) we have
\beq
G({\bf{J}})=\mint_{T_L}^{T_R} d\rho_{J} \dfrac{1}{4J\rho_{J}^2\sqrt{1+2K(J)\rho_{J}^2}}\left[\left(J_x-J\sqrt{1+2K(J)\rho_{J}^2}\right)^2+J_y^2\right] \, ,
\eeq
with $J_{\alpha}$ the $\alpha$ component of vector $\vJJ$. This results in
\beq
G({\bf{J}})=\dfrac{J_x}{2}\left(\dfrac{1}{T_R}-\dfrac{1}{T_L}\right)-J^2K(J)+\dfrac{J}{2}\left[\dfrac{\sqrt{1+2K(J)T_L^2}}{T_L}-\dfrac{\sqrt{1+2K(J)T_R^2}}{T_R} \right]
\label{fjmon}
\eeq
Notice that, for $\rho_{J}(x)$ to be monotone, $1+2K(J)\rho_{J}^2>0$. Thus, $K(J)>-(2T_L^2)^{-1}$. Integrating now
Eq. (\ref{optprofKMPmonot}) we obtain the following implicit equation for $\rho_{J}(x)$ in this regime
\begin{eqnarray}
2xJ \!=\!
\left\{ \! \begin{array}{cc}
{\displaystyle \frac{1}{\sqrt{2K(J)}} \ln \left[ \frac{ T_L + \sqrt{T_L^2 + \frac{1}{2K(J)}}}{ \rho_{J}(x) +
\sqrt{\rho_{J}(x)^2 + \frac{1}{2K(J)}}} \right]\, , } &
\, {\displaystyle K(J)>0 } \\ \\
{\displaystyle \frac{\sin^{-1}\left[T_L\sqrt{-2K(J)}\right] - \sin^{-1}\left[\rho_{J}(x)\sqrt{-2K(J)}\right]}{\sqrt{-2K(J)}} \ , } &
{\displaystyle -\frac{1}{2T_L^2}<K(J)<0 }
\end{array}
\right.
\label{profmonot}\hspace{-1.1cm}
\end{eqnarray}

Making $x=1$ and $\rho_{J}(x=1)=T_R$ in the previous equation, we obtain the implicit expression for the constant $K(J)$.
To get a feeling on how it depends on $J$, note that in the limit $K(J)\rightarrow (-1/2T_L^2)$, the
current $J\rightarrow J_0$,
while for $K(J)\rightarrow \infty$ one gets $J\rightarrow 0$. In
addition, from Eq. (\ref{profmonot}) we see that for $K(J)\rightarrow 0$ we find $J=(T_L-T_R)/2=\la \vJJ \ra$.

Sometimes it is interesting to work with the Legendre transform of the current LDF \cite{Derrida,Bertini,PabloPRL,PabloPRE,PabloAIP},
$\mu(\vla)=\max_{\vJJ}[G(\vJJ)+\vla\cdot\vJJ]$, where $\vla$ is a vector parameter conjugate to the current.
Using the previous results for $G(\vJJ)$ is is easy to show \cite{PabloAIP} that $\mu(\vla)=-K(\vla)\vJJ^*(\vla)^2$, where
$\vJJ^*(\vla)$ is the current associated with a given $\vla$, and the constant $K(\vla)=K(|\vJJ^*(\vla)|)$. The expression for $\vJJ^*(\vla)$
can be obtained from Eq. (\ref{profmonot}) above in the limit $x\to 1$. Finally, the optimal profile for a given $\vla$ is
just $\rho_{\vla}(x)=\rho_{|\vJJ^*(\vla)|}(x)$.

\subsection{Region II: Non-Monotonous Regime ($J>J_0$)}

In this case the optimal profile has a single maximum $\rho_J^*\equiv \rho_{J}(x=x^*)$ with $\rho_J^*=1/\sqrt{-2K(J)}$ and $-1/2T_L^2<K(J)<0$.
Splitting the integral in Eq. (\ref{ch1:ldf2}) at $x^*$, and using now Eq. (\ref{optprofKMPnomonot}) to change variables, we arrive at
\begin{eqnarray}
G({\bf{J}}) & = & \dfrac{J_x}{2}\left(\dfrac{1}{T_R}-\dfrac{1}{T_L} \right)
- \dfrac{J}{2}\Bigg[\dfrac{1}{T_L}\sqrt{1-\left(\dfrac{T_L}{\rho_J^*}\right)^2} +
\dfrac{1}{T_R}\sqrt{1-\left(\dfrac{T_R}{\rho_J^*}\right)^2} \nonumber \\
 & - & \dfrac{1}{2\rho_J^*}\left(\pi-\sin^{-1}\left(\dfrac{T_L}{\rho_J^*}\right)-\sin^{-1}\left(\dfrac{T_R}{\rho_J^*}\right) \right)\Bigg] \, .
\end{eqnarray}
Integrating Eq. (\ref{optprofKMPnomonot}) one gets an implicit equation for the non-monotone optimal profile
\begin{eqnarray}
2xJ \!=\!
\left\{ \! \begin{array}{cc}
{\displaystyle \rho_J^* \left[ \sin^{-1} \left( \frac{\rho_{J}(x)}{\rho_J^*} \right) - \sin^{-1} \left( \frac{T_L}{\rho_J^*} \right) \right] } &
\quad {\displaystyle \text{for } 0\le x<x^* } \\ \\
{\displaystyle 2J + \rho_J^* \left[ \sin^{-1} \left( \frac{T_R}{\rho_J^*} \right) - \sin^{-1} \left( \frac{\rho_{J}(x)}{\rho_J^*} \right) \right]  } &
\quad {\displaystyle \text{for } x^*<x\le 1 }
\end{array}
\right.
\label{profnomonot}
\end{eqnarray}
At $x=x^*$ both branches of the above equation must coincide, and this condition provides simple equations for both $x^*$ and $\rho_J^*$
\beq
J=\dfrac{\rho_J^*}{2}\left[\pi-\sin^{-1}\left(\dfrac{T_L}{\rho_J^*}\right)-\sin^{-1}\left(\dfrac{T_R}{\rho_J^*}\right)\right] \, \, \, ;
\, \, \, x^*=\dfrac{\dfrac{\pi}{2}-\sin^{-1}\left(\dfrac{T_L}{\rho_J^*} \right) }{\pi-\sin^{-1}\left(\dfrac{T_L}{\rho_J^*}\right)-\sin^{-1}\left(\dfrac{T_R}{\rho_J^*}\right)} \, .
\label{jnomon}
\eeq
Finally, as in the monotone regime, we can compute the Legendre transform of the current LDF, obtaining as before $\mu(\vla)=-K(\vla)\vJJ^*(\vla)^2$, where $\vJJ^*(\vla)$  and the constant $K(\vla)=K(|\vJJ^*(\vla)|)$ can be easily computed from the previous expressions \cite{PabloPRE}. Note that, in $\vla$-space, monotone profiles are expected for $|\vla+\vecep| \leq \frac{1}{2 T_R}\sqrt{1-\left(\frac{T_R}{T_L} \right)^2}$, where $\vecep=(\frac{1}{2}\left(T_L^{-1}-T_R^{-1} \right),0)$ is the driving force in this case,
while non-monotone profiles appear for $\frac{1}{2 T_R}\sqrt{1-\left(\frac{T_R}{T_L} \right)^2} \leq |\vla+\vecep| \leq \frac{1}{2}\left(\frac{1}{T_L}+\frac{1}{T_R} \right)$.

\begin{figure}
\centerline{
\includegraphics[width=9.5cm]{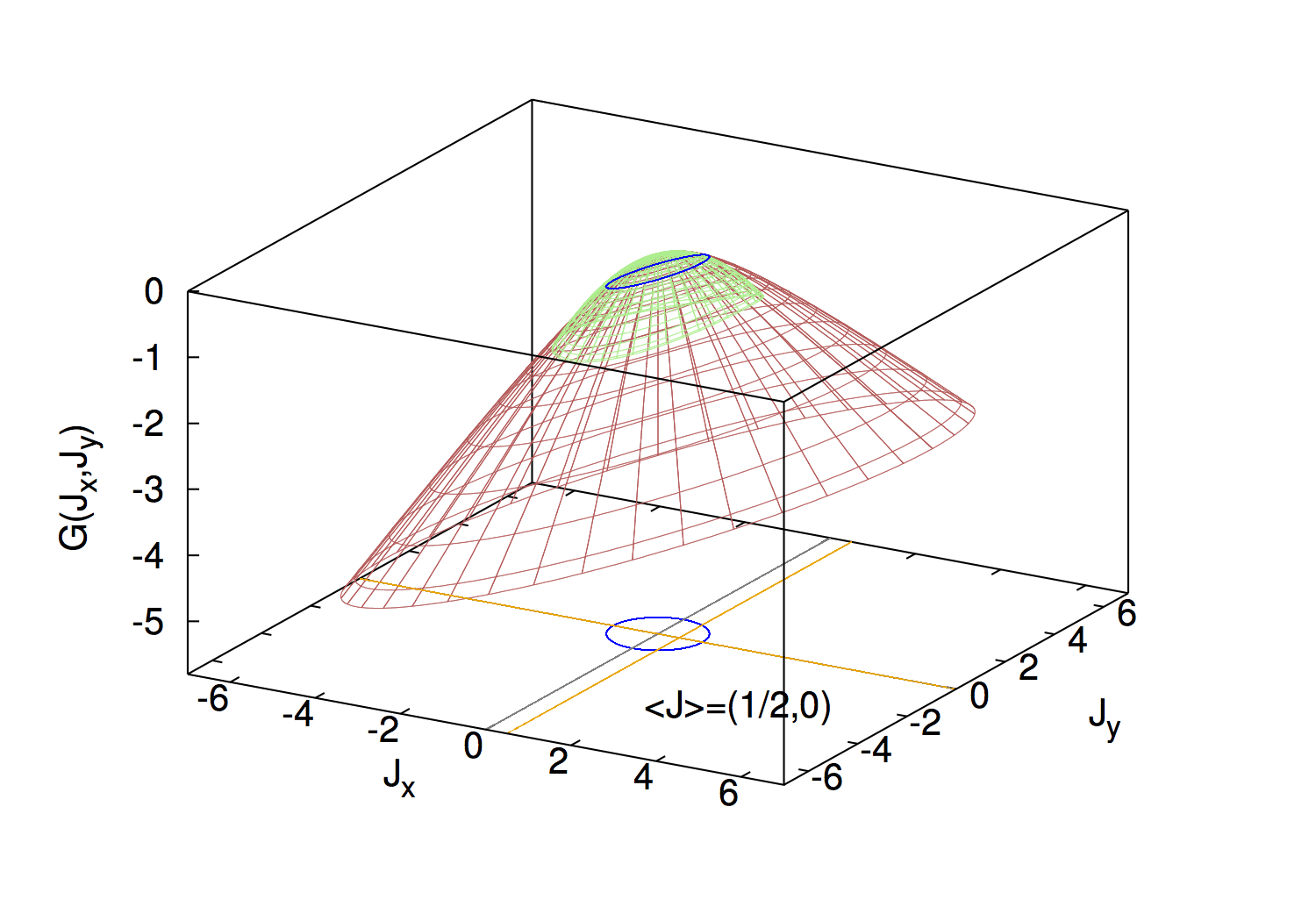}
\hspace{-1.5cm}
\includegraphics[width=9.5cm]{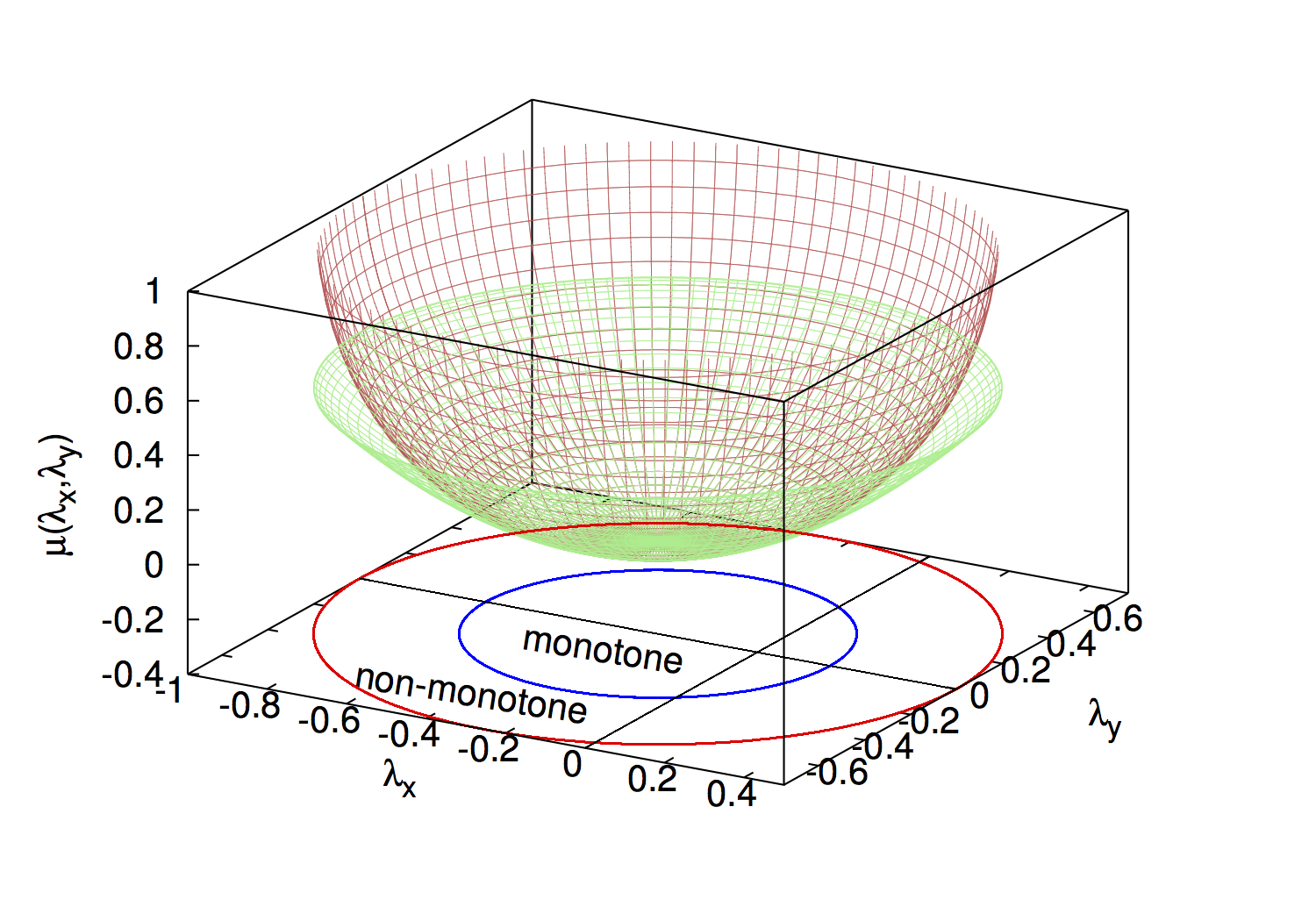}
}
\caption{(Color online) Left panel: $G({\bf{J}})$ for the 2D-KMP model for $T_L=2$ and $T_R=1$. The blue circle signals the crossover from monotone ($J<J_0\equiv\pi/3$) to non-monotone ($J>\pi/3$) optimal profiles. The green surface corresponds to the Gaussian approximation for small current fluctuations. Right panel: Legendre transform of both $G(\vJJ)$ (brown) and $G(\vJJ \rightarrow \la \vJJ \ra)$ (green), together with the projection in $\vla$-space of the crossover between monotonus and non-monotonous regime.}
\label{ldfpredic}
\end{figure}

Fig. \ref{ldfpredic} shows the predicited $G(\vJJ)$ (left panel) and its Legendre trasform (right panel) for the 2D-KMP model. Notice that the LDF is zero for $\vJJ=\la \vJJ \ra=((T_L-T_R)/2,0)$ and negative elsewhere. For small current fluctuations, $\vJJ\approx \la \vJJ\ra$, $G(\vJJ)$ obeys the following quadratic form
\beq
G(\vJJ)\approx-\dfrac{1}{2}\left(\dfrac{(J_x-(T_L-T_R)/2)^2}{\sigma_x^2}+\dfrac{J_y^2}{\sigma_y^2}\right)\, ,
\label{Gsmallq}
\eeq
with $\sigma_x^2=(T_L^2+T_LT_R+T_R^2)/3$ and $\sigma_y^2=T_L T_R$, resulting in Gaussian statistics for currents near the average as expected from the central limit theorem. A similar expansion for the Legendre transform yields
\beq
\mu(\vla)\approx \dfrac{\lambda_x}{2}\left[(T_L-T_R)+\sigma_x^2\lambda_x \right]+\dfrac{\sigma_y^2}{2}\lambda_y^2.
\label{musmalllambda}
\eeq
Notice that beyond this restricted Gaussian regime, current statistics is in general non-Gaussian. In particular, for large enough current deviations, $G(\vJJ)$ decays linearly, meaning that the probability of such fluctuations is \emph{exponentially} small in $J$ (rather than $J^2$).

Despite the complex structure of $G(\vJJ)$ in both regime I and II above, it can be easily checked that for any pair of isometric current vectors
$\vJJ$ and $\vJJ'$, such that $|\vJJ|=|\vJJ'|$, the current LDF obeys
\beq
G(\vJJ)-G(\vJJ')=\vecep \cdot (\vJJ-\vJJ') \, .
\label{IFR1}
\eeq
This relation, known as Isometric Fluctuation Relation, is not particular for the 2D KMP model but a general results for time-reversible diffusve systems, see \S\ref{sIFR}.

\section{Constants of motion}
\label{ap2}
A sufficient condition for the Isometric Fluctuation Relation (IFR), eqs. (\ref{IFT}) or (\ref{IFR1}), to hold is that
\be
\frac{\delta \vpi_1[\rho(\vrr)]}{\delta \rho(\vrr')} = 0 \, ,
\label{cond0}
\ee
with the functional $\vpi_1[\rho(\vrr)]$ defined in Eq. (\ref{ch1:defs1}) above. We have shown that condition (\ref{cond0}) follows from the time-reversibility of the dynamics, in the sense that the evolution operator in the Fokker-Planck formulation of Eq. (\ref{ch1:langevin}) obeys a local detailed balance condition, see Eq. (\ref{ch4:cond1}). Condition (\ref{cond0}) implies that $\vpi_1[\rho(\vrr)]$ is in fact a \emph{constant of motion}, $\vecep$, independent of the profile $\rho(\vrr)$. Therefore we can use an arbitrary profile $\rho(\vrr)$, compatible with boundary conditions, to compute $\vecep$. We now choose boundary conditions to be gradient-like in the $\hat{x}$-direction, with densities $\rho_L$ and $\rho_R$ at the left and right reservoirs, respectively, and periodic boundary conditions in all other directions. Given these boundaries, we now select a linear profile
\be
\rho(\vrr)=\rho_L + (\rho_R-\rho_L) x \, ,
\label{linprof}
\ee
to compute $\vecep$, with $x\in[0,1]$, and assume very general forms for the current and mobility functionals
\begin{eqnarray}
\vQ[\rho(\vrr)] & \equiv & D_{0,0}[\rho] \vnabla \rho + \sum_{n,m>0} D_{nm}[\rho]  (\vnabla^m \rho)^{2n} \vnabla \rho \, , \nonumber \\
\sigma[\rho(\vrr)] & \equiv & \sigma_{0,0}[\rho] + \sum_{n,m>0} \sigma_{nm}[\rho] (\vnabla^m \rho)^{2n} \, , \nonumber
\end{eqnarray}
where as a convention we denote as $F[\rho]$ a generic functional of the profile but not of its derivatives. It is now easy to show that $\vecep=\varepsilon \hat{x}+\vE$, with
\be
\varepsilon = \int_{\rho_L}^{\rho_R} d \rho \frac{D_{0,0}(\rho)+\sum_{n>0} D_{n1}(\rho)(\rho_R-\rho_L)^{2n}}
{\sigma_{0,0}(\rho)+\sum_{m>0} \sigma_{m1}(\rho)(\rho_R-\rho_L)^{2m}} \, ,
\label{epsilon}
\ee
and $\hat{x}$ the unit vector along the gradient direction. In a similar way, if the following condition holds
\be
\frac{\delta}{\delta \rho(\vrr')} \int \vQ[\rho(\vrr)] d\vrr = 0 \, ,
\label{cond2}
\ee
together with time-reversibility, Eq. (\ref{cond0}), the system can be shown to obey an extended Isometric Fluctuation Relation which links any current fluctuation $\vJJ$ in the presence of an external field $\vE$ with any other isometric current fluctuation $\vJJ'$ in the presence of an arbitrarily-rotated external field $\vE^*$, and reduces to the standard IFR for $\vE=\vE^*$, see Eq. (\ref{ch4:IFR2}) in the main text. Condition (\ref{cond2}) implies that $\vnu\equiv\int \vQ[\rho(\vrr)] d\vrr$ is another constant of motion, which can be now written as $\vnu=\nu\hat{x}$, with
\be
\nu = \int_{\rho_L}^{\rho_R} d \rho \left[D_{0,0}(\rho)+\sum_{n>0} D_{n1}(\rho)(\rho_R-\rho_L)^{2n}\right] \, ,
\label{nu}
\ee
As an example, for a diffusive system $\vQ[\rho(\vrr)]=-D[\rho] \vnabla \rho(\vrr)$, with $D[\rho]$ the diffusivity functional, and the above equations yield the familiar results
\begin{eqnarray}
\varepsilon & = & \int_{\rho_R}^{\rho_L}  \frac{D(\rho)}{\sigma(\rho)}d\rho \, , \nonumber \\
\phantom{aaa} \nonumber \\
\nu & = & \int_{\rho_R}^{\rho_L} D(\rho)d\rho \, , \nonumber
\end{eqnarray}
for a standard local mobility $\sigma[\rho]$.

\end{document}